\documentclass[12pt,notitlepage]{book}

\RequirePackage{lineno}
\usepackage{amssymb, amsmath}
\usepackage{mathpazo}
\usepackage{float,graphicx}
\usepackage[figuresright]{rotating}
\usepackage[usenames,dvipsnames]{color}
\def\Put(#1,#2)#3{\leavevmode\makebox(0,0){\put(#1,#2){#3}}}
\usepackage{microtype}
\usepackage[font=small,labelfont=bf,margin=15pt]{caption}
\usepackage{tabulary}
\usepackage{booktabs,multirow,dcolumn,bigdelim}

\usepackage[pdftex,plainpages=false,pdfpagelabels,backref,pdfborder={0 0 0}]{hyperref}
\usepackage{url}
\usepackage[toc,page,titletoc]{appendix}
\usepackage{afterpage}
\usepackage[scaled]{helvet}
\usepackage{sectsty}
\allsectionsfont{\normalfont\sffamily}
\usepackage{titletoc}
\titlecontents{chapter}
  [1.5em]
  {\linespread{0.9}\normalfont\sffamily\bfseries}
  {\contentslabel{1em}}
  {\hspace*{-2.3em}}
  {\mdseries\titlerule*[1pc]{.}\contentspage}
\titlecontents{section}
  [3.5em]
  {\linespread{0.9}\normalfont\sffamily}
  {\contentslabel{2.3em}}
  {\hspace*{-2.3em}}
  {\titlerule*[1pc]{.}\contentspage}
\titlecontents{subsection}
  [4.5em]
  {\linespread{0.9}\normalfont\sffamily}
  {\contentslabel{3em}}
  {\hspace*{-2.3em}}
  {\titlerule*[1pc]{.}\contentspage}
\titlecontents{subsubsection}
  [5.5em]
  {\linespread{0.9}\normalfont\sffamily}
  {\contentslabel{2.3em}}
  {\hspace*{-2.3em}}
  {\titlerule*[1pc]{.}\contentspage}
\usepackage[text={6.5in,8.75in},headheight=15pt,centering]{geometry}
\usepackage{xspace}
\graphicspath{{figs/}}

\DeclareGraphicsExtensions{.pdf,.png}
\newcommand{\filenamedot}{.}
\usepackage{fancyhdr}
\fancypagestyle{plain}{%
  \fancyhf{}
  \fancyfoot[RO,LE]{\sffamily \thepage}
}
\usepackage{eurosym}

\newcommand {\pp}{\mbox{$p$$+$$p$}\xspace}

\newcommand{\geant}{\mbox{\sc Geant4}\xspace}

\newcommand{\pythia}{\mbox{\sc Pythia}\xspace}
\newcommand{\rapgap}{\mbox{\sc Rapgap}\xspace}
\newcommand{\milou}{\mbox{\sc Milou}\xspace}

\newcommand{\egoing}{\mbox{electron-going}\xspace}
\newcommand{\hgoing}{\mbox{hadron-going}\xspace}
\newcommand{\egodir}{electron-going direction\xspace}
\newcommand{\hgodir}{hadron-going direction\xspace}

\newcommand{\lyxdot}{.}
\usepackage{subfig}

\begin{document}


\frontmatter

\pagestyle{empty}

\renewcommand*\familydefault{\sfdefault}
{\sffamily
\vspace{4in}
\begin{center}
  \huge
   Concept for an Electron Ion Collider (EIC) detector built around the BaBar solenoid
\end{center}

\vspace{1cm}

\begin{figure}[H]
  \begin{center}
    \includegraphics[trim = 310 0 325 0, clip, width=\linewidth]{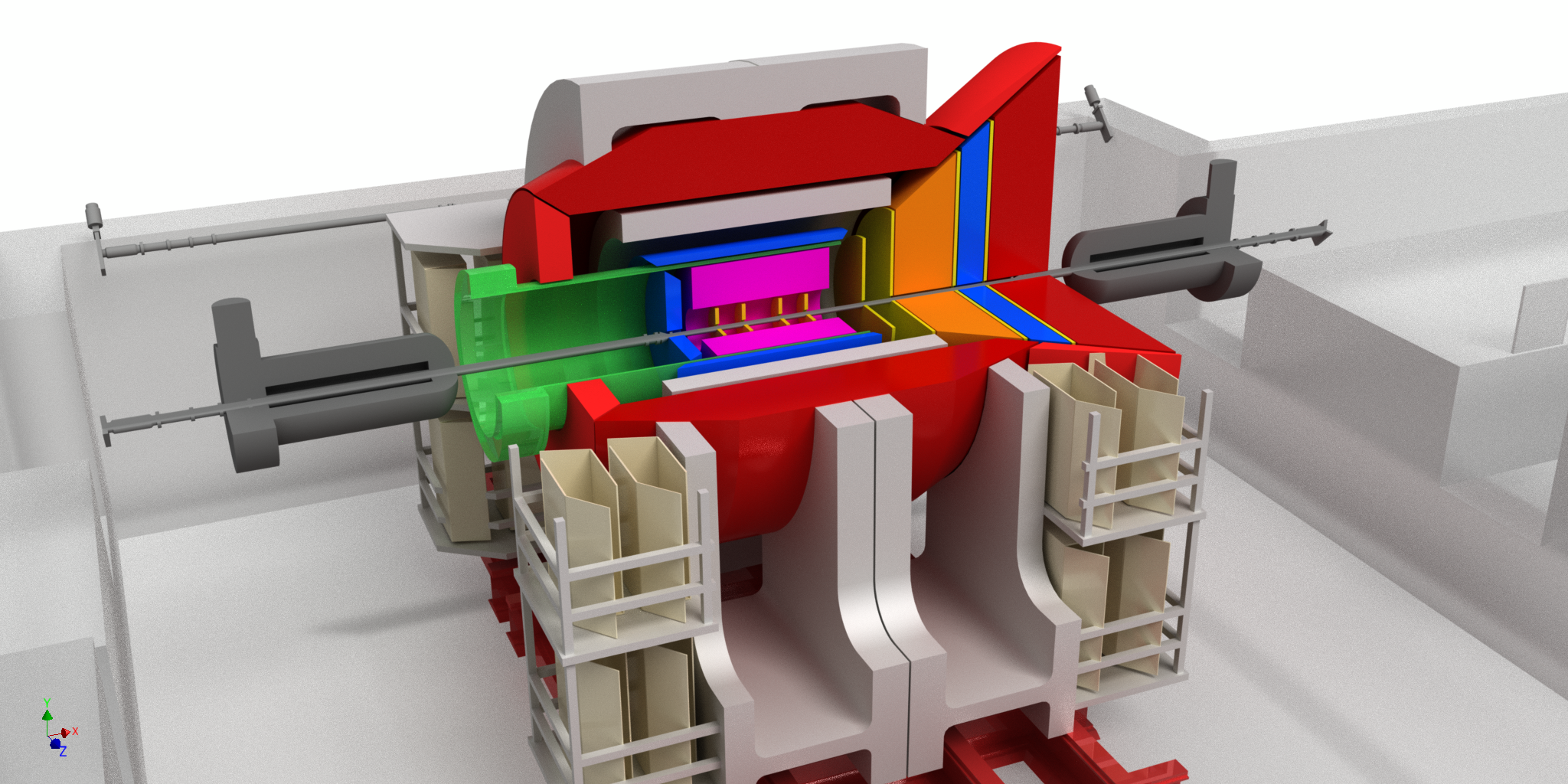}
  \end{center}
\end{figure}

\vspace{1cm}

\begin{center}
  \Large
  The PHENIX Collaboration \\
  February 3, 2014
\end{center}

}

\vfill
\renewcommand*\familydefault{\rmdefault}

\clearpage

\pagestyle{fancy}
\chapter*{Executive Summary}
\label{executive_summary}
\setcounter{page}{1}

The PHENIX collaboration presents here a concept for a detector at a
future Electron Ion Collider (EIC).  The exact performance
specifications for an EIC in terms of energy and luminosity are being
optimized.  This document responds to a specific charge to evaluate
the physics performance at an EIC at Brookhaven National Laboratory
(BNL) with a potential turn-on date of 2025 is envisioned with an
electron beam energy up to 10 GeV, hadron beam energies up to 255 GeV
for protons and 100 GeV/nucleon for gold ions, and design luminosities
of $10^{33}$ cm$^{-2}$s$^{-1}$ for 10 GeV on 255 GeV $e$$+$$p$
collisions.  The EIC detector proposed here, referred to as ePHENIX,
will have excellent performance for a broad range of exciting EIC
physics measurements, providing powerful investigations not currently
available that will dramatically advance our understanding of how
quantum chromodynamics binds the proton and forms nuclear
matter.Though not detailed in this study, this detector concept has
increased physics reach and capabilities for significantly higher
electron beam energies and luminosities with modest design
augmentation.

In 2013, the PHENIX collaboration and Brookhaven National Laboratory
submitted a proposal for an extensive update to the PHENIX detector.
This upgrade, referred to as sPHENIX, consists of new large acceptance
electromagnetic and hadronic calorimetry built around the
superconducting solenoid recently acquired from the decommissioned
BaBar experiment at SLAC.  sPHENIX will make key new measurements of
probes of the strongly coupled quark gluon plasma (sQGP) and allow for
fundamental tests of our picture of its inner
workings~\cite{Aidala:2012nz}.  From the beginning, it was realized
that the sPHENIX detector design, with its large bore superconducting
solenoid, midrapidity calorimetry, open geometry, coupled with the
existing investment in infrastructure in the PHENIX interaction
region, provides an excellent foundation for an EIC detector.  With
this in mind, EIC design considerations for the sPHENIX proposal have
been incorporated from the start~\cite{decadalPlan}.

\begin{figure}[p]
  \centering
  \includegraphics[trim = 280 150 240 140, clip, width=0.75\linewidth]{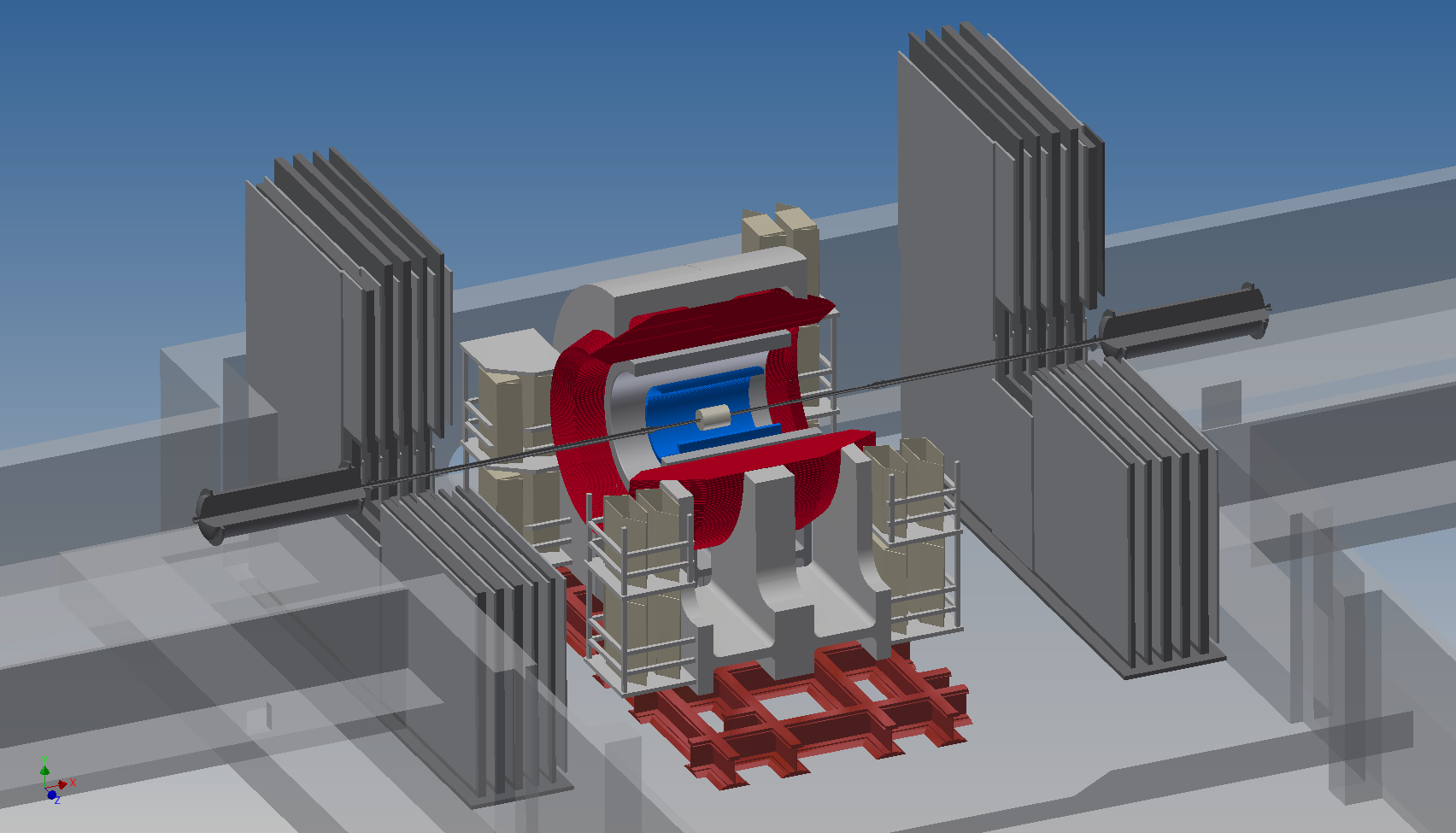}
  \vskip 0.1cm
  \includegraphics[width=0.75\linewidth]{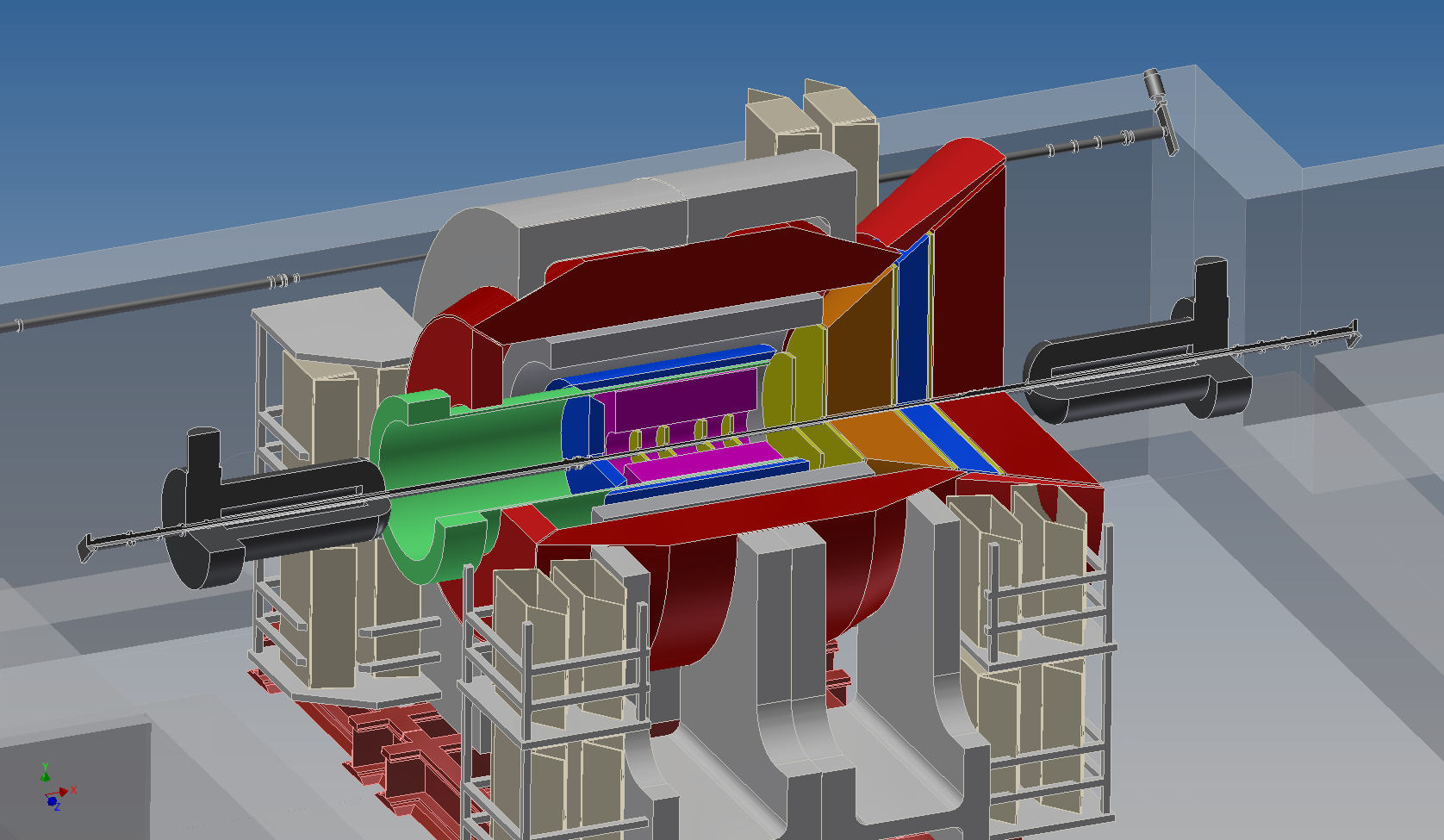}
  \caption{The evolution of the sPHENIX detector, with its focus on
    jets and hard probes in heavy-ion collisions, into ePHENIX, with
    additional capabilities supporting its focus on $e$$+$$p$ and
    $e$$+$A collisions. (top) The sPHENIX detector in the existing
    PHENIX experimental hall. (bottom) The ePHENIX detector, in the
    same hall, showing the reuse of the superconducting solenoid and
    the electromagnetic and hadronic calorimeter system. The eRHIC
    focusing quadrupoles, each located 4.5 m from the interaction
    point, and the height of the beam pipe above the concrete floor,
    set the dominant physical constraints on the allowable dimensions
    of ePHENIX.}
  \label{fig:ePHENIX}
\end{figure}

A full engineering rendering of the ePHENIX detector --- showing how
ePHENIX builds upon sPHENIX --- is shown in Figure~\ref{fig:ePHENIX}.
In addition to fully utilizing the sPHENIX superconducting solenoid
and barrel calorimetry, ePHENIX adds new detectors in the barrel and
\egoing and \hgoing directions.  In the \egoing direction a crystal
calorimeter is added for electron identification and precision
resolution.  A compact time projection chamber, augmented by
additional forward and backward angle GEM detectors, provides full
tracking coverage.  In the \hgoing direction, electromagnetic and
hadronic calorimetry sits behind the tracking detectors.  Critical
particle identification capabilities are provided in the central
rapidity region by a barrel DIRC and in the \hgoing direction by a gas
RICH and an aerogel RICH.  The sPHENIX upgrade could be ready for
physics data taking in 2020 and the ePHENIX additions available for
the earliest start of EIC physics in 2025.

The physics case for an EIC is documented in depth in the EIC White
Paper~\cite{Accardi:2012hwp}.  An EIC with 5--10 GeV electron beam
energies will enable major scientific advances in at least three main
areas: 1) Detailed imaging of the spin and momentum structure of the
nucleon; 2) Investigation of the onset of gluon saturation in heavy
nuclei; and 3) Study of hadronization in cold nuclear matter.  Again,
for higher energies not detailed in this document, an augmented
ePHENIX would extend this physics significantly including
investigations of fundamental symmetries.

In this document we review each area with a focus on the connection to
detector acceptance and performance requirements.  We consider each
subsystem in sufficient detail to be able to map out the performance
using both parametrized and full \geant simulations.  We find a broad
suite of observables where ePHENIX has excellent capabilities.

The ePHENIX detector capably addresses much of the physics enabled at
this EIC machine.  We believe we have struck a strong balance between
capabilities and costs for ePHENIX, but there remain clear targets for
augmenting those capabilities---for instance, by adding a silicon
vertex detector to enable measurements of open charm observables
(e.g., $F_2^c$).  In addition, there is a possibility to upgrade eRHIC
to higher energy electron beams at a future date, and we believe
ePHENIX provides an excellent base upon which an upgraded detector
capable of exploiting the physics potential of those collisions could
be built. There is also the potential, if one can realize appropriate
instrumentation in the \hgoing direction while \pp and $p$$+$A
collisions are still available in RHIC, to pursue a rich program of
forward physics measurements.

The PHENIX collaboration itself has outstanding detector expertise and
technical support as a base for the construction of an EIC detector.
Nonetheless, we view ePHENIX as a fundamentally new collaboration that
would require and welcome the addition of new institutions bringing
with them additional design and detector expertise, physics insights,
and scientific leadership.

This document is organized as follows.
Chapter~\ref{chap:physics_goals} illustrates the wide spectrum of EIC
physics that can be addressed.
Chapter~\ref{chap:ephenix_detector_requirements} describes the
detector requirements that follow from that physics and which drive
the ePHENIX design.  Chapter~\ref{chap:ephenix_detector_concept}
details the ePHENIX detector concept and shows its performance for key
measurements.

\clearpage

\resetlinenumber

\setcounter{tocdepth}{2}
\tableofcontents

\clearpage

\mainmatter

\renewcommand{\thepage}{\arabic{page}}
\setcounter{chapter}{0}
\setcounter{page}{1}

\chapter[EIC Physics]{Physics at an Electron-Ion Collider}
\label{chap:physics_goals}

The 2007 Nuclear Physics Long Range Plan~\cite{NSAC_LRP:2007} states
that the Electron-Ion Collider (EIC) embodies ``the vision for
reaching the next QCD frontier.''  In this Chapter we review the
primary physics goals as detailed in the EIC White
Paper~\cite{Accardi:2012hwp} and the broad physics program that can be
carried out with the ePHENIX detector.

\section{Fundamental questions addressed by the EIC}
\label{sec:eic-quest}

The EIC is designed to address several important question that are described in detail in the recent EIC White Paper~\cite{Accardi:2012hwp}.   Quoting
from the White Paper, these questions are reproduced here:
\begin{itemize}
\item {\bf How are the sea quarks and gluons, and their spins,
distributed in space and momentum inside the nucleon?} How are these
quark and gluon distributions correlated with overall nucleon
properties, such as spin direction?  What is the role of the orbital
motion of sea quarks and gluons in building the nucleon spin?
\item {\bf Where does the saturation of gluon densities set in?} Is
there a simple boundary that separates this region from that of more
dilute quark-gluon matter? If so, how do the distributions of quarks
and gluons change as one crosses the boundary? Does this saturation
produce matter of universal properties in the nucleon and all nuclei
viewed at nearly the speed of light?
\item {\bf How does the nuclear environment affect the distribution of
quarks and gluons and their interactions in nuclei?}  How does the
transverse spatial distribution of gluons compare to that in the
nucleon? How does nuclear matter respond to a fast moving color charge
passing through it?   What drives the time scale for color
neutralization and eventual hadronization?
\end{itemize}

The White Paper describes in detail the ``golden'' measurements in inclusive
Deep Inelastic Scattering (DIS), Semi-Inclusive DIS (SIDIS), and exclusive 
scattering at a future $e$$+$$p$ and $e$$+$A collider which will address the above 
questions employing a perfect detector.

\section{eRHIC: realizing the Electron-Ion Collider}

The accelerator requirements for an EIC that can answer the questions listed
above are spelled out in the EIC White Paper~\cite{Accardi:2012hwp}.  Two possible designs are presented based on current facilities:  (1) the 
eRHIC design, which adds a Energy Recovery LINAC to the existing RHIC complex 
at Brookhaven National Laboratory (BNL) which can accelerate polarized protons 
up to 250~GeV and ions such as gold up to 100~GeV/nucleon, and (2) the 
ELectron-Ion Collider (ELIC) design, which uses the 12~GeV Upgrade of CEBAF at Jefferson Laboratory with a 
new electron and ion collider complex.  

For the purposes of this document we focus on eRHIC with the following
design parameters:
\begin{itemize}
\item{A polarized electron beam with energy up to 10~GeV and polarization of
70\%,}
\item{A polarized proton beam with energy up to 250~GeV and polarization of 70\%,}
\item{An ion beam which can run a range of nuclei from deuteron to gold and 
uranium with energy up to 100~GeV/nucleon for gold,}
\item{Luminosity with a 10~GeV electron beam of $10^{33}$~cm$^{-2}$s$^{-1}$ 
for $e$$+$$p$ with 250~GeV proton beam energy, and 
 $6\times10^{32}$~cm$^{-2}$s$^{-1}$ for $e$$+$A with 100~GeV ion beams.} 
\end{itemize}

\section{Physics deliverables of ePHENIX}

The three fundamental and compelling questions in QCD to be addressed by the EIC discussed in
Section \ref{sec:eic-quest} can be broken down in to five golden measurements suggested in the 
EIC White Paper~\cite{Accardi:2012hwp}.

The first three relate to using the proton as a laboratory for fundamental QCD studies.
 
\begin{itemize}

\item {\bf The longitudinal spin of the proton:}   
With the good resolution calorimetry and tracking
in ePHENIX, Inclusive DIS measurements in polarized 
$e$$+$$p$ collisions will decisively determine the gluon and quark spin 
contributions to the proton spin.  
Further, planned particle identification capabilities will 
allow ePHENIX to pin down the spin contributions from the different 
quark flavors.

\item {\bf Transverse motion of quarks and gluons in the proton:} With the 
excellent particle identification capabilities of ePHENIX and the high luminosity of eRHIC, 
unparalleled SIDIS measurements will be possible, and enable
us to explore and understand how the intrinsic motion of partons in the 
nucleon is correlated with the nucleon or parton spin.

\item {\bf Tomographic imaging of the proton:}  The large acceptance of  
ePHENIX for tracking and calorimetry, far forward proton and neutron
detector capabilities,  the high luminosity of eRHIC and the phase space 
accessible in a collider geometry enables ePHENIX to significantly 
extend the kinematic coverage of exclusive measurements such as Deeply Virtual 
Compton Scattering (DVCS).  
With these, detailed images of how (sea) quarks and gluons are 
distributed in the proton will become possible for the first time.

\end{itemize}

The following two relate to extending these techniques to the heaviest stable nuclei.

\begin{itemize}

\item {\bf Hadronization and its modification in nuclear matter:}  With 
ePHENIX PID and the versatility of eRHIC to collide many different ions, 
measurements of identified hadrons in $e$$+$$p$ and $e$$+$A will allow 
precise study of how quarks hadronize in vacuum and in nuclear matter.

\item {\bf QCD matter at extreme gluon density:}  ePHENIX will enable   
measurements of diffractive and total DIS cross-sections 
in $e$$+$A and $e$$+$$p$.  Since the diffractive cross section is viewed as a double gluon
exchange process, the comparison of diffraction to total cross section in $e$$+$A and 
$e$$+$$p$ is a very sensitive indicator of the gluon saturation region. ePHENIX would
be an ideal detector to explore and study this with high precision. 

\end{itemize}

Below we discuss each of these points in more detail and with specific details on the
ePHENIX capabilities.

\subsection{The proton as a laboratory for QCD}

\begin{figure}
\centering
\includegraphics[width=0.8\textwidth]{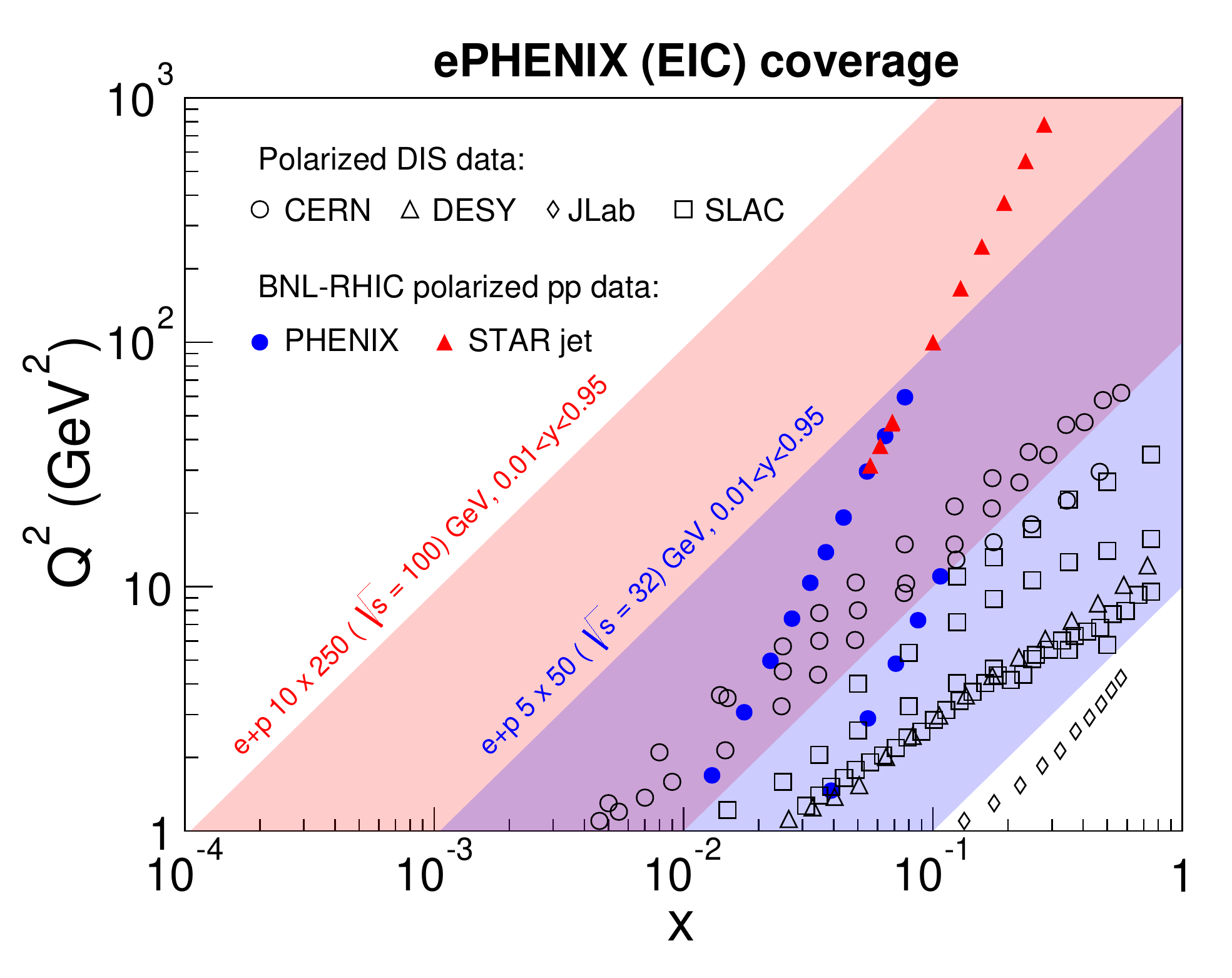}
\caption{\label{fig:kin_helicity}{Kinematic coverage of ePHENIX for two beam 
energy configurations, 10$\times$250~GeV and 5$\times$50~GeV, which show the range of eRHIC capabilities.  
Also shown are data from current polarized fixed target DIS experiments and RHIC $p$$+$$p$ collisions.}} 
\end{figure}

Deep Inelastic Scattering experiments over the last several decades 
have greatly enhanced our understanding of the proton substructure.  
Measurements with colliding beams at H1 and ZEUS at HERA have mapped out 
the momentum distributions of quarks and gluons, and shown that the 
gluons carry roughly half of the proton momentum.  Fixed target experiments,
with polarized nucleons and leptons at SLAC, CERN, DESY and JLab have 
revealed new surprises about proton structure, finding that only a small 
fraction of the proton spin comes from the quark spin and that there is 
significant 
correlation between the intrinsic motion of quarks and the nucleon spin.  
Measurements at both fixed target and colliders have started to image the 
proton through exclusive measurements.  

eRHIC will greatly enhance the kinematic coverage for DIS with polarized 
beams, as shown in Figure~\ref{fig:kin_helicity}.  With the capabilities of 
ePHENIX, we will significantly extend our understanding of the proton.  The 
gluon and flavor dependent sea quark spin contributions to the proton spin 
will be determined, as will the possible orbital angular momentum 
contributions.  The spatial and momentum distributions of (sea) quarks and 
gluons can be mapped, giving a multidimensional description of the proton.

\subsubsection{Longitudinal spin of the proton}
\label{sec:longitudinal_spin}

\begin{figure}
  \centering
  \includegraphics[width=0.75\textwidth]{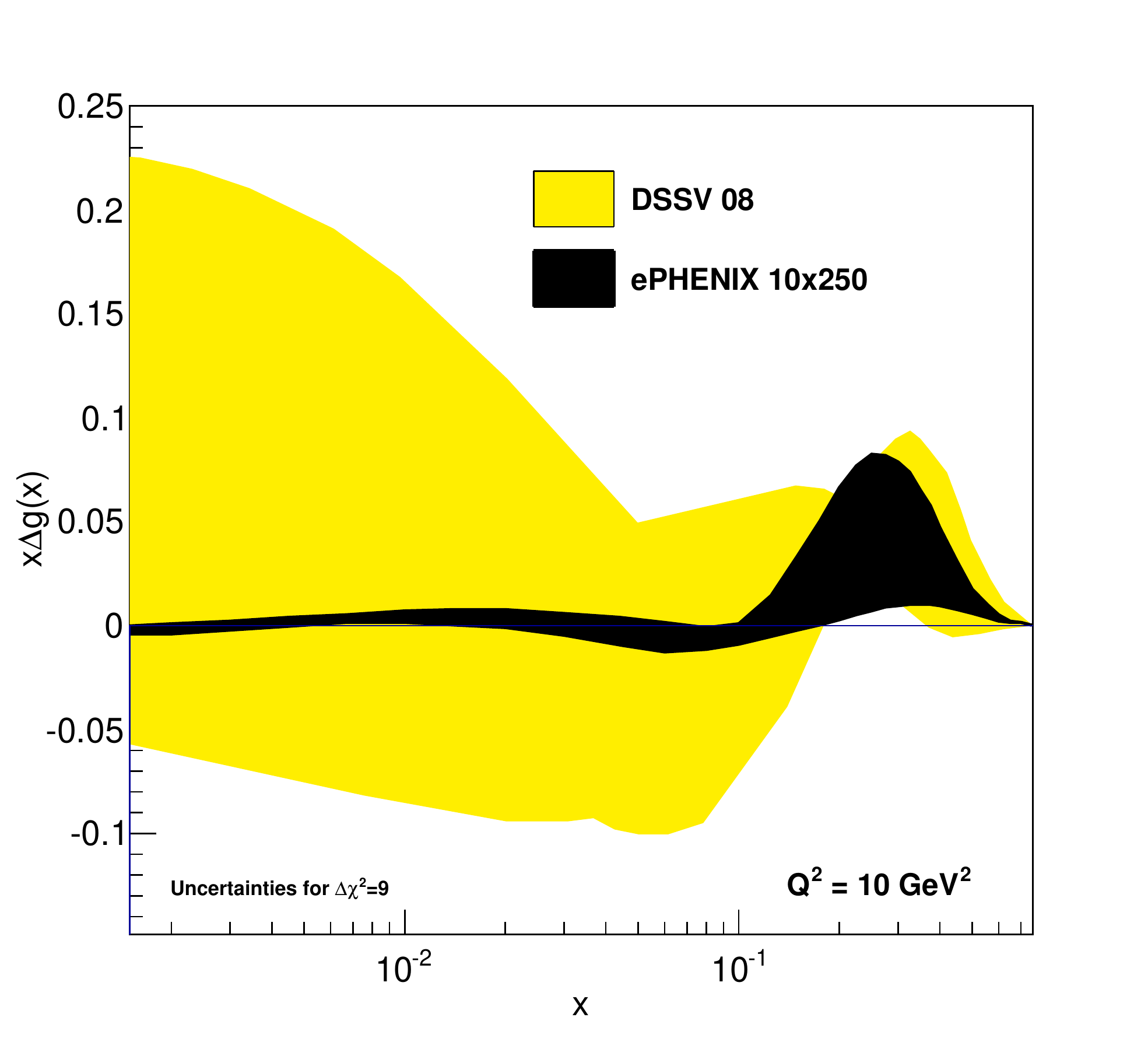}
  \caption{The projected reduction in the uncertainty (black) on the
    gluon longitudinal spin distribution based on simulated \pythia
    events corresponding to an integrated luminosity of 10~fb$^{-1}$
    at the 10~GeV $\times$ 250~GeV beam energy configuration.  A
    $1\%$ systematic uncertainty in beam and target polarization is
    applied.  The yellow area shows the uncertainty from current data
    based on the analysis in Ref.~\cite{deFlorian:2009vb}.}
  \label{fig:deltag}
\end{figure}

The puzzle of the proton spin, to which the quark spin only contributes 
roughly a third, has spurred two decades of study.  Measurements from fixed
target polarized DIS have determined the quark contribution, but are less 
sensitive to the gluon due to the small kinematic coverage.  Current RHIC 
measurements indicate that the gluon spin contribution may be comparable 
or even larger than the quark spin contribution, but 
due to the limited coverage at low longitudinal momentum fraction, $x$, 
large uncertainty remains, as is shown in Figure~\ref{fig:deltag} (yellow band).

Determining the gluon longitudinal spin contribution is a primary goal of the 
EIC and of ePHENIX, and will be possible due to the large reach in $x$ and 
four-momentum transfer squared, $Q^2$.  Figure~\ref{fig:deltag} shows the 
expected 
impact from ePHENIX measurements of inclusive DIS on the uncertainty of the 
gluon helicity distribution as a function of $x$.  

With the ePHENIX particle identification (PID) detectors, measurements of pions and
kaons will greatly improve on the determination of the sea quark longitudinal
spin distribution as well, including that of the strange quark, $\Delta s$,
which has been of particular interest in the last few decades,
because of the contradictory results obtained from different data.
Current global analyses use hyperon beta decay to constrain $\Delta s$,
which indicates a negative value for the full integral over $x$.  Fixed target SIDIS
measurements of kaon asymmetries, which directly probe $\Delta s$, though
at low values of $Q^2$ and in a limited $x$ range, find a positive contribution for
$x>0.01$.  eRHIC provides data over a wide $x$ and $Q^2$ range. Further, 
ePHENIX will provide excellent particle ID capability
to identify kaons and allow direct measurements of strangeness spin
contribution to the nucleon down to $\sim2\times10^{-4}$.

\subsubsection{Transverse motion of quarks and gluons in the proton}
\label{sec:transverse_motion}

\begin{figure}
\begin{center}
\begin{minipage}[b]{0.38\textwidth} \centering
{\hskip -0.1in}
\includegraphics[width=0.96\textwidth]{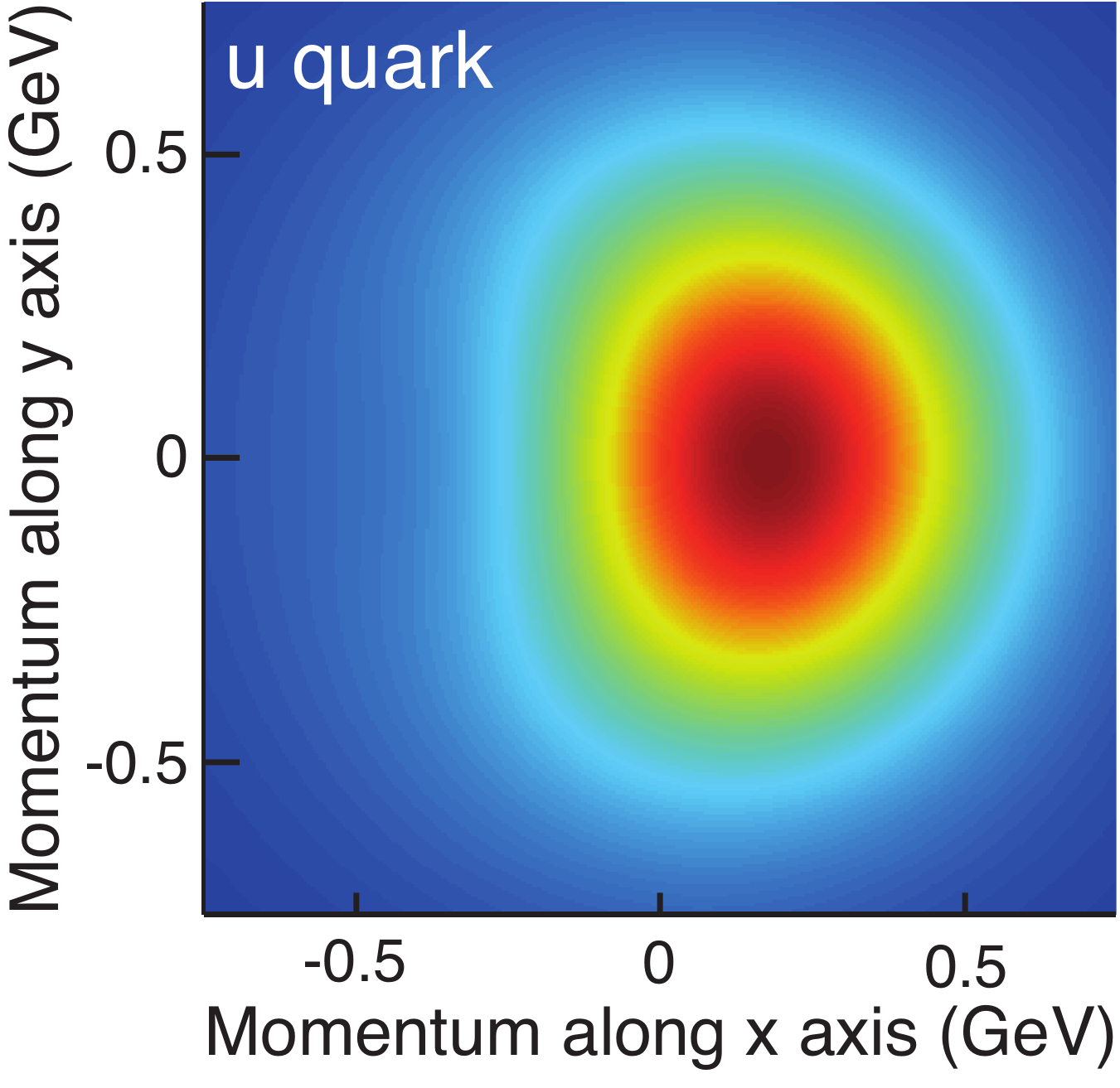}
\end{minipage} 
\hskip 0.05in
\begin{minipage}[b]{0.60\textwidth} \centering
\includegraphics[width=0.98\textwidth]{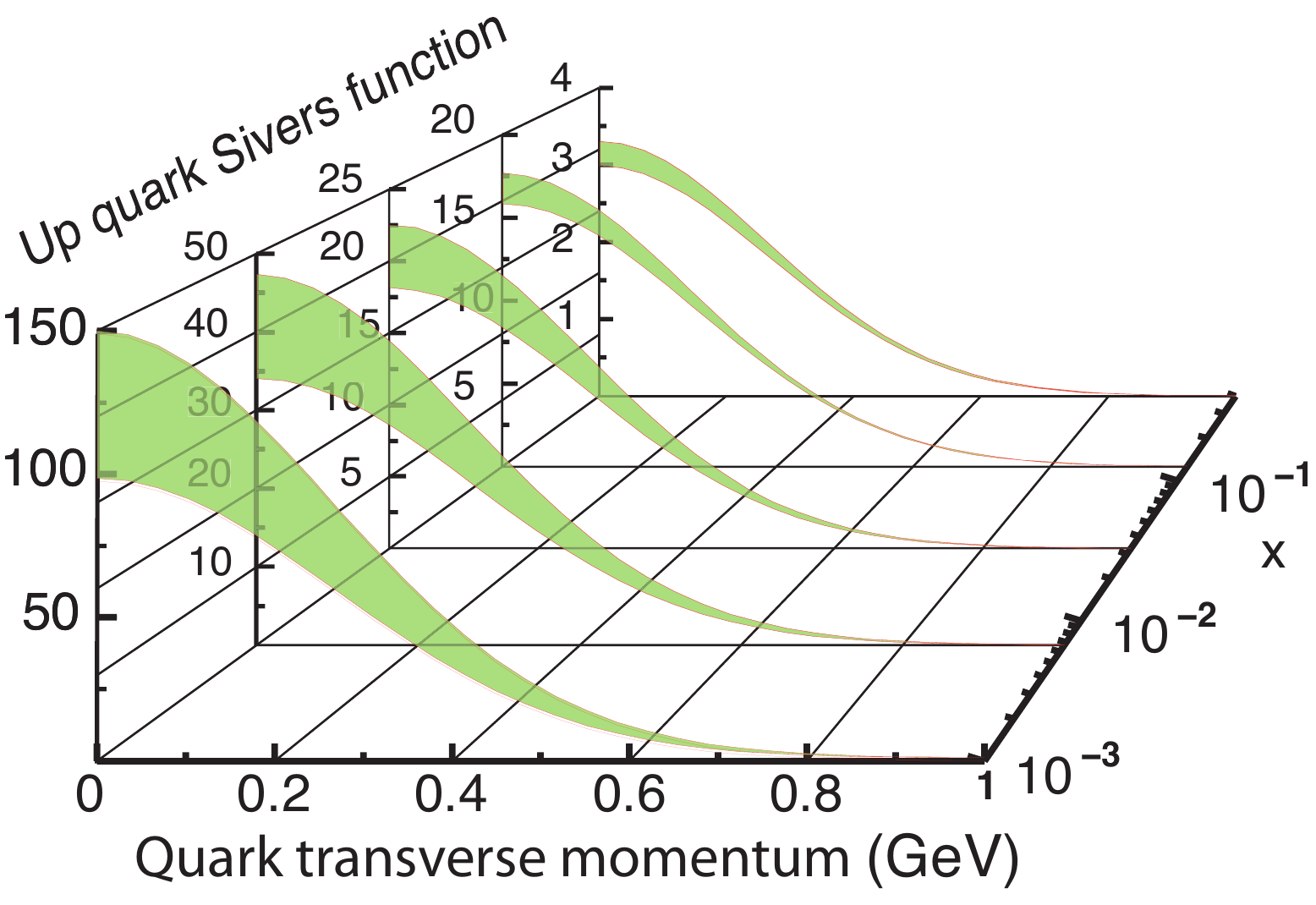}
\end{minipage}
\end{center} 
\vskip -0.05in
\caption{\label{fig:tmds}{ [Reproduced from Ref.~\cite{Accardi:2012hwp}.]
(left) The transverse-momentum
distribution of an up quark with longitudinal momentum fraction $x=0.1$
in a transversely polarized proton moving in the z-direction, while
polarized in the y-direction. The color code indicates the probability
of finding the up quarks.  (right) The transverse-momentum
profile of the up quark Sivers function at five $x$ values accessible
with the kinematics avialable at eRHIC, 
and corresponding statistical uncertainties.}
}
\end{figure}

Large transverse spin asymmetries measured in fixed target 
SIDIS in the past decade have spurred significant 
theoretical work.   These asymmetries relate to the 
transversity distribution, the correlation between the transverse spin of the 
proton and a transversely polarized quark in it, and Transverse Momentum 
Distributions (TMDs), such as the Sivers or Boer-Mulders distributions, which 
describe correlations between either the proton or quark spin and the quark 
intrinsic motion, specifically the transverse momentum of the quark.  With 
measurements of identified pions and kaons, these asymmetries give a 2$+$1 dimensional 
description of the spin and momentum distributions of different quark flavors 
in the proton, such as is shown in Figure~\ref{fig:tmds}.

Current measurements, however, are only able to probe a small 
region in $x$ and $Q^2$, limiting the description to the valence quark region.
Understanding of how the sea quarks and gluons contribute requires a larger 
kinematic range, such as provided at eRHIC. With the PID capabilities of 
ePHENIX, asymmetry measurements with transversely polarized nucleons and 
electrons in SIDIS will enable the study of these TMDs over most of this 
range, significantly expanding our knowledge of the proton structure.  
The 
constraint on the Sivers distributions was discussed in the EIC White 
Paper~\cite{Accardi:2012hwp}, with the expectations shown in Figure~\ref{fig:tmds}.
For the first time, determination of the Sivers distribution over a 
wide range in $x$ will be possible, including the low $x$ region where gluons 
dominate.

The transversity distribution, when coupled with the Collins fragmentation asymmetry, 
would result in an azimuthal asymmetry in the hadron production. This
has been called the Collins effect, and is a measurement that goes to the
heart of establishing the transversity distribution in a proton \cite{Anselmino:2007fs}. 
Measurement over the wide kinematic region would not only allow us to
measure transversity, but the wide $x$-coverage possible at eRHIC would 
afford the first reliable measurement of the tensor charge of the proton
(the integral over $x$ of the transversity distribution). No other currently
operational or planned facility can do this.

\subsubsection{Tomographic imaging of the proton}
\label{sec:tomographic_imaging}

Hard exclusive processes such as the Deeply Virtual Compton Scattering (DVCS) and
Deeply Virtual Vector Meson production (DVVM) involve interactions between the 
virtual photon and the partons in the proton without breaking the proton, resulting 
in the production of a real photon in DVCS or a vector meson in DVVM processes.
Just as elastic lepton-nucleon scattering gives information on the spatial 
distribution of the electric charge and magnetization in the nucleon, DVCS and DVVM 
processes probe the transverse distribution of quarks, anti-quarks and gluons. 
This information is encoded in generalized parton distributions (GPDs), 
which quantify the distributions of quarks and gluons in terms
of their positions in the transverse plane and longitudinal momentum fraction, 
providing 2$+$1 dimensional imaging of the nucleon. 
Measurements with polarized beams enable studies of spin-orbit correlations 
of quarks and gluons in the nucleon, by correlating the shift in the parton 
transverse distribution and proton transverse polarization. 
It is intuitively connected with orbital angular momentum carried 
by partons in the nucleon and hence of great interest in addressing the 
nucleon spin puzzle (nucleon spin decomposition)~\cite{Ji:1996ek}.

The existing data on GPDs from fixed target experiments cover only a 
limited kinematical range of $t$ (the squared momentum transfer to the 
proton), medium to high $x$ and low $Q^2$. 
The $t$ is connected through the Fourier transform with the impact parameter 
range probed.
While data from HERA collider experiments (ZEUS and H1) covered lower $x$ 
and a wide range in $Q^2$, they are statistically limited.  
Furthermore, the HERA proton beams were unpolarized, so ZEUS and H1 
were not able to study the proton-spin dependence in these measurements. 
With its large acceptance, excellent detection capabilities, 
high luminosity and broad range of energies of the polarized 
proton/helium beams available at eRHIC, ePHENIX will provide high precision 
data over a wide range of $x$, $Q^2$ and $t$. 
The wide range in $t$ possible at eRHIC is of crucial importance, 
and will be achieved by integrating Roman Pot detectors in the accelerator 
lattice from the outset. 
Similar measurements performed with ion beams will allow analogous imaging 
of nuclei, allowing the first look at the parton distributions inside 
the nuclei.

The EIC White Paper demonstrates the precision that can be achieved 
in such a program with Deeply Virtual Compton Scattering (DVCS) 
and exclusive $J/\psi$ production. 
The detector requirements for such measurements discussed in the
White Paper and what we propose as ePHENIX are similar. 
For such, we expect ePHENIX will be able to make high impact measurements 
of GPDs.




\subsection{Nucleus as a laboratory for QCD}

Electron scattering interactions from nuclei allow key tests of the modification of
parton distribution functions in nuclei of various sizes.  The EIC has the unprecedented energy
reach to probe deep into the low-$x$ quark and gluon region where there are predictions
of significant non-linear evolution effects and possibly the realization of a universal
state of the QCD vacuum at high gluon density.  
In addition, rather than looking at the modified number of deep inelastic scatterings, one can
study via SIDIS the changes in the process of a highly virtual struck quark to color neutralize and eventually
hadronize when in the presence of a nuclear medium.  

\subsubsection{Hadronization and its modification in nuclear matter}
\label{sec:hadronization}

\begin{figure}[h]
  \centering
  \includegraphics[width=0.49\linewidth]{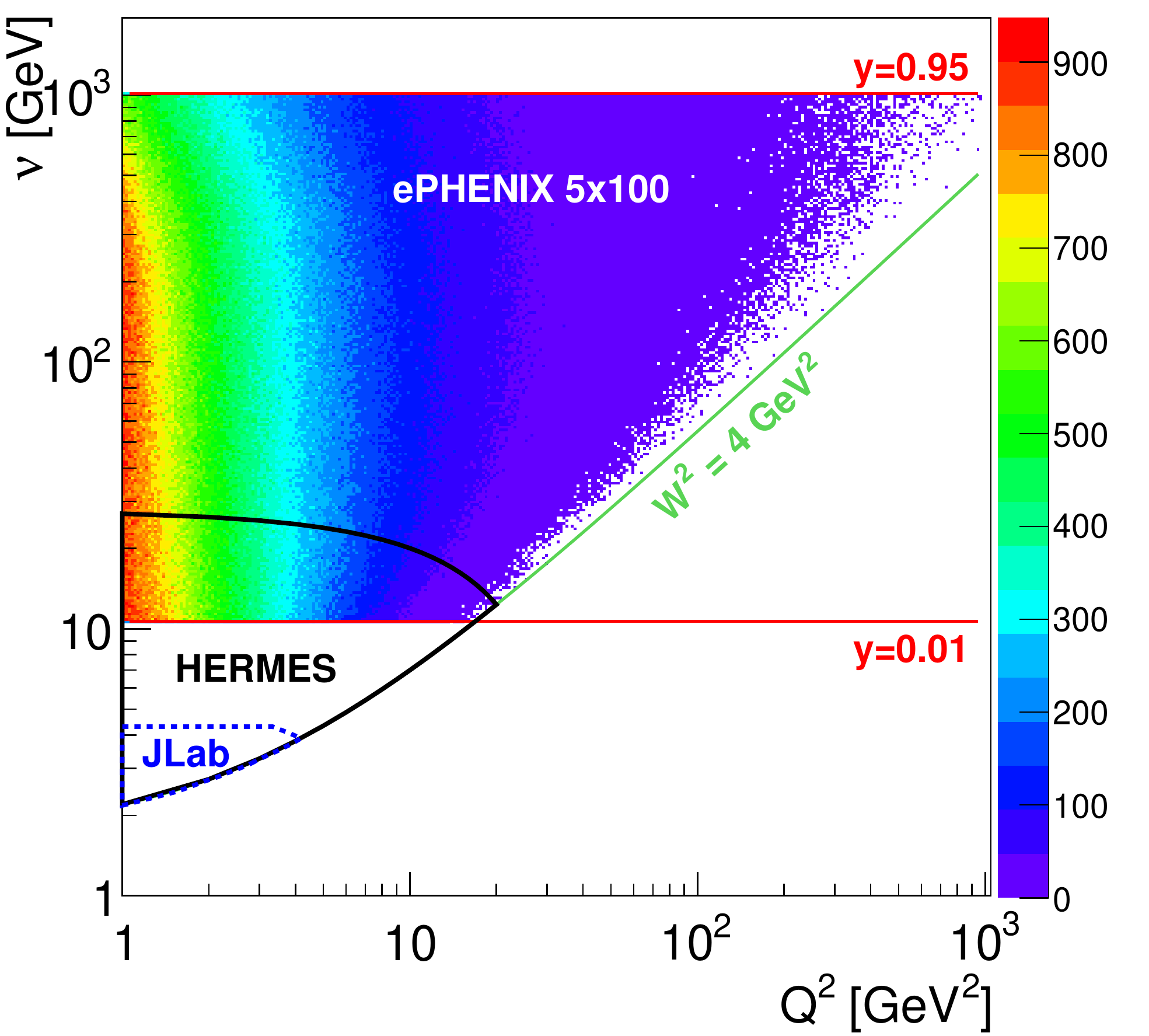}
  \includegraphics[width=0.49\linewidth]{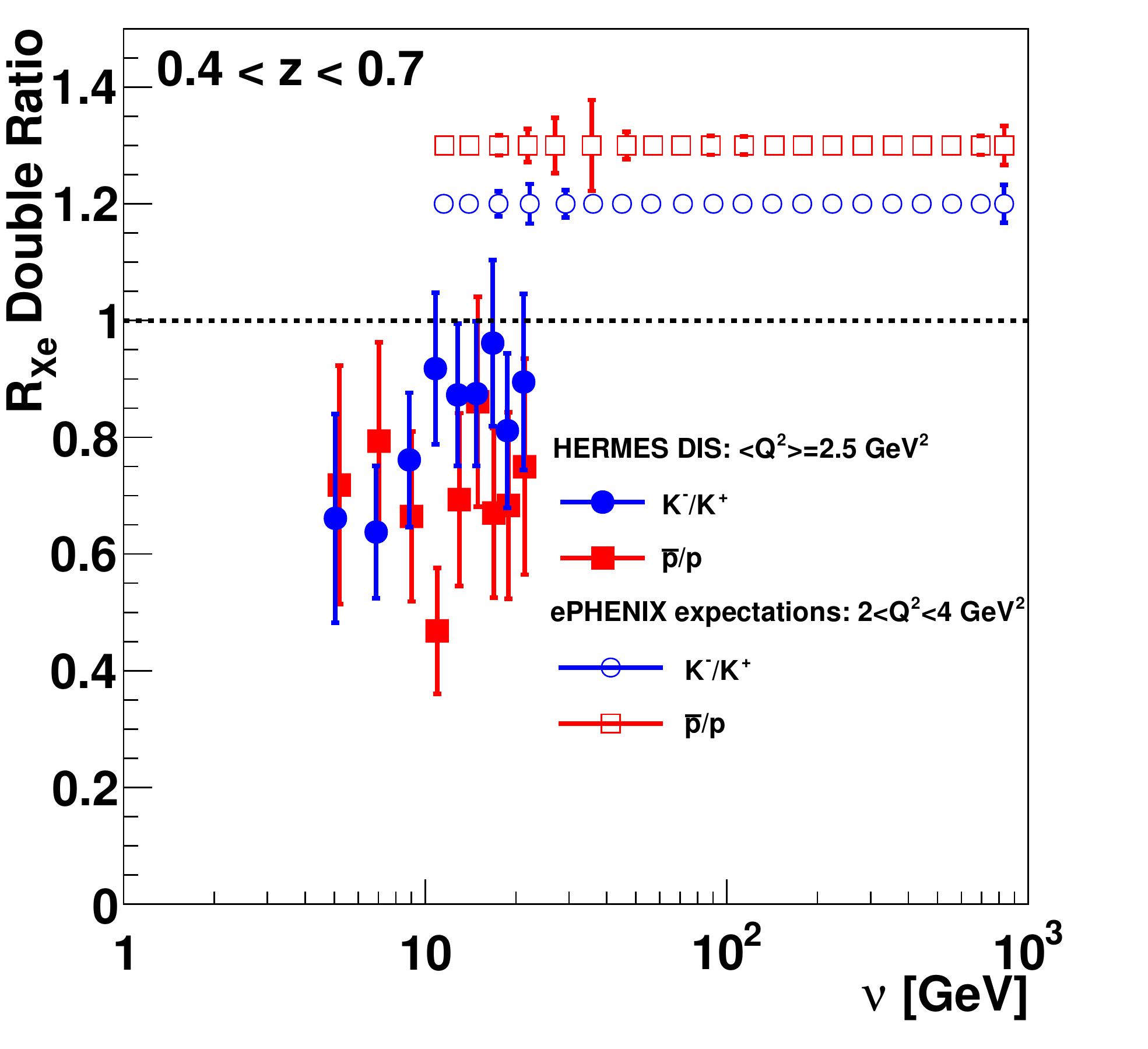}
  \caption{(left) Shown is the very large virtuality $Q^{2}$ and $\nu$
    coverage for ePHENIX (EIC) measurements with collisions of 5~GeV
    electrons on 100~GeV/nucleon heavy nuclei.  The z-axis color scale 
    shows the relative distribution of events from 
    the \pythia event generator.  Also
    shown are the kinematic reach for the CLAS experiment at
    JLab~\cite{Daniel:2011nq} and for the HERMES
    results~\cite{Airapetian:2007vu}.  (right) Experimental data from
    HERMES~\cite{Airapetian:2007vu} on the modified fragmentation from
    xenon targets ($R_{Xe}$) in the range $0.4 < z < 0.7$ and with
    average $\left<Q^{2}\right> = 2.5$~GeV$^{2}$.  The filled points are the
    double ratio for antiprotons relative to protons (red) and for
    $K^{-}$ relative to $K^{+}$ (blue).  ePHENIX will measure with
    precision the modified fragmentation distribution with excellent
    $\pi, K, p$ particle identification over a very broad range of
    $Q^{2}$ and $\nu$.  The open symbols show the expected statistical 
    precision for ePHENIX with its particle identification capabilities
    for one bin in $Q^2$, $2<Q^2<4$~GeV$^2$ based on 2~fb$^{-1}$ at the 
    5~GeV $\times$ 100~GeV beam energy configuration.}
  \label{fig:hadronization}
\end{figure}

Deep inelastic scattering with heavy nuclear targets provides an
effective stop watch and meter stick with which one can measure the
color neutralization and hadronization times, and understand important
details of partonic interactions with the nucleus.  By varying the
size of the nuclear target (at eRHIC all the way up to uranium) and
changing key DIS parameters ($Q^{2}, \nu, z, p_{T}^{2}, \phi$) one can
calibrate this watch and meter stick.  Figure~\ref{fig:hadronization}
shows the kinematic reach for 5~GeV electrons scattering from
100~GeV/nucleon heavy nuclei in terms of the initial virtuality
$Q^{2}$ and the energy of the struck quark in the nuclear rest frame
$\nu$.  Earlier experiments with fixed targets have measured very
interesting modifications in apparent fragmentation functions, and yet
those results are limited to small values of $Q^{2}$ and $\nu$.  In
the case of the published HERMES results~\cite{Airapetian:2007vu} in 
Fig.~\ref{fig:hadronization}, one
observes a dramatic decrease in the number of high-$z$ hadrons (those
with a large fraction of the struck quark momentum) in scattering from
nuclear targets.  There are many possible explanations of the
experimental results, including parton energy loss due to multiple
scattering in the nucleus and induced gluon radiation --- a similar
mechanism has been used to explain the ``jet quenching'' phenomena
discovered in heavy ion collisions at RHIC.  Other theoretical
frameworks predict a strong correlation between a short color
neutralization timescale and high-$z$ resulting processes.  An
excellent review of the various theoretical approaches is given in
Ref.~\cite{Boer:2011fh}.  Figure~\ref{fig:hadronization} also shows the 
expected statistical precision with the ePHENIX PID capabilities 
over the full $\nu$ range in one $Q^2$ bin.

If the struck quark remains an undressed color charge while it
traverses the nucleus, one might expect that the ratio of final state
hadrons ($\pi^{+}, K^{+}, p$ and their anti-particles) would show the
same degree of nuclear modification.  Shown in the right panel of
Figure~\ref{fig:hadronization} are the double ratios of modifications
$R_{Xe}$ with a xenon target for antiprotons to protons and $K^{-}$ to
$K^{+}$.  It is notable that there is a larger suppression for the
hadrons with a larger cross section with nucleons (e.g.
$\sigma_{\overline{p}+N} > \sigma_{p+N}$ and $\sigma_{K^{-}+N} >
\sigma_{K^{+}+N}$).  If this is due to hadronization occurring within
the nucleus, then inelastic collisions can result in the differential
attenuation.  How does this attenuation vary with the energy of the
struck quark?  The EIC realization has the enormous reach in the
energy of the struck quark $\nu$ at fixed $Q^{2}$ to measure the full
evolution with high statistics. As demonstrated in this document,
ePHENIX will have excellent $\pi, K, p$ particle identification to
make exactly this measurement with high statistics.  In addition, one
can vary the virtuality which is also expected to play a significant
role in the length scale probed in the nucleus and thus rate of
initial radiation.

Tests with charm mesons via displaced vertex measurements are not in the initial suite of ePHENIX capabilities,
and could be added with a later inner thin silicon detector.  Measurements of the interactions of charm quarks with the nucleus
would be quite interesting in the context of suppressed radiation due to the ``dead-cone'' effect.  However, the relation
to kinematic variables $z$ and $\nu$ may depend on the balance of DIS events from intrinsic charm as opposed to photon-gluon fusion 
reactions resulting in $c\overline{c}$ pair production.

\subsubsection{QCD matter at extreme gluon density}
\label{sec:saturation}

A key goal of any future EIC is to explore the gluonic matter at low $x$, 
where it is anticipated that the density of gluons will saturate as the rate 
of gluon recombination balances that of gluon splitting.  In fact, there are
well known modifications to the quark distribution functions in nuclei that
have  significant $x$ dependence:  high $x$ Fermi motion effects, then the EMC suppression,
anti-shadowing enhancement, and finally nuclear shadowing at the lowest $x$.  
The ePHENIX detector, combined with the large kinematic reach of an $e$$+$A collider, 
is in an excellent position to map this physics out in the gluon sector.

\begin{figure}[h!] 
  \begin{center}
    \includegraphics[trim = 0 0 -20 0, width=0.6\textwidth]{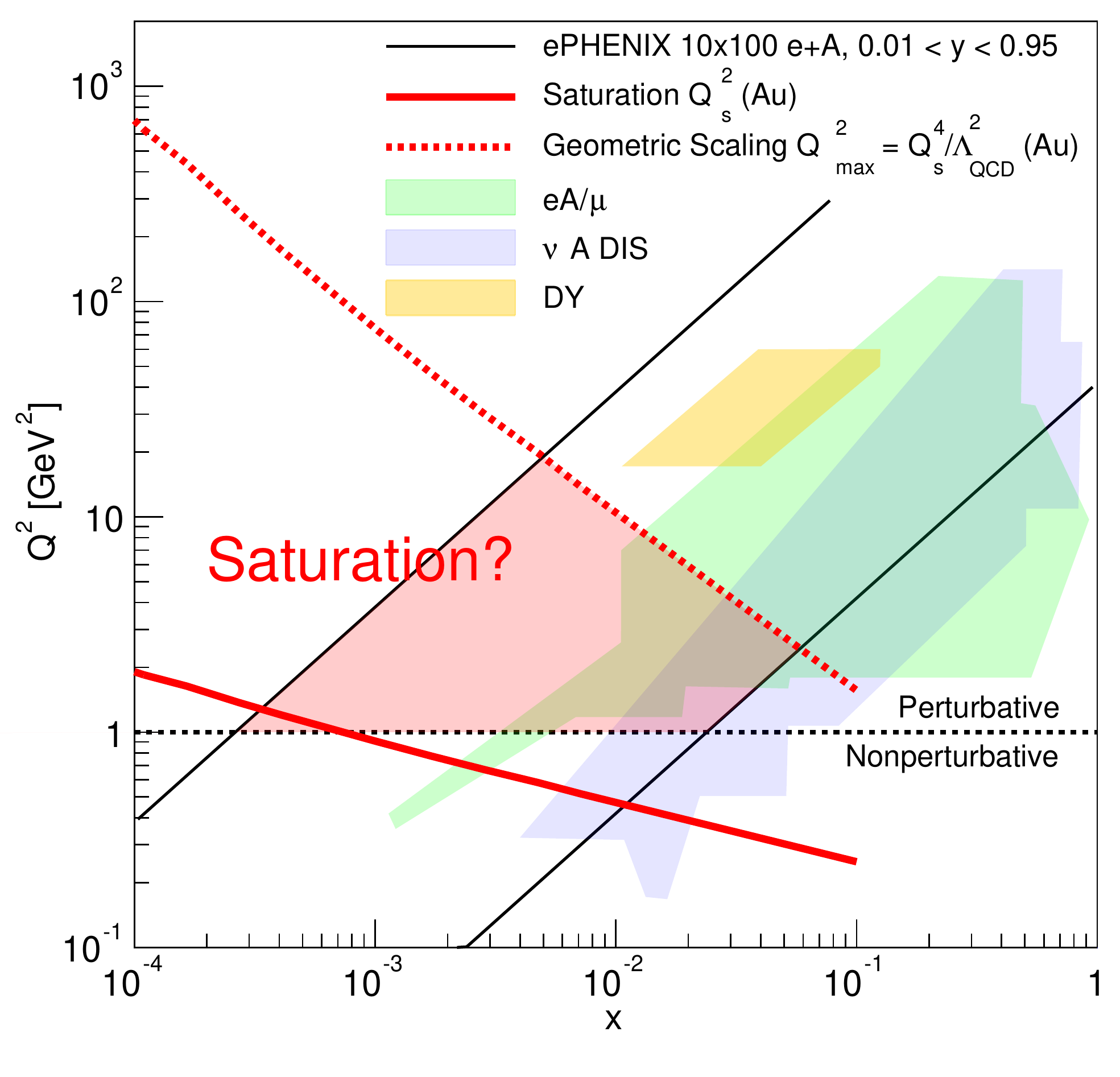}
  \end{center} 
  \caption{\label{fig:kin-wf} Shown is the coverage in $x$ and $Q^{2}$ for the EIC and the ePHENIX detector
    for 10 GeV electrons on 100 GeV/nucleon heavy nuclei.  The two black lines indicate the kinematic coverage
    with selections on the inelasticity $0.01 < y < 0.95$ (which might be slightly reduced depending on the final
    electron purity at low momentum).  Also shown are the kinematic coverage by previous experiments in $e$$+$A and $\nu$$+$A
    DIS and also Drell-Yan measurements.  The red solid line is an estimate of the $x$ dependence for the saturation scale
    $Q_{s}^{2}$.  The region where this universal saturated matter dictates the physics is estimated to extend over the
    geometric scaling region up to $Q^{2}_{max} = Q^{4}_{s}/\Lambda^{2}_{QCD}$ shown by the red dashed line~\cite{Iancu:2002tr}. 
}
\end{figure} 

The lowest $x$ regime with saturated gluon densities is unique to QCD,
as gluons carry the QCD charge, ``color'', and so interact with
themselves.  In order to explore this saturation region, one must
probe nuclear matter at high center-of-mass energy, so as to reach as
low in $x$ as possible while still in the perturbative QCD regime
(i.e., $Q^2>1$~GeV$^2$).  Generally, a saturation scale, $Q_s$, is
defined to indicate the onset of saturation (where the gluon splitting
and recombination balance each other), with $Q_s$ falling as $x$
increases.  In reality the point at which recombination starts to
balance the gluon splitting is a range in $x$ and $Q^2$ and so making
measurements over a wide range in $x$ and $Q^2$ is necessary to fully
understand these effects.

eRHIC will have a significantly lower center-of-mass energy than 
HERA, and so cannot improve upon the minimum $x$ 
probed with measurements in $e$$+$$p$.  However, eRHIC will also be 
capable of accelerating heavy ions in $e$$+$A collisions.  As 
the $x$ probed is related to the resolution of the probe, collisions 
at the same $Q^2$ can resolve significantly lower $x$ due to the 
larger extent of the nucleus: the partons in the highly accelerated
nucleus are probed coherently. This effectively reduces the $x$ 
probed in $e$$+$A collisions by a factor of $A^{\frac{1}{3}}$, with 
$A$ the atomic weight, as this is proportional to the size of the 
nucleus.  At the energies planned for eRHIC, based on measurements 
in $p(d)$$+$A, one expects saturation effects in inclusive DIS in $e$$+$A.

\begin{figure}[t]
  \begin{center}
    \includegraphics[width=0.6\textwidth]{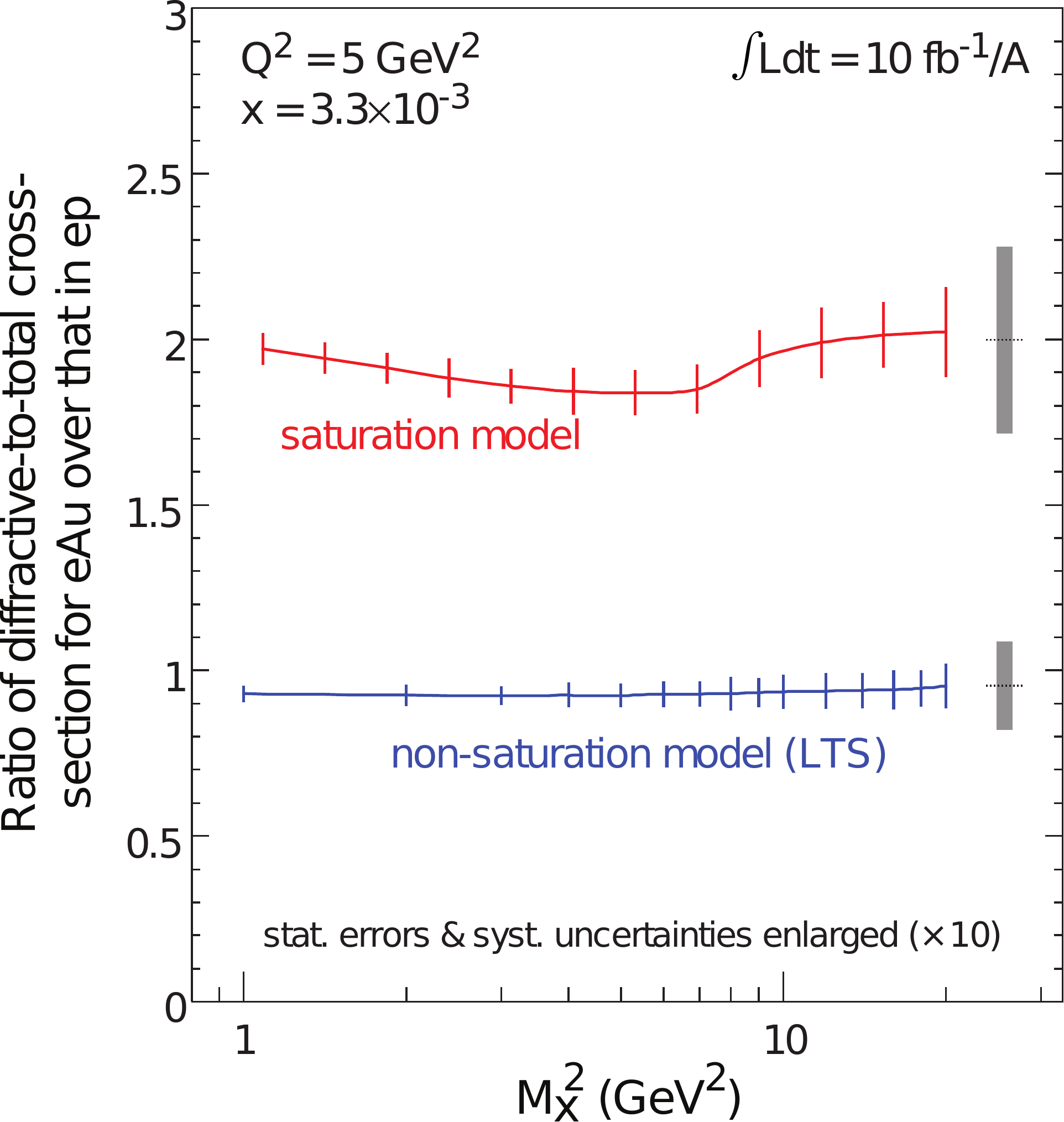}
  \end{center} 
  \caption{\label{fig:sat_diff}{[Reproduced from Ref.~\cite{Accardi:2012hwp}.]   Ratio of diffractive-to-total 
      cross-section for $e$$+$Au normalized to $e$$+$$p$ plotted as a function 
      of the squared mass of the hadronic final state, $M^2_X$.  The expected
      uncertainties for 10~fb$^{-1}$ are scaled by a factor of 10 to be visible.
      The ePHENIX detector will have similar capabilities as was assumed for 
      this plot, and will achieve similar precision.
    }  }
\end{figure} 

Figure~\ref{fig:kin-wf} 
shows the $x$ and $Q^2$ coverage of ePHENIX
for the 10~GeV~$\times$~100~GeV/nucleon configuration compared 
with the current fixed target data.  Two red lines are drawn, 
one (solid) showing expectations of $Q^2_s$ in $e$$+$Au and the other (dashed) 
showing the expected turn on of geometric scaling, which relates to the 
saturation scale by $Q^2_{max} = Q^4_s/\Lambda^2_{QCD}$.  The shaded red region
is where ePHENIX can search for saturation effects.

As described in the EIC White Paper~\cite{Accardi:2012hwp}, it can 
be even more effective to explore this region of dense
gluonic matter with diffractive physics, where at least two gluons are
exchanged in the interaction.  Therefore, a primary measurement to 
probe saturation effects at eRHIC will be comparing the diffractive-to-total 
cross-section from $e$$+$$p$ and $e$$+$A.  The ratio of these cross-sections 
will directly relate to the size of any saturation effects.  
Figure~\ref{fig:sat_diff}, taken from the EIC white paper~\cite{Accardi:2012hwp}, 
shows the prediction of one saturation model for this cross-section ratio 
with and without saturation, indicating large possible 
effects.  Note that the statistical and systematic uncertainties in this plot 
are scaled up by a factor of 10 in order to be visible.  This measurement 
relies on measuring events with a large rapidity gap, which is the 
signature of diffractive events due to the fact that the hadron remains 
intact after the scattering (though in the case of ions, the nucleus 
may still break up).  The ePHENIX detector will have wide calorimetric coverage, and so will be 
able to make a measurement of the ratio of diffractive-to-total 
cross-sections with 
comparable precision as shown in Figure~\ref{fig:sat_diff}.

\chapter{Detector Requirements}
\label{chap:ephenix_detector_requirements}

The detector requirements for Deep Inelastic Scattering measurements
are well established by previous DIS experiments (H1, ZEUS, HERMES,
COMPASS, etc.), by EIC group
studies~\cite{Accardi:2012hwp,Boer:2011fh}, and through work by the
eRHIC Task Force~\cite{EIC-TF}. Table~\ref{table:dreq}
summarizes these basic requirements and how ePHENIX would meet them.
After a brief overview of the relevant kinematic variables, detailed
studies are presented in this chapter.

\begin{table}[hbtp]
\begin{center}
\caption{Detector requirements}
\label{table:dreq}
\begin{tabular}{p{0.49\linewidth}|p{0.49\linewidth}}
\toprule
Detector requirements & Detector solution \\
\midrule
{\bf Electron-ID:}  \newline
High purity (~99\%) identification of the scattered lepton over hadron 
and photon background \newline {\it Important for \egodir and barrel acceptance}& 
Electromagnetic Calorimetry and charged particle tracking \newline
Minimum material budget before EMCal \newline
Good energy and tracking resolution for $E/p$ matching \\
\midrule

{\bf Resolution in $x$ and $Q^2$:}  \newline
Excellent momentum and angle resolution of the scattered lepton to provide high 
survival probability (~80\%) in each ($x$,$Q^2$) bin (important for unfolding) 
\newline {\it Important for \egodir and barrel acceptance} &
High resolution EMCal and tracking in \egodir \newline
Good (tracking) momentum resolution for $E_e'<10$~GeV in barrel \newline
Good (EMCal) energy resolution for $E_e'>10$~GeV in barrel  \\
\midrule

{\bf Hadron identification:}  \newline
$>90\%$ efficiency and $>95\%$ purity 
&
In barrel acceptance: DIRC for $p_h<4$ GeV/c  \newline
In \hgodir:  Aerogel for lower momentum and gas RICH for higher momentum \\
\midrule

{\bf Wide acceptance for leptons and photons in DVCS:}   \newline
Ability to measure DVCS lepton and photon within $-4<\eta<4$ &
EMCal and tracking with good resolution over for lepton and photon measurements covering $-4<\eta<4$ \\
\midrule

{\bf Electron/Photon separation:}   \newline
Separate DVCS photon and electron in \egoing direction &
High granularity EMCal in \egoing direction \\
\midrule

{\bf Measurement of scattered proton in exclusive processes} &
Roman pots in \hgodir \\
\midrule

{\bf "Rapidity gap" measurement capabilities:}   \newline
Measure particles in $-2<\eta<4$ for diffractive event identification &
Hadronic calorimetry covering $-1<\eta<5$, and EMCal covering $-4<\eta<4$\\
\midrule
{\bf Forward Zero-Degree calorimetry:}  \newline
Measure neutrons from nucleus breakup in diffractive $e$$+$A events &
Zero-Degree calorimeter in \hgoing direction planned, in coordination with CAD \\
\bottomrule
\end{tabular}
\end{center}
\end{table}

The suggested ePHENIX detector configuration is shown in Figure~\ref{fig:ePHENIX}.
It is built around the sPHENIX detector,
which is a superconducting solenoid and electromagnetic
and hadronic calorimeter in the central region ($-1<\eta<1$ for pseudorapidity $\eta$).
This proposal would add to that detector the following detector subsystems:

\begin{description}
\item[\egodir ($-4<\eta<-1$):] High resolution Crystal EMCal with GEM tracking.
\item[Barrel ($-1<\eta<1$):] Compact-TPC for low mass tracking and PID for momentum $p<4$ GeV/c with DIRC
\item[\hgodir ($1<\eta<4$):] Hadronic and Electromagentic calorimeters, GEM trackers, and Aerogel-based ($1<\eta<2$) and gas-based RICH for PID up to momentum $p\sim50$~GeV. 
\item[Far-Forward in \hgodir:] Roman Pots and Zero-Degree Calorimeter.
\end{description}

\section{Kinematics}
\label{s:kinematics}

In DIS, a lepton is scattered off a target hadron
via the exchange of a virtual boson, which for electron beam energy $E_e<10$~GeV
can always be taken as a virtual photon.  Defining the four-momenta of the incoming and 
scattered electron and the incoming proton as $k$, $k'$ and $p$ respectively, we can
define the following Lorentz invariant quantities:
\begin{alignat}{3}
s   &\equiv& (k + p)^2& &=& 4E_eE_p \label{eq:rts}  \\
Q^2 &\equiv& -q^2 = -(k-k')^2& &=& 2 E_{e}E_{e}'\left(1 - cos\theta\right) \label{eq:Q2}\\
  y &\equiv& \frac{p \cdot q}{k \cdot p}& &=& 1 - \frac{E_{e}'}{E_e} + \frac{Q^2}{4E_e^2} \label{eq:y}\\
  x &\equiv& \frac{Q^2}{2p\cdot q}& &=& \frac{Q^2}{ys} \label{eq:x} \\
\nu &\equiv& \frac{p \cdot q}{M}&  &=& \frac{Q^2}{2Mx} \label{eq:nu}
\end{alignat}
where $s$ is the center-of-mass energy squared, $q$ is the 4-momentum 
transferred from scattered electron and $Q^2$ is the virtuality of the photon 
which gives the resolution scale of the scattering, $y$ is the inelasticity 
of the scattering and $x$ is Bjorken $x$, the fractional momentum carried by 
the struck parton.  Here, we have also written these in the lab frame in 
terms of the measured scattering angle, $\theta$ and the energies of the 
proton and incoming and scattered electron, $E_p$, $E_e$ and $E_e'$, 
respectively, under the approximation that the electron and proton mass are 
small compared to the beam energies.

For inclusive DIS, where only the kinematics of the scattered lepton are measured, 
Eq.~\ref{eq:rts}--\ref{eq:nu} fully describe the event.  For SIDIS, in which a final state
hadron is also measured, additional variables are needed.  The fraction of the scattered parton's
momentum carried by the hadron is defined as
\begin{equation}
\label{eq:z}
z \equiv \frac{p_h \cdot p}{q\cdot p}
\end{equation}
where $p_h$ is the four-momentum of the measured hadron.  Further, we can define 
$p_{h\perp}$ as the transverse momentum of the hadron w.r.t. the virtual photon, in the 
center-of-mass frame of the proton (or ion) and virtual photon.  

For exclusive processes, in addition to the scattered lepton, 
the final state photon in DVCS or meson in Deeply Virtual Meson Production 
as well as the scattered proton are measured.  
In this case, another kinematic variable is introduced -- 
the squared momentum transfer to the proton, $t$, 
defined as
\begin{equation}
\label{eq:t}
t \equiv (p' - p)^2
\end{equation}
where $p'$ is the four-momentum of the scattered proton.  


\section{Inclusive DIS and scattered electron measurements}

In inclusive DIS, where only the kinematics of the scattered electron are 
necessary, the primary requirements of any detector are electron 
identification and sufficient resolution in $x$ and $Q^2$, 
which in turn mandates good energy and angle resolution for the scattered 
electron measurements (Eq.~\ref{eq:Q2}--\ref{eq:x}).  

\subsection{Electron Identification}

\begin{figure}[ht]
\begin{center}
\includegraphics[trim=0 0 0 370,clip,width=0.75\textwidth]{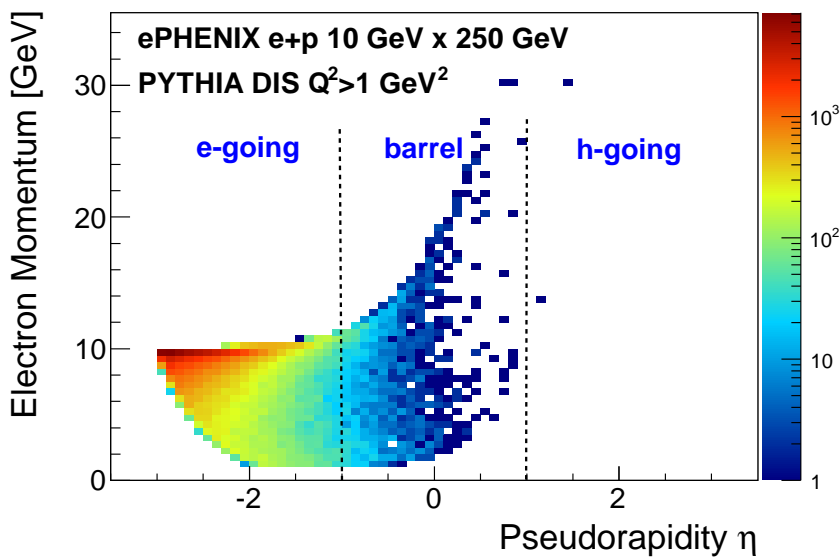}
\end{center}
\caption{Shown is the distribution of scattered electrons in
  pseudorapidity and energy.  The results are from \pythia DIS
  simulations for $e$$+$$p$ collisions with $10~\mathrm{GeV} \times
  250~\mathrm{GeV}$ beam energies.  The events are selected as DIS
  with $Q^{2} > 1$~GeV$^{2}$.  }
\label{fig:eeta}
\end{figure}

In collider geometry, the DIS electrons are scattered mainly in the \egodir
and central rapidities (barrel acceptance), see Figure~\ref{fig:eeta}. 
Central rapidity selects scatterings with higher
$Q^2$ and higher $x$ (due to its correlation with $Q^2$).
The higher the electron beam energy, the more scattering there is in 
the \egodir. The energy of the scattered electron varies
in the range from zero up to the electron beam energy and even to higher values
for electrons detected in the barrel acceptance, see Figure~\ref{fig:eeta}.

\begin{figure}[ht]
\begin{center}
\includegraphics[width=0.95\textwidth]{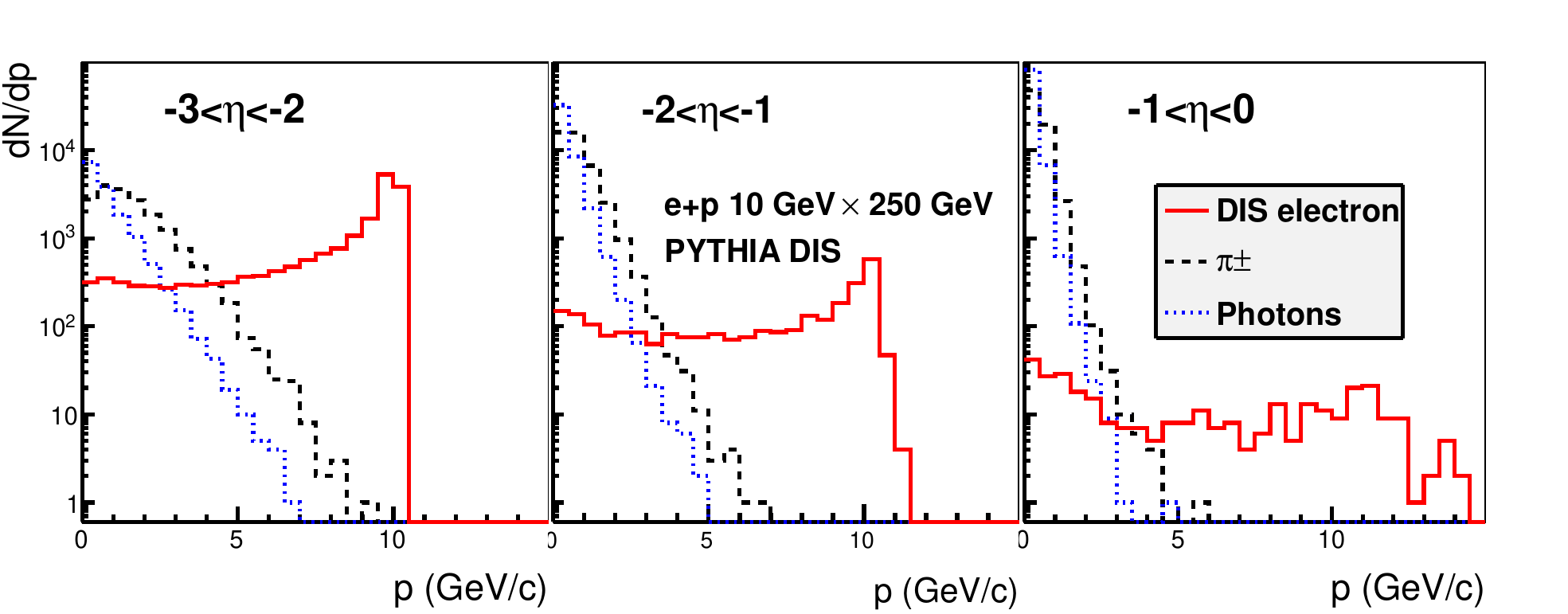}
\end{center}
\caption{For $10~\mathrm{GeV} \times 250~\mathrm{GeV}$ beam energy configuration:
Momentum spectra for scattered electron (red), charged pions (black) and
photons (blue).
}
\label{fig:spectra}
\end{figure}

Collider kinematics allow clear separation of the scattered electrons from
other DIS fragments --- hadrons and their decay products --- which are detected
preferably in the \hgodir, leaving much softer spectra in the central region and
the \egodir. Figure~\ref{fig:spectra}
shows scattered electron momentum spectra along
with photon (mainly from hadron decays) and charged pion spectra.  For the 10~GeV
electron beam, hadronic and photonic backgrounds are small above $\sim5$~GeV/c, but 
increase rapidly at lower momenta.

\begin{figure}[ht]
\begin{center}
\includegraphics[width=0.95\textwidth]{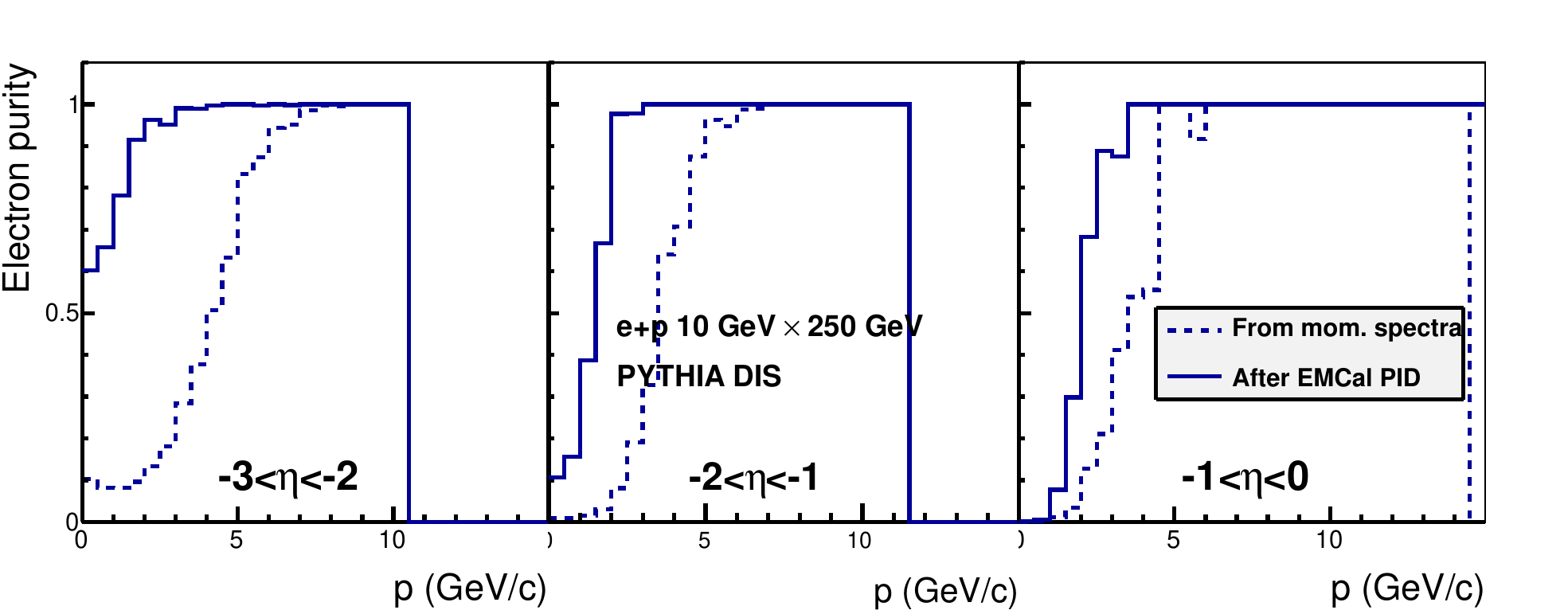}
\end{center}
\caption{For 10 GeV $\times$ 250 GeV beam energy configuration:
The fraction of charged particles from DIS electrons 
before electron identification (dotted) and after identification 
with the EMCal response and E/p matching (solid).
}
\label{fig:epurity}
\end{figure}

\begin{figure}[ht]
\begin{center}
\includegraphics[width=0.75\textwidth]{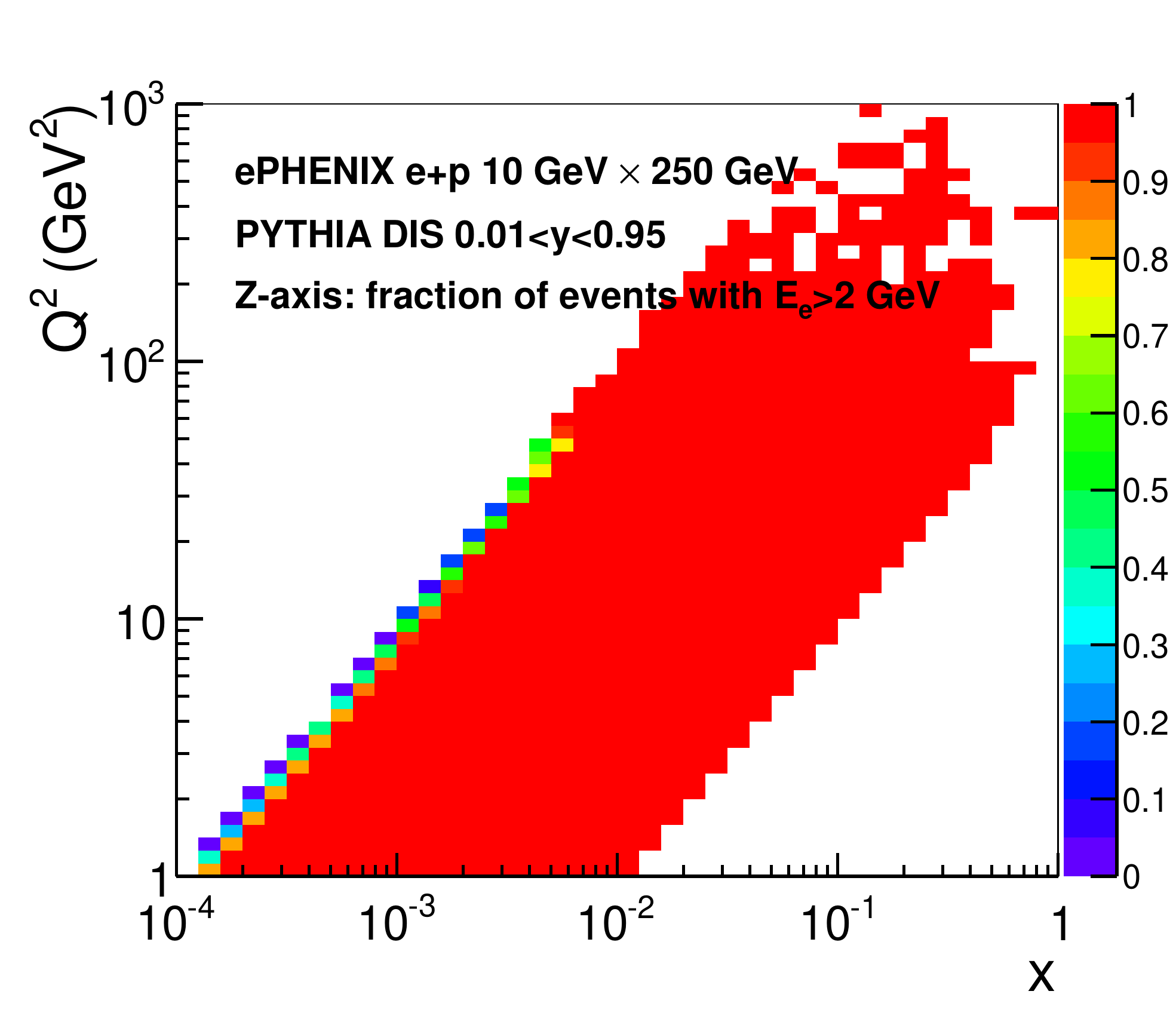}
\end{center}
\caption{For 10 GeV $\times$ 250 GeV beam energy configuration:
The color axis indicates the fraction of events in $(x,Q^2)$ space
surviving after a $>2$~GeV energy cut on the DIS scattered electron. 
}
\label{fig:kin_2gev}
\end{figure}

The different response of the EMCal to hadrons and electrons,
along with a direct comparison of energy deposited in the EMCal
and momentum measured in the tracking system (i.e., $E/p$ matching)
provides a significant suppression of hadronic background
in DIS scattered electron measurements:
from a factor of 20--30 at momenta near 1 GeV/c to a factor of greater than
100 for momenta above 3~GeV/$c$.
Figure~\ref{fig:epurity} shows the effectiveness of electron identification 
with the EMCal and tracking, 
providing high purity for DIS scattered electron measurements
at momenta $>$3 GeV/$c$ for the 10~GeV electron beam 
(and $>$1.5 GeV/$c$ for the 5~GeV electron beam).
The evaluations above are done with a parametrized
response of the EMCal to hadrons and electrons, and EMCal and tracking
resolutions described in Sections~\ref{sec:EMCAL}
and \ref{sec:VertexTracking}.
Further enhanced electron identification is expected from the use of the 
transverse shower profile.  
We are also studying possible electron identification improvement with 
longitudinal segmentation in the crystal calorimeter in the \egodir.
These are expected to move the detector capabilities for high purity electron 
identification down to 2 GeV/c (1 GeV/c) for 10 GeV (5 GeV) electron beam, 
which only marginally limits the $(x,Q^2)$ space probed in our measurements, 
see Figure~\ref{fig:kin_2gev}.


Photon conversion in material between the collision point and the tracker 
(mainly beam pipe, with thickness as small as $0.3\%$ of radiation length) 
is not expected to contribute sizable background. 
Moreover, conversion electron-positron pairs will be well identified by our 
tracking system in the magnetic field and additionally suppressed by E/p 
matching cut. 
A detailed GEANT simulation study is ongoing to quantify this effect.

\subsection{Resolution in $x$ and $Q^2$  and bin survival probability}
\label{sec:resolution}

Measurements of the scattered electron energy and polar angle impact
the DIS kinematic reconstruction, Eq.~\ref{eq:Q2}--\ref{eq:x}.
Unfolding techniques are generally used to correct for smearing in
$(x,Q^2)$ due to detector effects, and the effectiveness of this
technique depends on the degree to which events migrate from their
true $(x,Q^2)$ bin to another.  This migration can be characterized by
the likelihood of an event remaining in its true $(x,Q^2)$ bin --- the
bin survival probability.

The energy resolution $\sigma_E$ is directly propagated to $\sigma_{Q^2}$, 
so that $\sigma_{Q^2}/Q^2=\sigma_E/E$. The EMCal energy and tracking
momentum resolutions will provide excellent precision for $Q^2$
measurements.  Conversely, the $\sigma_x$ resolution is magnified by a
factor of $1/y$ as $\sigma_x/x=1/y \cdot \sigma_E/E$, and so the energy
resolution in this approach effectively defines the limit of our
kinematic reach at low y.

\begin{figure}[ht]
  \begin{center}
    \includegraphics[width=0.95\textwidth]{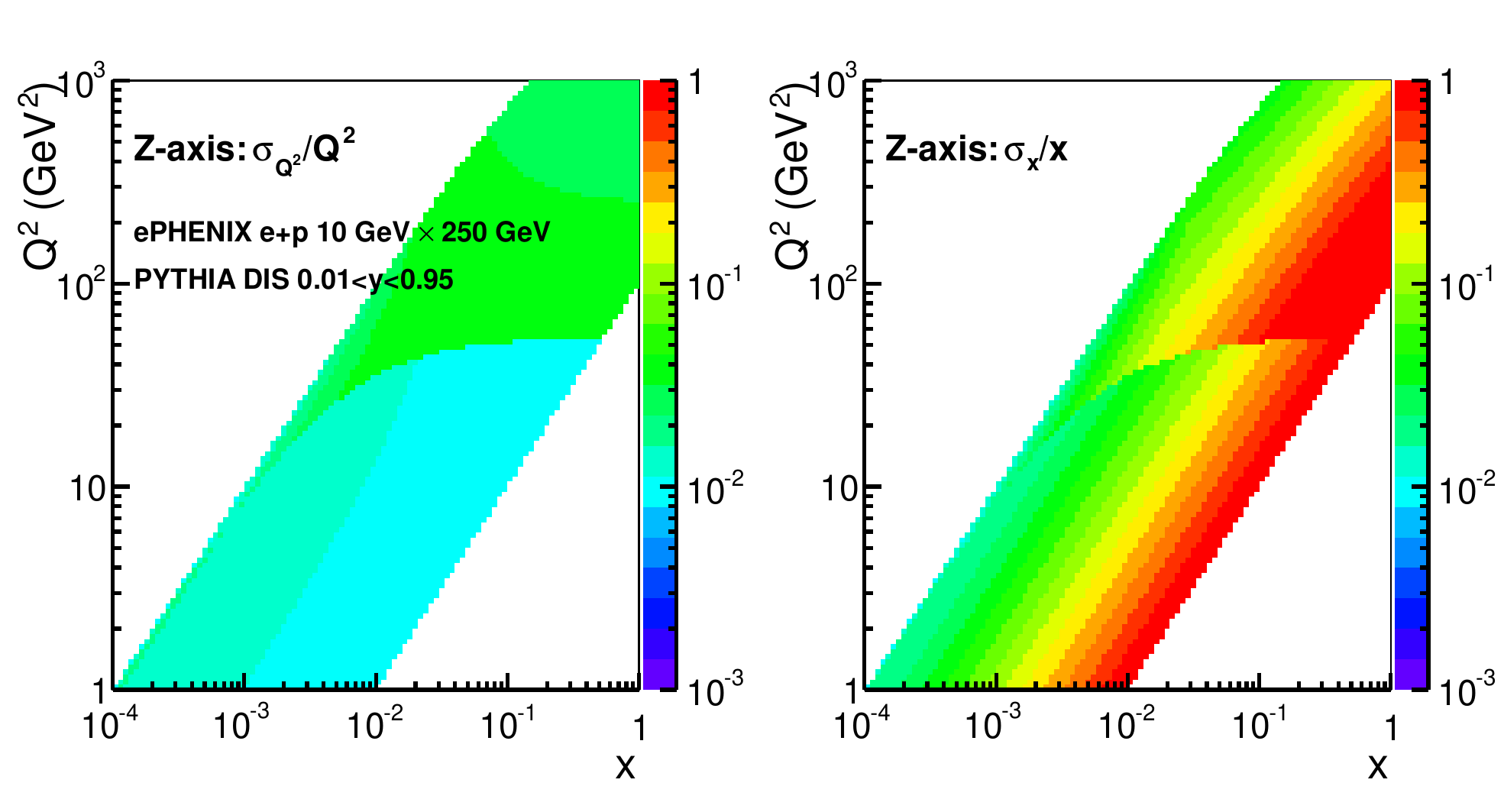}
  \end{center}
  \caption{For $10~\mathrm{GeV} \times 250~\mathrm{GeV}$ beam energy configuration: the
    relative resolution for $Q^2$ (left) and $x$ (right) as a function
    of ($x,$$Q^2$).  }
\label{fig:res}
\end{figure}

\begin{figure}[ht]
\begin{center}
\includegraphics[width=0.75\textwidth]{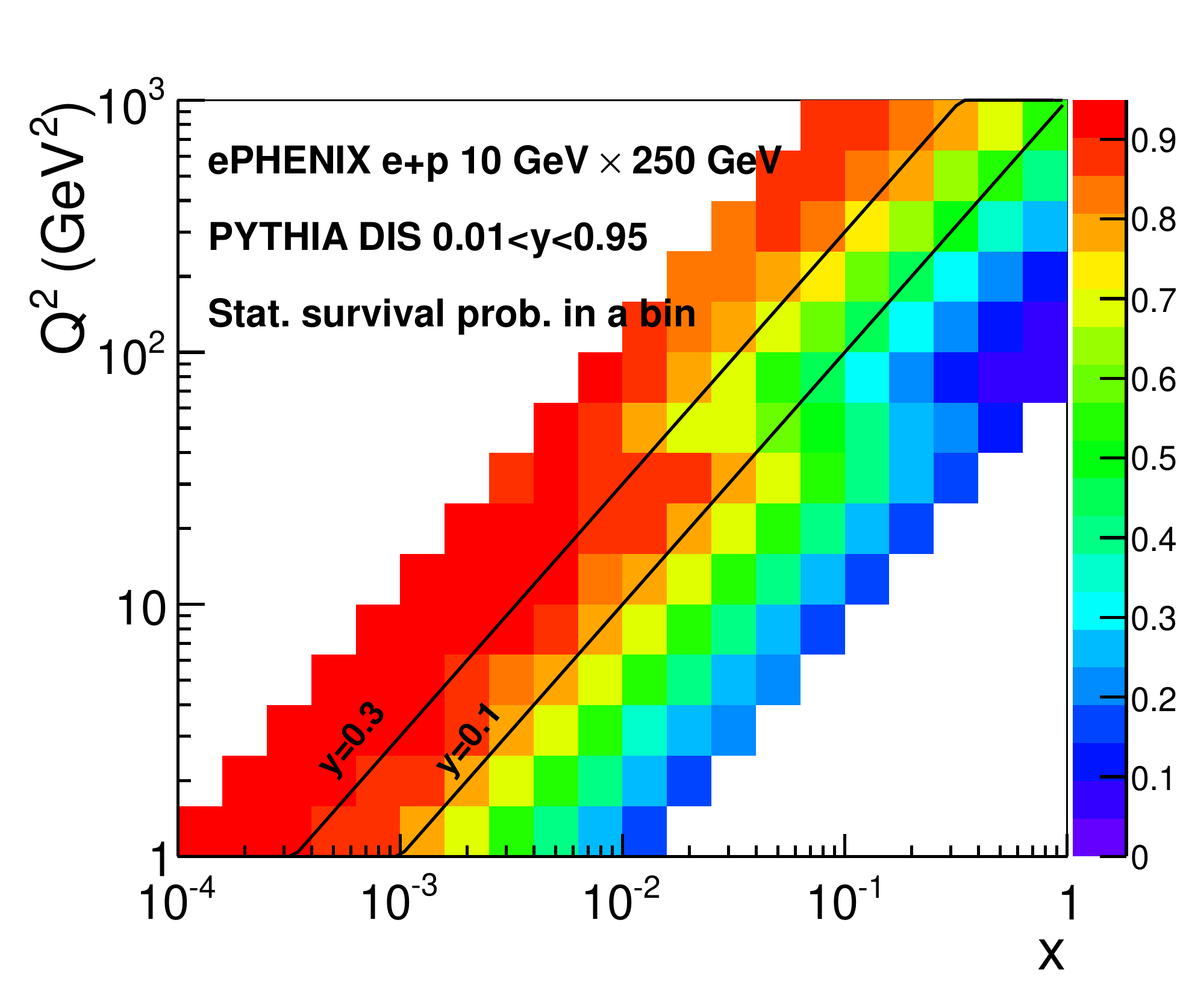}
\end{center}
\caption{For $10~\mathrm{GeV} \times 250~\mathrm{GeV}$ beam energy
  configuration: Statistics survivability in ($x$,$Q^2$) bins.}
\label{fig:res2}
\end{figure}

Figure~\ref{fig:res} shows the relative resolution in $Q^2$ and $x$
measurements using the standard ``electron'' method, in which the 
scattered electron is measured.  While the $Q^2$ relative
uncertainty, $\sigma_{Q^2}/Q^2$, is better than 10\% over whole
$x$-$Q^2$ acceptance, the relative uncertainty on $x$, $\sigma_{x}/x$,
clearly demonstrates its $y$-dependence (the same $y$ points are on
the diagonal, as from Eq.~\ref{eq:x}, $Q^2=syx$). The step in resolution
around $Q^2=50$~GeV$^2$ in
these plots corresponds to the transition from the \egodir to the barrel
acceptance, which differ mainly in the resolution of the different
electromagnetic calorimeters covering those two regions of the
acceptance.  All of this translates to the statistics survival probability
in a bin shown in Figure~\ref{fig:res2}, which is calculated for five bins per
decade in each of $x$ and $Q^2$.  The survival probability is
$>80\%$ for $y>0.1$ in the \egodir and for $y>0.3$ in the barrel acceptance.




The effect of the polar angle resolution $\theta$ in 
Eq.~\ref{eq:Q2}--\ref{eq:x}, is the biggest for forward scattering 
(small $\theta$). It was found that crystal EMCal position resolution 
(better than 3~mm for $>1$~GeV electrons, see Chapter~\ref{sec:CrystalEMCAL}) 
provides enough precision for scattered electron angle measurements, 
so that it affects the statistics migration in bins on 
Figure~\ref{fig:res2} only marginally.

The Jacquet-Blondel method using the hadronic final state is an
alternative approach to reconstruct DIS kinematics.  Its resolution
for inelasticity $y$, and hence for $x$, is nearly flat, so it
provides much better precision for $x$ determination than the
``electron'' method, in the region with small $y$. It is also
better in the higher $Q^2$ region corresponding to the barrel
acceptance, where the resolution of the ``electron'' method is
limited by the EMCal resolution.

The Jacquet-Blondel method requires the measurement of all final state
hadrons produced in $e$$+$$p$ or $e$$+$A scattering.  A study with
the \pythia generator shows that the precision of this approach does
not deteriorate if the hadron detection capabilities are limited to
$|\eta| < 4$. This method provides relative precision for the
measurement of $x$ of better than 20\%, which satisfies the bin
statistics migration criteria discussed above.  It was found that for
$y<0.3$ the precision of this approach deteriorates only slightly when
hadron measurements are limited to the barrel and forward acceptance
$-1<\eta<4$ (the acceptances we plan to equip with hadron
identification capabilities, see
Chapter~\ref{sec:HadronPIDDetectors}). As was shown above,
measurements at higher $y$ are well provided by the ``electron'' method.

Therefore, combining the electron and hadronic final state measurements provides
precise determination of basic kinematic variable $x$, $y$ and $Q^2$
in the whole kinematical space.

QED radiative effects (radiation of real or virtual photons) are
another source of smearing which is usually corrected with unfolding
techniques. Unlike energy-momentum resolutions which introduces
Gaussian-like smearing, radiative corrections are tail-like. 
They can be responsible for as much as 10--20\% of statistics migrating away 
from a bin, and dominate over energy-momentum smearing at higher $y$
(compare to Figure~\ref{fig:res2}).


\section{Semi-inclusive DIS and hadron ID}

As was discussed in Chapter~\ref{chap:physics_goals}, measurements of
hadrons in SIDIS events are necessary to determine both the (sea)quark
separated helicity distributions and TMDs.  It is also important for
understanding the hadronization process in nuclear matter.  For these
measurements, one needs to identify the hadron, particularly in the
case of pions and kaons.  In this section, we discuss the kinematic
ranges of interest for pions, kaons and protons, and in Chapter
\ref{chap:ephenix_detector_concept}, we discuss technology choices which can
effectively make these measurements.

\begin{figure}[ht]
  \centering
  \includegraphics[trim=0 0 20 370, clip, width=0.8\textwidth]{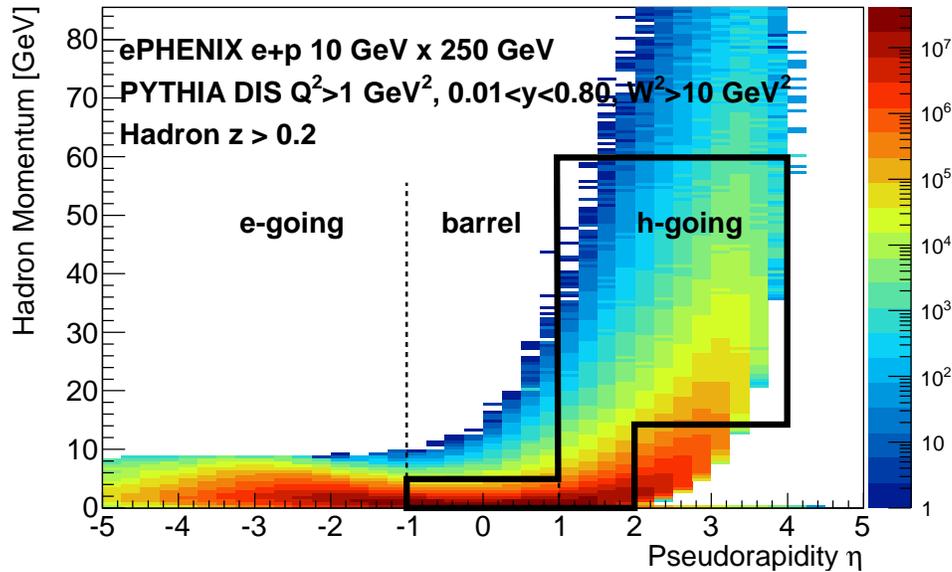}
  \caption{Shown is the distribution of hadrons from DIS events 
    in $e$$+$$p$ as a
    function of momentum and pseudorapidity, based on \pythia
    simulations of the $10~\mathrm{GeV} \times 250~\mathrm{GeV}$ beam
    energy configuration.  The black outline indicates the
    pseudorapidity and momentum range covered for kaons by the planned
    PID detectors in ePHENIX.}
  \label{fig:250x10_mometa}
\end{figure}

Figure~\ref{fig:250x10_mometa} shows the yields of positively charged
hadrons as a function of momentum and pseudorapidity for the 10~GeV 
$\times$ 250~GeV beam configuration.  A minimum $z$
cut of $z > 0.2$ to remove soft physics effects and beam remnant is
applied.  For $\eta<0$, the hadron momenta are limited by the electron 
beam momentum, while in the \hgodir, the hadron
momenta extend almost to the full proton beam energy.
The results are similar for other beam energy configurations.

\begin{figure}[ht]
\centering
\includegraphics[width=0.48\textwidth]{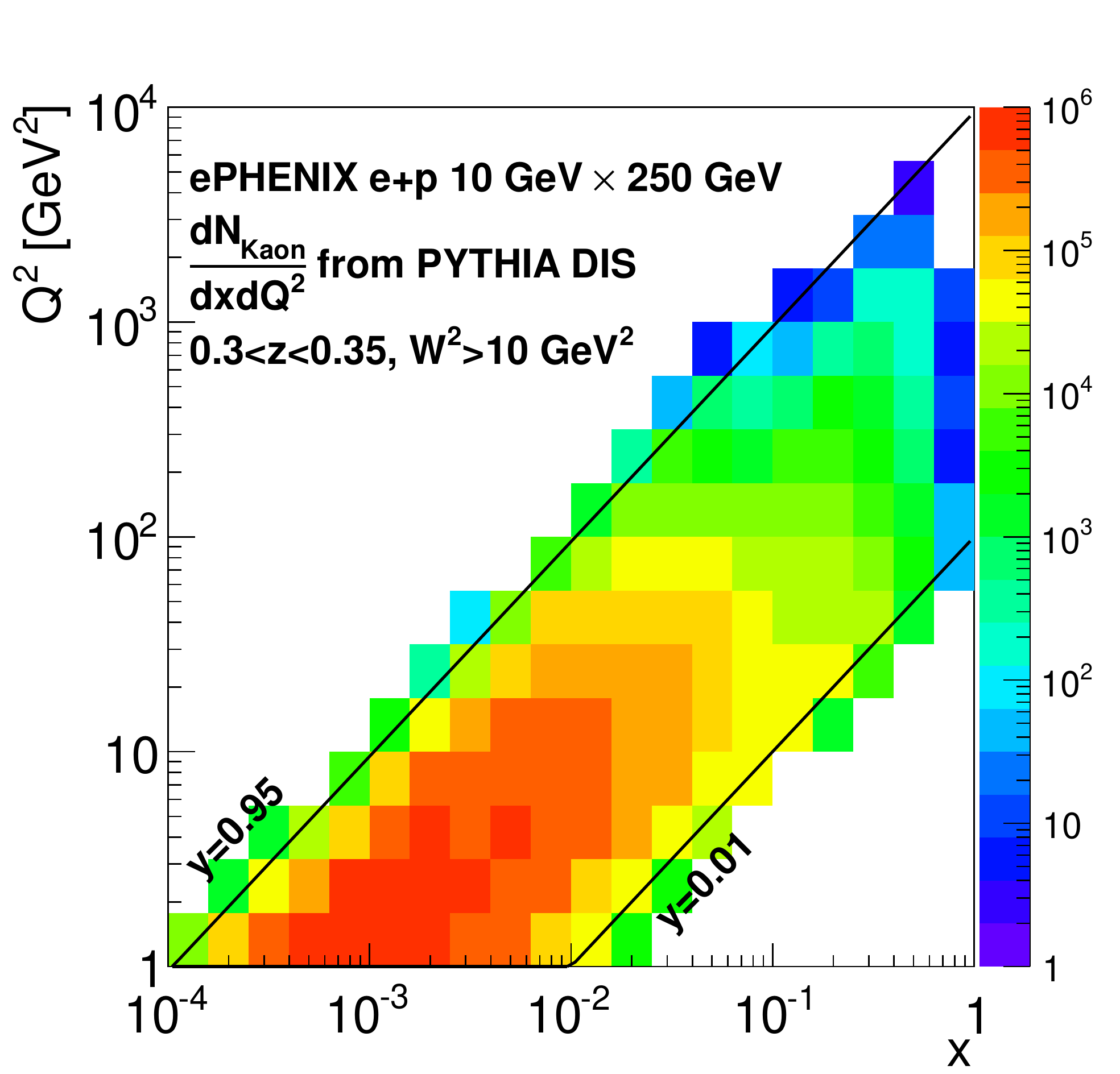}
\includegraphics[width=0.48\textwidth]{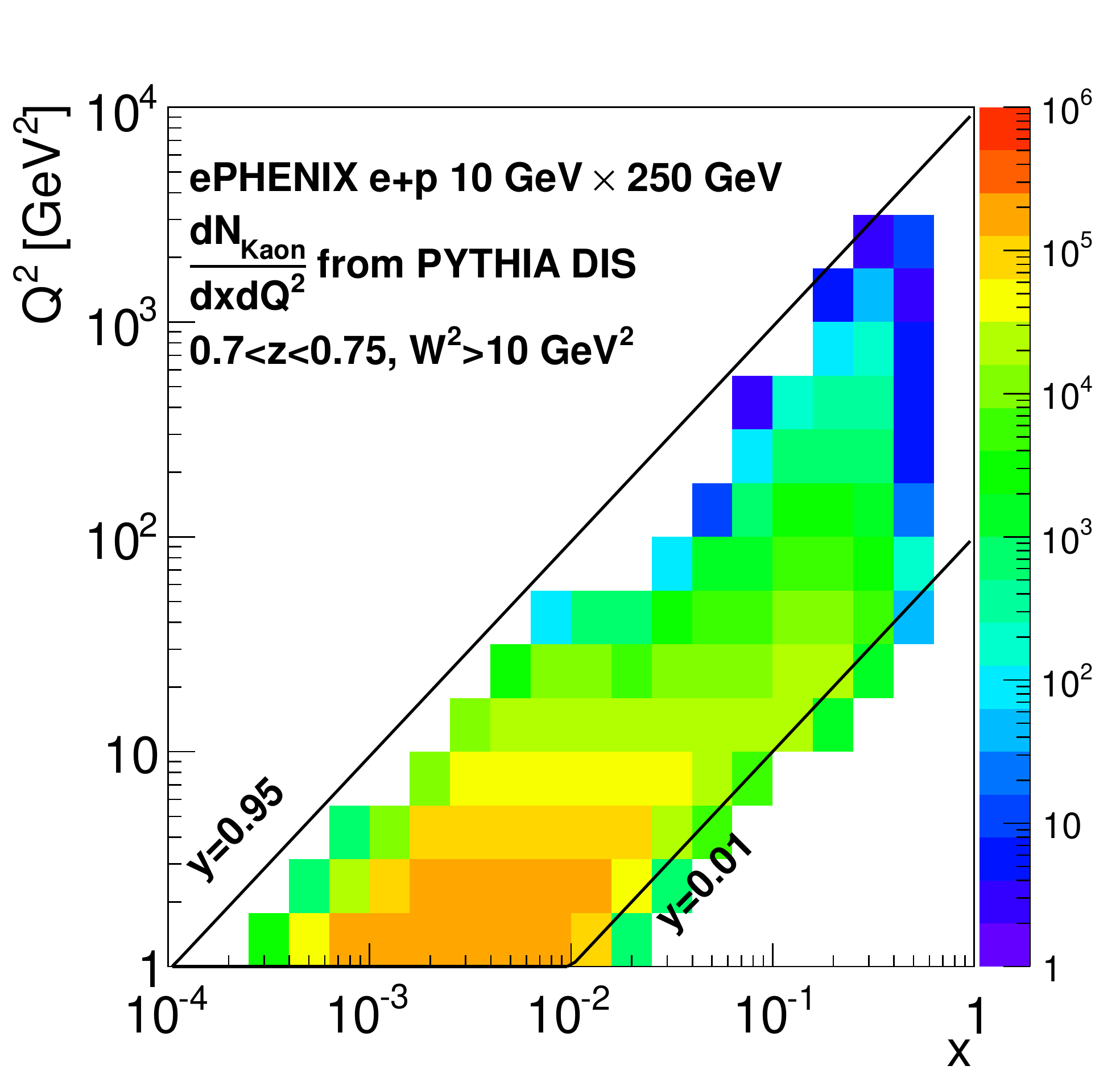}
\caption{$x$ and $Q^2$ distribution of events with kaons which can be
  identified with the ePHENIX PID detectors in expected binning at 
  (left) low and (right) high $z$. }
\label{fig:SIDISxQ2}
\end{figure}

As was stated above, ePHENIX will have three PID systems: (1) a DIRC
covering $|\eta| < 1$ providing $\pi$-$K$ separation below
3.5--4~GeV/$c$ (depending on purity and efficiency requirements), (2)
an aerogel based RICH covering $1<\eta<2$ providing $\pi$-$K$
($K$-$p$) separation below 6 (10)~GeV/$c$ and (3) a gas based RICH
covering $1<\eta<4$ providing $\pi$-$K$ separation for
$3<p<50$~GeV/$c$ and $K-p$ separation for $15<p<60$~GeV/$c$ (depending
on the balance between efficiency and purity chosen).  Based on these
numbers, the PID for kaons would cover the $\eta$ and $p$ region
outlined in black in Figure~\ref{fig:250x10_mometa}.  The resulting
ePHENIX $x$ and $Q^2$ coverage for SIDIS events with an identified 
kaon is shown in Figure~\ref{fig:SIDISxQ2},
for low ($0.30 < z <0.35$) and high ($0.70<z<0.75$) $z$ bins, along
with lines indicating the accessible DIS $y$ range ($0.01<y<0.95$).

\begin{figure}[ht]
  \centering
  \includegraphics[width=0.95\textwidth]{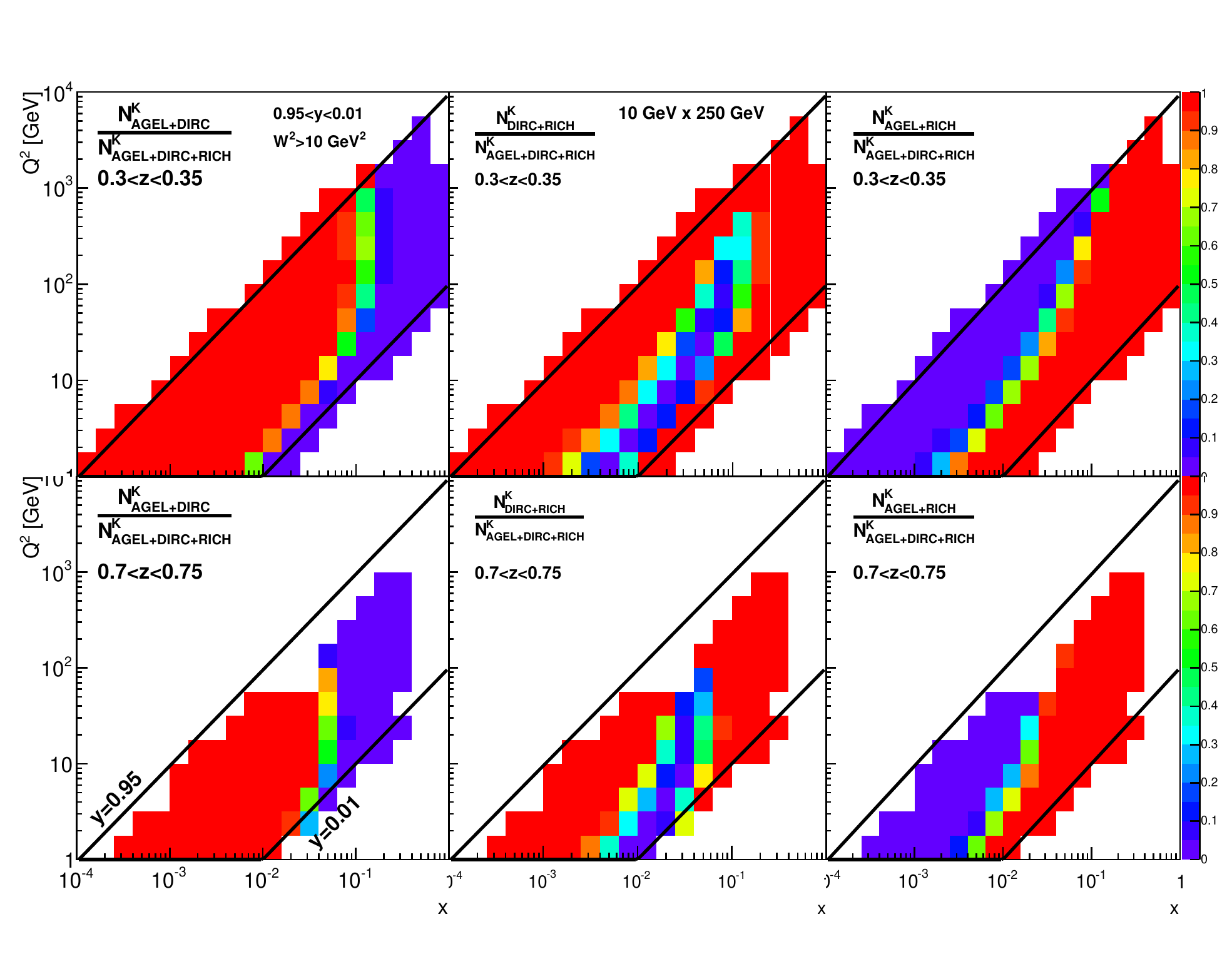}
  \caption{Efficiency as a function of $x$ and $Q^2$ of kaon
    identification when comparing to baseline ePHENIX design with a
    DIRC, RICH and Aerogel when one of these subsystems is removed.
    The top three plots are for low $z$ ($0.3<z<0.35$) and the bottom
    three are for high $z$ ($0.7<z<0.75$).  Also shown are lines
    indicating different values of $y$.}
  \label{fig:kaon_pid_impact}
\end{figure}

Figure~\ref{fig:kaon_pid_impact} shows the impact on the $x$ and $Q^2$ 
coverage of removing one of the three PID detectors planned for ePHENIX 
at low and high $z$.  The plots show the ratio of kaon yields when using only 
two PID detectors to those with all three detectors (i.e., standard 
ePHENIX).  If the gas-based RICH detector is removed (left), the high $x$ 
reach, particularly at high $Q^2$, is lost.  If the aerogel-based RICH is removed 
(middle), sensitivity to the region of moderate $x$, $Q^2$ and $y$ is lost.  Finally, 
if the DIRC is removed, significant kinematic coverage at low $x$, as well as moderate
$x$ and high $Q^2$ is lost.  To achieve a wide $x$ and $Q^2$ coverage, all three
detectors are necessary. Extending the aerogel-based RICH to $\eta>2$ does not 
extend the kinematic coverage; the momentum range covered by such a detector 
corresponds to very low values of $y$.

\section{Exclusive DIS}

\begin{figure}[ht]
  \begin{center}
    \includegraphics[width=0.75\textwidth]{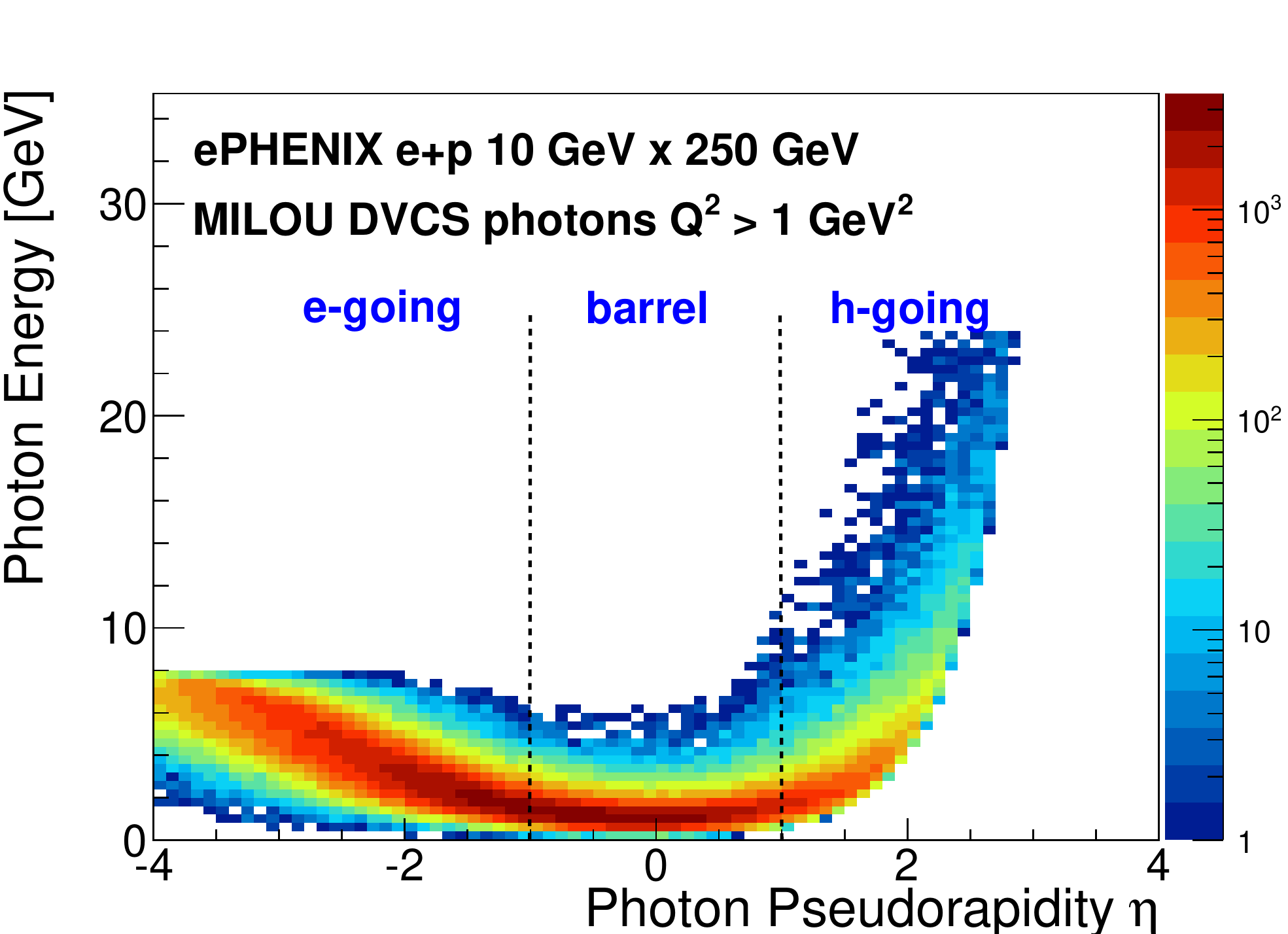}
  \end{center}
  \caption{For the $10~\mathrm{GeV} \times 250~\mathrm{GeV}$ beam
    energy configuration: DVCS photon energy vs pseudorapidity
    distribution; the $z$-axis scale shows the relative distribution
    of events from the \milou event generator.  }
  \label{fig:dvcs_eeta}
\end{figure}

Among exclusive processes, Deeply Virtual Compton Scattering
(DVCS) is of special interest (see
Chapter~\ref{sec:tomographic_imaging}). The produced DVCS photon
energy versus pseudorapidity distribution is shown in
Figure~\ref{fig:dvcs_eeta}.  Most of the photons fall in the \egodir and
the barrel (central rapidity) acceptance.  The photon energy for $-1<\eta<1$ 
varies in the range $\sim1$--4~GeV/$c$ and is nearly independent of the 
beam energy in the range considered for eRHIC.  Photons in the \egodir are
more correlated with the electron beam and have energy from
1~GeV up to electron beam energy.

\begin{figure}[ht]
  \begin{center}
    \includegraphics[width=0.95\textwidth]{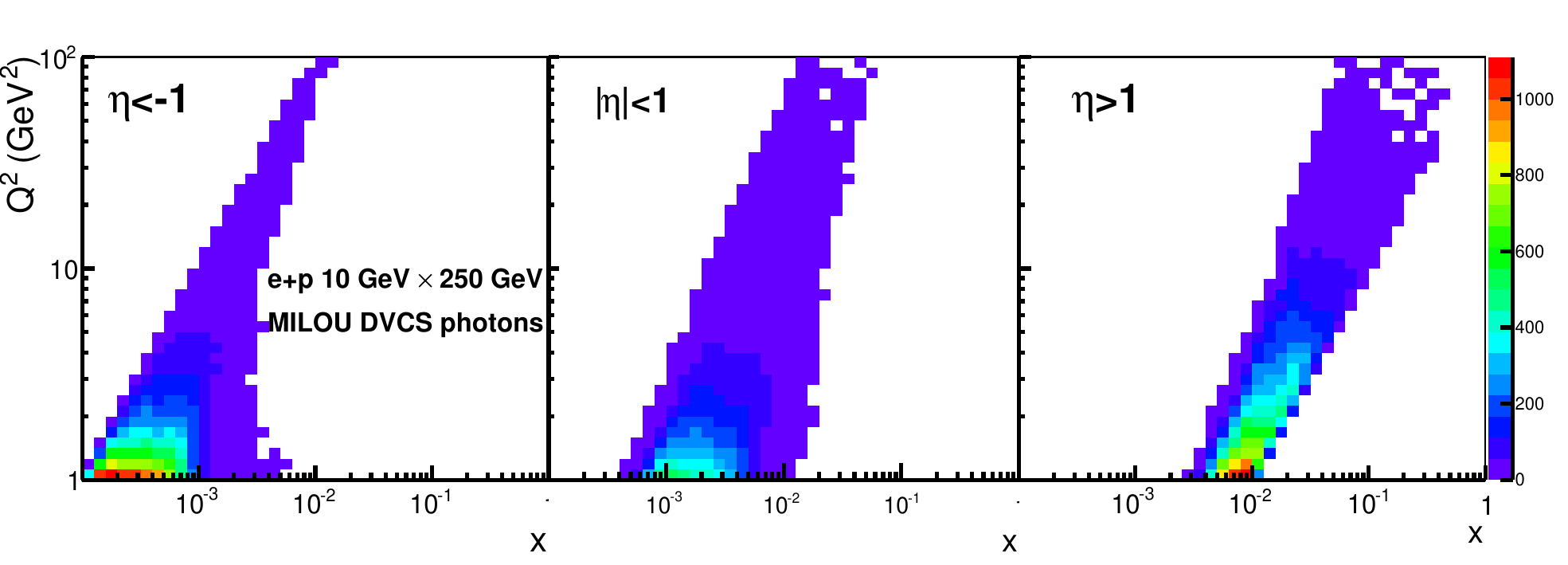}
  \end{center}
  \caption{For $10~\mathrm{GeV} \times 250~\mathrm{GeV}$ beam energy
    configuration: $x$-$Q^2$ coverage for DVCS events with photon
    detected in the \egodir, $\eta<-1$ (left), or central rapidities,
    $|\eta|<1$ (middle) and \hgodir, $\eta>1$ (right). The $z$-axis scale
    shows relative distribution of events from the \milou event
    generator.}
  \label{fig:dvcs_q2x}
\end{figure}

Figure~\ref{fig:dvcs_q2x} shows the $x$-$Q^2$ range covered by
DVCS measurements for different rapidity ranges, emphasizing the
importance of measurements over a wide rapidity range. Wide
kinematical coverage is also important for separating DVCS events from
Bethe-Heitler (BH) events (when a photon is radiated from the initial
or final state lepton), which share the same final state.  This can be done by 
utilizing the different kinematic distributions of DVCS and BH
photons (e.g., in rapidity and inelasticity $y$).  The planned EMCal
and tracking cover $|\eta| < 4$ (Chapter~\ref{sec:EMCAL} and
~\ref{sec:VertexTracking}) and will provide excellent capabilities for
DVCS measurements.

To ensure the reliable separation of electromagnetic showers in the
EMCal from the scattered electron and the DVCS photon, sufficient EMCal
granularity is necessary. The minimal angle separation between the 
electron and the photon is reached for electrons with the smallest scattering 
angle (i.e., the smallest $Q^2$) and is inversely proportional to
electron beam energy.  For a 10~GeV electron beam and $Q^2>1$~GeV$^2$,
the minimum angle is $\sim 0.1$ rad.  The proposed crystal 
EMCal in the \egodir, with granularity $\sim 0.02$ rad 
(see Chapter~\ref{sec:CrystalEMCAL}), will provide the necessary 
electron and photon shower separation.

It is also important to ensure the exclusiveness of the DVCS measurements, 
and so it is highly
desirable to reconstruct the scattered beam proton. The proton
scattering angle is inversely proportional to proton beam energy and
varies from 0 to 5~mrad for 250~GeV proton beam and four-momentum
transfer $-t<1$~GeV$^2$.  It can be detected with the planned "Roman Pots"
detectors located along the beam line (See Chapter~\ref{sec:beamline}).  

\section{Diffractive measurements}

\begin{figure}[ht]
\begin{center}
\includegraphics[width=0.95\textwidth]{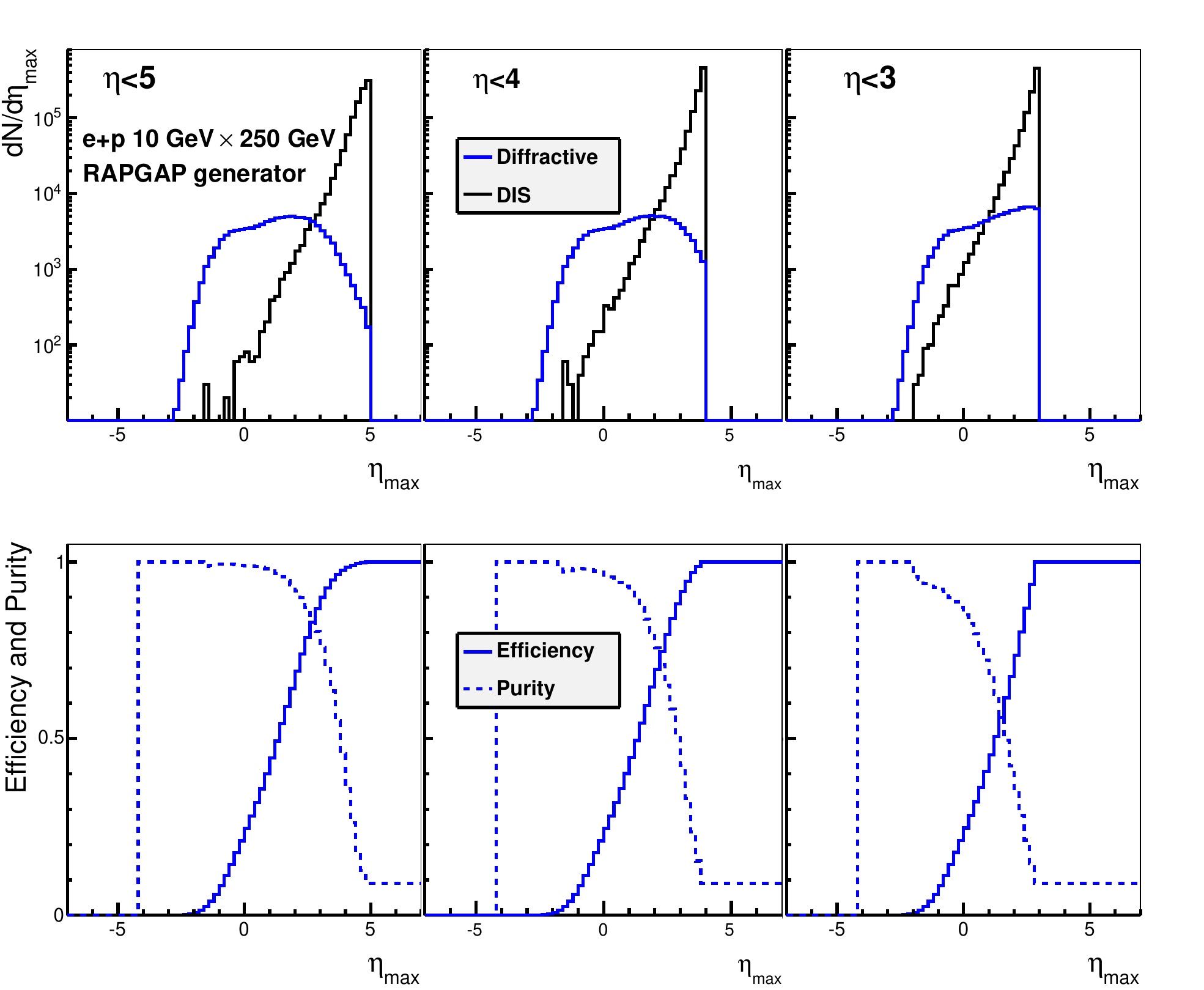}
\end{center}
\caption{For the $10~\mathrm{GeV} \times 100~\mathrm{GeV}$ beam energy
  configuration: Top: Pseudorapidity distribution for the most forward
  going particle in DIS events (black) and in diffractive events
  (blue); Bottom: Efficiency (dashed) and purity (solid) for
  diffractive event identification as a function of pseudorapidity cut
  defining the rapidity gap, for different detector acceptance:
  $|\eta|<5$ (left), $|\eta|<4$ (middle), $|\eta|<3$ (right).
  Obtained using the \rapgap generator developed at HERA and tuned to
  H1 and ZEUS data.}
\label{fig:diff}
\end{figure}

Diffractive event measurements play an important role in nucleon and
nucleus imaging. They are particularly sensitive to the gluon distribution
in nuclei and hence to gluon saturation phenomena.  Diffractive events
are characterized by a rapidity gap, i.e. an angular region in the
direction of the scattered proton or nucleus devoid of other
particles.  Figure~\ref{fig:diff} shows the pseudorapidity
distribution for the most forward going particle in DIS events and in
diffractive events.  Extending the forward acceptance of the detector
to $\eta = 4$ and beyond is important if one is to have good
capability using the rapidity gap method for detecting diffractive
events and to separate them from DIS processes.

The planned ePHENIX EMCal and tracking coverage of $|\eta| < 4$ and
hadronic calorimetry coverage of $-1 <\eta < 5$ are expected to provide
excellent identification capabilities for diffractive events.  In addition,
to separate coherent (the nucleus remains intact) and incoherent (the
nucleus excites and breaks up) diffractive events, we plan to place a
zero degree calorimeter after the first RHIC dipole magnet 
(see Chapter~\ref{sec:beamline}), which is
expected to be very efficient at detecting nuclear break-up by
measuring the emitted neutrons.

\chapter{Detector Concept}
\label{chap:ephenix_detector_concept}

\begin{figure}[h]
  \centering
  \includegraphics[width=0.8\linewidth]{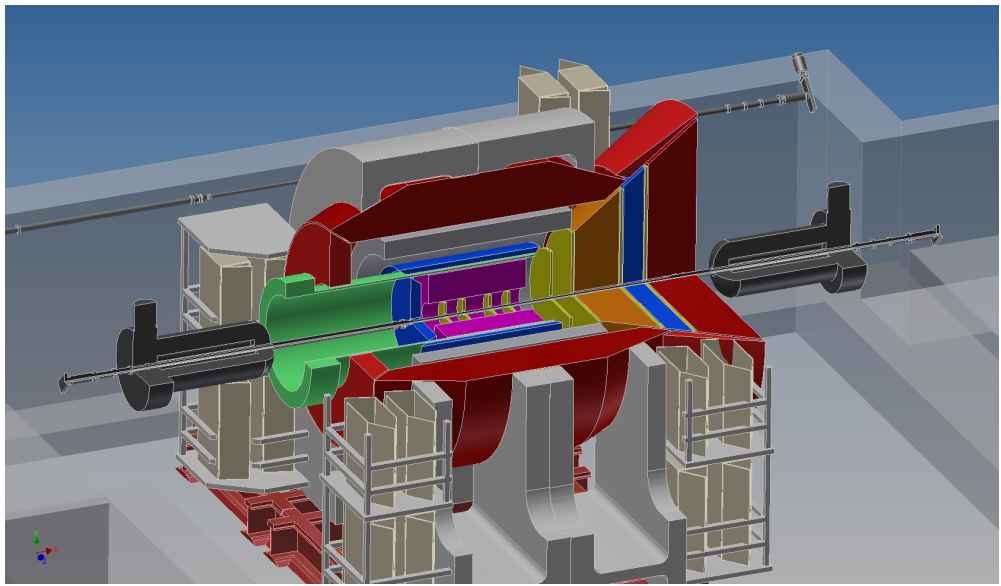}
  \caption{Engineering rendering of ePHENIX in the PHENIX experimental
    hall.  The drawing shows the location of the final eRHIC focusing
    quadrupoles as well as the electron bypass beamline behind the
    detector.}
  \label{fig:ephenix_drawing}
\end{figure}

A full engineering rendering of ePHENIX is shown in
Figure~\ref{fig:ephenix_drawing}.  The drawing shows the ePHENIX
detector in the existing PHENIX experimental hall and illustrates the
reuse of the superconducting solenoid and the electromagnetic and
hadronic calorimeter system of sPHENIX.  The rendering also shows the final eRHIC
focusing quadrupoles, each located 4.5 m from the interaction point (IP).
Those magnets and the height of the beam pipe above the concrete
floor, set the dominant physical constraints on the allowable
dimensions of ePHENIX. This Chapter will describe the ePHENIX detector
concept in terms of its component subdetectors and their expected
performance.

\begin{figure}[ht]
  \begin{center}
    \includegraphics[trim = 0 90 0 0, clip, width=1.0\textwidth]{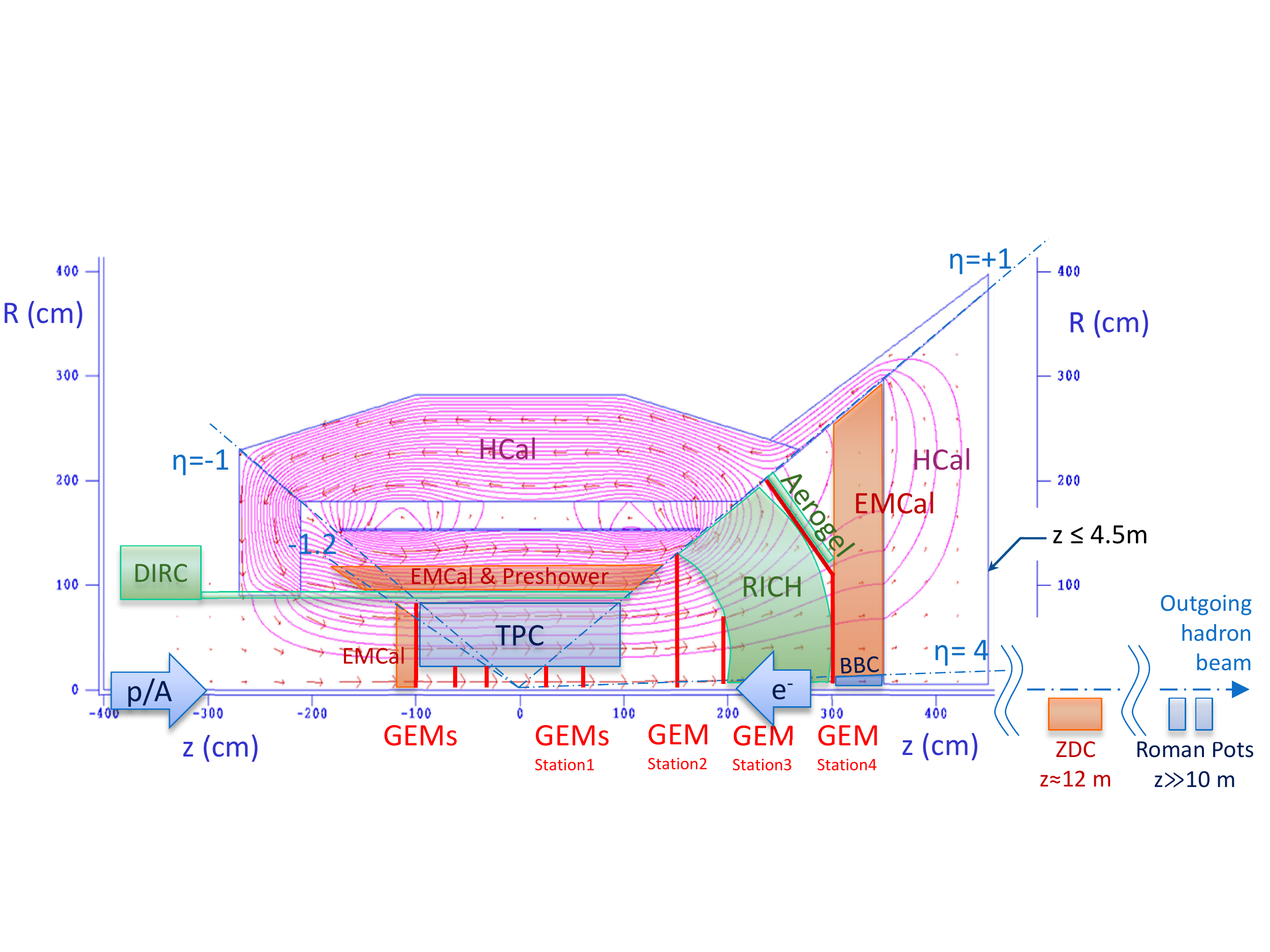}
  \end{center}
  \caption{A cross section through the top-half of the ePHENIX
    detector concept, showing the location of the superconducting
    solenoid, the barrel calorimeter system, the EMCal in the \egoing
    direction and the system of tracking, particle identification
    detectors and calorimeters in the \hgoing direction.  Forward
    detectors are also shown along the outgoing hadron beamline.
    The magenta curves are contour
    lines of magnetic field potential as determined using the 2D
    magnetic field solver, POISSON.}
  \label{fig:straw-man}
\end{figure}

The ePHENIX detector consists of a superconducting solenoid with
excellent tracking and particle identification capabilities covering a
large pseudorapidity range, as shown in Figure~\ref{fig:straw-man}.  It
builds upon an excellent foundation provided by the proposed sPHENIX
upgrade~\cite{Aidala:2012nz} detailed in the MIE proposal submitted to
the DOE Office of Nuclear Physics by Brookhaven National Laboratory in
April 2013.  The strong sPHENIX focus on jets for studying the
strongly-coupled quark-gluon plasma in $p$$+$$p$, $p$/$d$$+$A and
A$+$A is enabled by excellent electromagnetic and hadronic calorimetry
in the central region ($|\eta|<1$).

The C-AD Interaction Region (IR) design at the time the Letter of Intent charge was issued
had the final focusing quadrupoles of the accelerator positioned
$\pm4.5$~m from the IP and employed a ``crab crossing''
to maintain high luminosity while allowing the electron and hadron
beams to intersect at an angle of 10~mr (see Figure~\ref{fig:ZDC}).
The ePHENIX detector concept shown in Figure~\ref{fig:ephenix_drawing}
and Figure~\ref{fig:straw-man} respects these constraints.  For
instance, the hadronic calorimeter in the \hgoing direction fits
within the 4.5~m constraint imposed by the accelerator magnets, and
the detector is aligned so that the electron beam travels along the
symmetry axis of the magnetic field.  Clearly, the progress of ePHENIX
from concept to final design will be done in close consultation with
C-AD to ensure that the design of IR and the design of the detector
remain synchronized.

\begin{figure}[ht]
  \begin{center}
    \includegraphics[trim = 0 50 0 0, clip, width=1.0\linewidth]{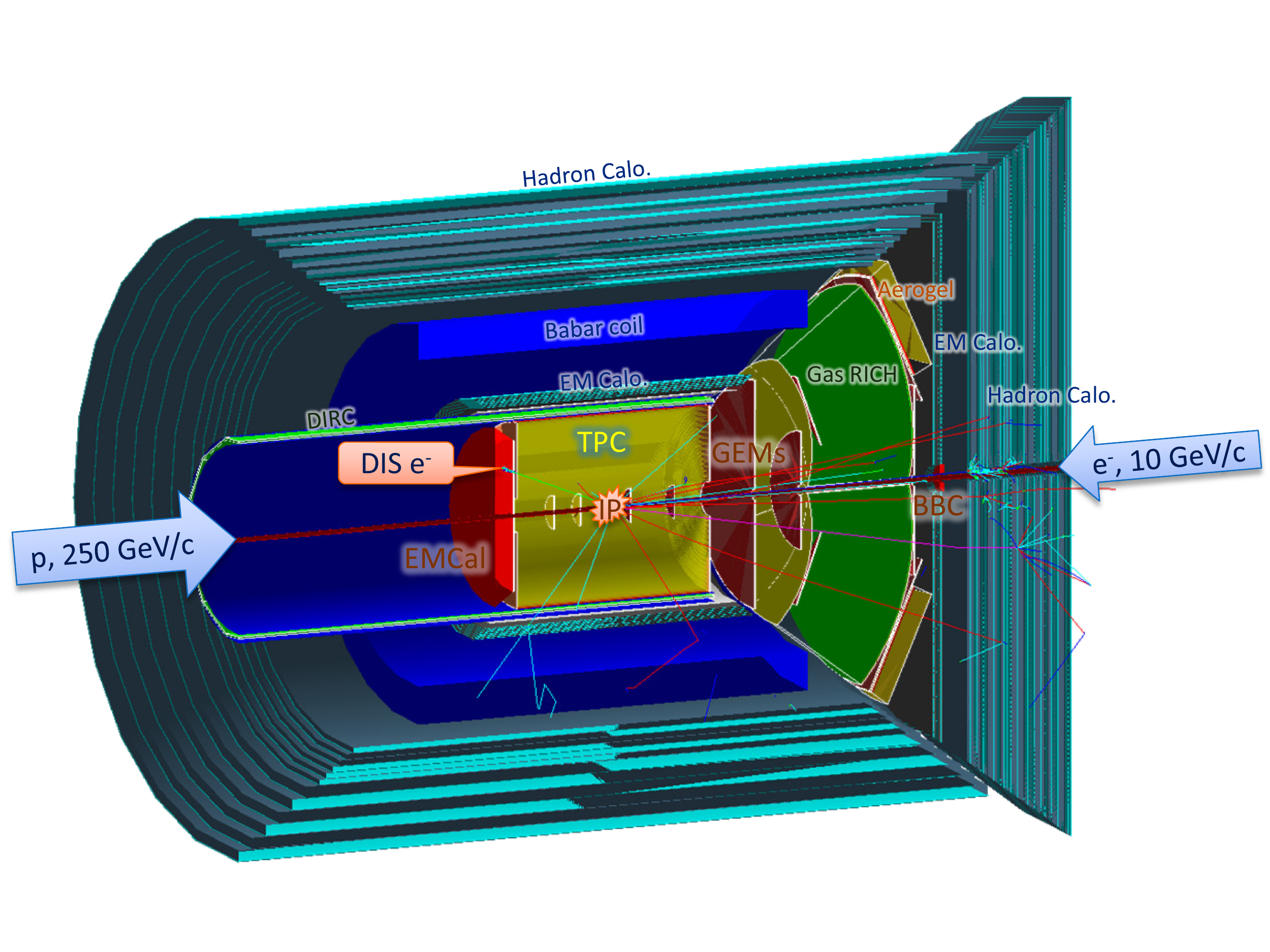}
  \end{center}
  \caption{The response of the ePHENIX detector to a single event, as
    determined using \geant.  The field map in this simulation was
    determined using the 2D magnetic field solver OPERA.  These same
    OPERA calculations were used to verify and validate the
    calculations underlying the magnetic field lines shown in
    Figure~\ref{fig:straw-man}.}
  \label{fig:geant4}
\end{figure}

We have an extensive \geant description of the ePHENIX detector, based
on the same software framework as used in PHENIX and sPHENIX, which
enables ready use of many existing PHENIX software analysis tools.  An
example of running a DIS event through the \geant detector description
is shown in Figure~\ref{fig:geant4}.

The DOE funded sPHENIX subsystems which will be reused in ePHENIX are:
\begin{description}
\item[Superconducting solenoid:] The sPHENIX detector concept reuses
  the BaBar superconducting solenoid to provide a 1.5~Tesla
  longitudinal tracking magnetic field. Its field is shaped in the
  forward directions with an updated yoke design in the ePHENIX
  detector as discussed in Section~\ref{sec:MagneticSystem}.

\item[Electromagnetic calorimeter:] A tungsten-scintillator sampling
  electromagnetic calorimeter with silicon photomultipliers (SiPMs)
  enables a compact barrel calorimeter positioned inside the bore of
  the superconducting solenoid.  The calorimeter system provides full
  azimuthal coverage for $|\eta|<1$ with an energy resolution of
  $\sim12\%/\sqrt{E}$. The readout is segmented into towers measuring
  roughly $\Delta\eta\times\Delta\phi\sim0.024\times0.024$.

\item[Hadronic calorimeter:] A $5\lambda_{\mbox{int}}$-depth hadron
  calorimeter surrounds the solenoid. An iron-plate and scintillator
  sampling design provides an energy resolution of better than
  $\sim100\%/\sqrt{E}$ with full azimuthal coverage. It also serves as
  part of the magnetic flux return for the solenoid.
\end{description}

In addition, new subsystems will be added to the ePHENIX detector,
which will be further discussed in this chapter. These subsystems
include:
\begin{description}
\item[Electron going direction:] GEM detectors~\cite{Sauli:1997qp,
    Abbon:2007pq} and lead-tungstate crystal electromagnetic
  calorimeters
  \item[Central barrel:] Fast, compact TPC tracker and DIRC
  \item[Hadron going direction:] GEM tracking system, gas-based RICH, aerogel-based RICH, beam-beam counter (BBC),
    electromagnetic and hadron calorimeter
  \item[Beam line of hadron-going direction:] Roman pot detectors and
    a zero-degree calorimeter
\end{description}

\section{Magnet system}
\label{sec:MagneticSystem}

\renewcommand{\arraystretch}{1.9}
\addtolength{\tabcolsep}{-0.5pt}
\begin{table}
\caption{Main characteristics of BaBar solenoid~\cite{Bell:1999vf}}
\centering
\begin{tabular}{ll}
\toprule
Central Induction & 1.5~T\tabularnewline
Winding structure & 2~layers, $2/3$ higher current density at both ends\tabularnewline
Winding axial length & 3512~mm\tabularnewline
Winding mean radius & 1530~mm\tabularnewline
BaBar operation current & 4596~A, 33\% of critical current\tabularnewline
Total turns & 1067\tabularnewline
\bottomrule
\end{tabular}
\label{tab:Main-characteristics-of-BaBar}
\end{table}

As with sPHENIX, ePHENIX is based around the BaBar superconducting
solenoid~\cite{Bell:1999vf} with no modifications to its inner
structure. The major specifications for its coil are listed in
Table~\ref{tab:Main-characteristics-of-BaBar}.  A notable feature of
the BaBar magnet is that the current density of the solenoid can be
varied along its length, i.e., lower current density in the central
region and higher current density at both ends. This is accomplished
by using narrower windings (5~mm) for the last 1~m at both ends. The
central winding uses 8.4~mm-width coils~\cite{Bell:1999vf}. The main
purpose of the graded current density is to maintain a high field
uniformity in the bore of the solenoid, which is also a benefit for
ePHENIX.  This design feature enhances the momentum analyzing power in
both the \egoing and \hgoing directions.

A magnetic flux return system, consisting of the forward
steel/scintillator hadron calorimeter, a flaring steel lampshade, and
a steel endcap not only returns the flux generated by the solenoid,
but shapes the field in order to aid the momentum determination for
particles in the \hgoing and \egoing directions.  As shown
in~Figure~\ref{fig:straw-man}, the flux return system consists of the
following major components:

\begin{itemize}
\item Forward steel/scintillator hadron calorimeter, at $z=3.5$ to $4.5$~m
  \item Steel flux shaping lampshade, along the $\eta\sim1$ line
  \item Barrel steel/scintillator hadron calorimeter, from $r=1.8$ to $2.8$~m
  \item Steel end cap, at $z=-2.1$ to
    $-2.7$~m and $r>90$~cm
\end{itemize}

The magnetic field lines were calculated and cross checked using three
different 2D magnetic field solvers (POISSION, FEM, and OPERA) and are
shown in Figure~\ref{fig:straw-man}. In the central region, a 1.5~Tesla
central field along the electron beam direction is produced. The field
strength variation within the central tracking volume is less than
$\pm3\%$.

\section{Vertex and Tracking}
\label{sec:VertexTracking}

\begin{figure}
  \centering
  \includegraphics[trim = 33 0 24 19, clip, width=0.95\textwidth]{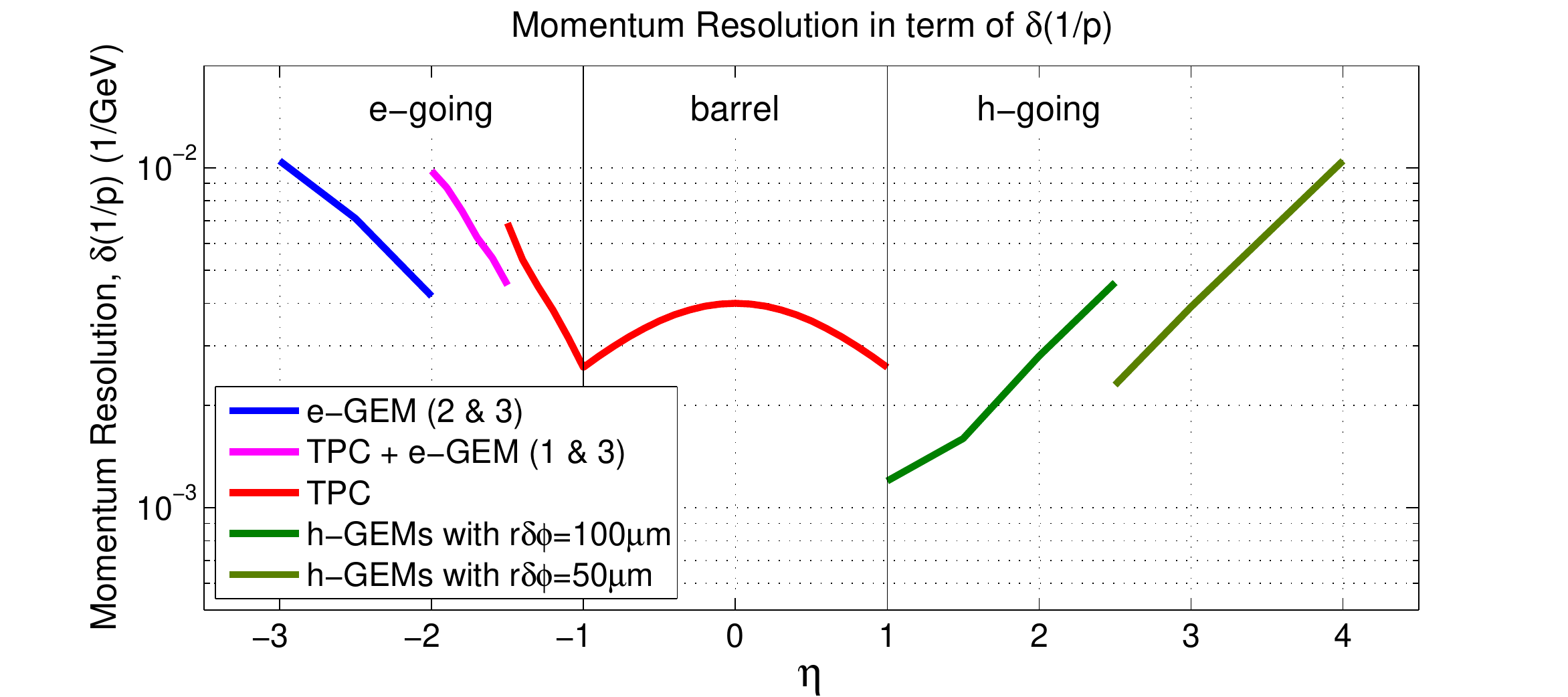}

  \caption{Momentum resolution over the full pseudorapidity coverage
    of the planned tracking system in the high momentum limit.
    Multiple scattering contribution to the relative momentum resolution 
    (not shown on the plot) was studied with GEANT4 simulation, 
    and found to vary from below 1\% at low pseudorapidity 
    to $\sim$3\% at $|\eta|$=3.}
  \label{fig:momentum_resolution}
\end{figure}

The $z$-location of the primary event vertex will be determined using
a timing system enabling a precision of $\Delta z \leq 5$~mm. The
ePHENIX tracking system utilizes a combination of GEM and TPC trackers
to cover the pseudorapidity range of $-3<\eta<4$. The momentum
resolution for the full device is summarized in
Figure~\ref{fig:momentum_resolution}.

\subsection {Event vertex measurement}

The vertex information is used for the determination of photon
kinematics and for assisting the track fitting.  Precise vertex
information is important for momentum determination in the \egoing
direction, where tight space constraints limit the possible number of
tracking planes.  The location of the vertex will be measured by:

\begin{itemize}
\item For non-exclusive processes, we propose to identify the
  $z$\nobreakdash-location for the vertex using timing information
  from a BBC detector in the \hgodir in
  coincidence with the electron beam RF timing. The BBC detector
  covers $\eta = 4$--5 at $z = 3.0$~m. A timing resolution of
  30~ps or better enables the measurement of the vertex with
  resolution of $\Delta z=5$mm.  It leads to a sub-dominant error for
  the momentum determination for the \egoing direction ($\delta p/p =
  2$\%).  This timing resolution can be provided by the existing
  technology of Multigap Resistive Plate Chamber
  (MRPC)~\cite{An:2008zzc} or by microchannel plate detectors (MCP)
  photomultiplier~\cite{Adams:2013} with a thin quartz \v{C}erenkov
  radiator, a technology which is under active current development.

\item We plan to measure the average transverse beam position by
  accumulating tracking information over the course of a one hour run.
  The statistical precision for the beam center determination is
  expected to be much smaller than the distribution of the transverse
  collision profile ($\sigma_{x,y}\sim 80$~$\mu$m), and therefore
  a negligible contribution to the uncertainty for event-by-event
  vertex determination.

\end{itemize}

\subsection {Tracking in the central region, $-1<\eta<1$}

A fast, compact Time Projection Chamber (TPC) will be used for
tracking in the central region, occupying the central tracking volume
of $r = 15$--80~cm and $|z| < 95$~cm and covering $-1 < \eta < 1$. A
TPC will provide multiple high resolution space point measurements
with a minimal amount of mass and multiple scattering.  The design is
based on a GEM readout TPC, similar to a number of TPCs that have
either already been built or are currently under design. For example,
the LEGS TPC~\cite{Geronimo:2005} utilized a fine chevron-type
readout pattern with a pad
size of 2~mm $\times$ 5 mm and achieved a spatial resolution $\sim$
200~$\mu$m. The use of such a readout pattern
helps minimize the total channel count for the electronics and hence
the total cost. The GEM TPC upgrade for
ALICE~\cite{ALICE_TPC:2012,Ketzer:2013laa} and the large GEM readout
TPC for ILC~\cite{Abe:2010aa,Schade:2011zz} are other examples of
large GEM TPCs that have recently been studied.

It is assumed that the TPC will have a single high voltage plane at $z
= 0$~cm and be read out on both ends, resulting in a maximum drift
distance $\sim 95$~cm. It will use a gas mixture with a fast drift
time, such as 80\% argon, 10\% CF$_4$ and 10\% CO$_2$, which, at an
electrical field of 650~V/m, achieves a drift speed $\sim
10$~cm/$\mu$s, and would result in a maximum drift time of 10~$\mu$s.
With a position resolution of $\sigma(r\Delta \phi) =300$~$\mu$m and
65 readout rows, the expected transverse momentum resolution would be $\delta
(1/p_{T}) = 0.4\%/($GeV$/c)$ for high momentum tracks.

\subsection {Tracking in \hgodir, $\eta>1$}

The design of the magnetic flux return enables tracking in the \hgoing
direction in the main and fringe fields of the BaBar magnet. Compared
to a compact solenoid with no current density gradient, the BaBar
magnet system improves the momentum analyzing power for forward tracks
by about a factor of four due to two main factors: 1) the BaBar magnet
has a length of 3.5~m, which provides a longer path length for
magnetic bending; 2) the higher current density at the ends of the
solenoid improves the magnetic field component transverse to forward
tracks, and therefore provides higher analyzing power.

The tracking system at high $\eta$ in the \hgoing direction utilizes four stations of GEMs.

\begin{itemize}
\item Station~1 consists of two planes with complementary $\eta$
  coverages. They are located at $z = 17$ and 60~cm, respectively,
  covering a radius of $r = 2$--15~cm.
\item Stations~2--4 are at $z = 150$, 200, 300~cm, respectively,
  covering $\eta = 1$--4.
\end{itemize}

The readout planes for these devices are optimized to preserve high
position resolution in the azimuthal direction ($\sim 200$~$\mu$m in
$r \delta \phi$ using a chevron-type readout with a pad size similar to the
central TPC) and $\sim$10--100 mm in $\delta r$, while minimizing the
readout channel cost. However, the $r$-$\phi$ resolution can be improved
to be better than 100~$\mu$m, even for tracks at larger angles (up to
45~degrees), by the use of mini-drift GEM detectors, in which a small
track segment, or vector, is measured for each track at each measuring
station.  These detectors, which are currently under development
~\cite{EIC_Tracking_RD:2013}, would provide improved position
resolution with less material and lower cost than multiple stations of
planar GEM detectors. For this letter, we assumed that a high resolution
GEM readout pattern (1~mm wide chevron-type readout) with a  $r \delta \phi \sim 50$~$\mu$m
for the inner tracking region ($\eta>2.5$).
For the outer tracking region ($1<\eta<2.5$), mini-drift GEM
with 2~mm chevron-type readout provide $r \delta \phi \sim 100$~$\mu$m.
The momentum resolution is estimated in Figure~\ref{fig:momentum_resolution}.

It should be noted that the size of the GEM trackers for Stations 2--4
are quite large ($\sim 5$--20~m$^2$).  It is currently challenging to
produce such large GEM foils and to do so at an affordable cost.
However, there has been substantial progress in this area in recent
years at CERN due to the need for large area GEM detectors for the CMS
Forward Upgrade~\cite{CMS_GEM:2012}. CERN has developed a single mask
etching technology which allows fabrication of very large area GEMs
(up to 2~m $\times$ 0.5~m), and they plan to transfer this technology
to various commercial partners (such as Tech Etch in the US, which
supplied the GEM foils for the STAR Forward GEM Detector).  We
anticipate being able to procure such large area GEM detectors by the
time they are needed for EIC.

\subsection {Tracking in the \egodir, $\eta<-1$}

The electron direction tracking is designed to fit in the space
limited by the DIRC ($R < 80$~cm) and the electromagnetic calorimeter
($z > -100$~cm). Three GEM tracking stations, located at $z = 30$, 55
and 98~cm, are used in combination with the TPC and vertex information
to determine the momentum vector.

\begin{itemize}
  \item For $\eta = -1.5$ to $-1$, TPC track segment and vertex are used
  \item For $\eta = -2.0$ to $-1.5$, vertex, TPC track segment, GEM station 1 and 3 are used.
  \item For $\eta = -3.0$ to $-2.0$, vertex, GEM station 2 and 3 are used.
\end{itemize}

Similar to the \hgoing direction,
the position resolution for these detectors is $r\Delta\phi$ 50~$\mu$m for $-3<\eta<-2$
using 1~mm wide chevron-type readout. For $-2<\eta<-1$, the mini-drift GEM technology~\cite{EIC_Tracking_RD:2013}
and 2~mm wide chevron-type readout provide $r \delta \phi \sim 100$~$\mu$m.
The radial resolution is $\delta r = 1$~cm (stations 1 and 2) and
$\delta r = 10$~cm (station 3). As shown in
Figure~\ref{fig:momentum_resolution}, a momentum resolution of $\Delta
p/p < 5\%$ can be achieved for tracks of $p<4$~GeV$/c$ and
$-1<\eta<-3$, which is sufficient for the calorimeter
$E$\nobreakdash-$p$ matching cut for the electron
identification.
For DIS kinematics reconstruction the tracking radial resolution is not
crucial as enough precision for scattered electron polar angle measurements
will be provided by the EMCal, see Section~\ref{sec:resolution}.

\section{Electromagnetic calorimeters}
\label{sec:EMCAL}

The ePHENIX detector will have full electromagnetic calorimeter
coverage over $-4<\eta<4$.
The sPHENIX barrel electromagnetic calorimeters will also be used in
ePHENIX, covering $-1<\eta<1$ with an energy resolution of
$\sim12\%/\sqrt{E}$.
In addition, crystal and lead-scintillator
electromagnetic calorimeter are planned for the \egoing and \hgoing
direction, respectively.
Optimization of the design of the barrel and endcap calorimeters
will aim for uniform response in the overlap region between $-1.2<\eta<-1$.

\subsection {Crystal Electromagnetic calorimeter}
\label{sec:CrystalEMCAL}

\begin{figure}
\centering
\includegraphics[width=0.4\textwidth]{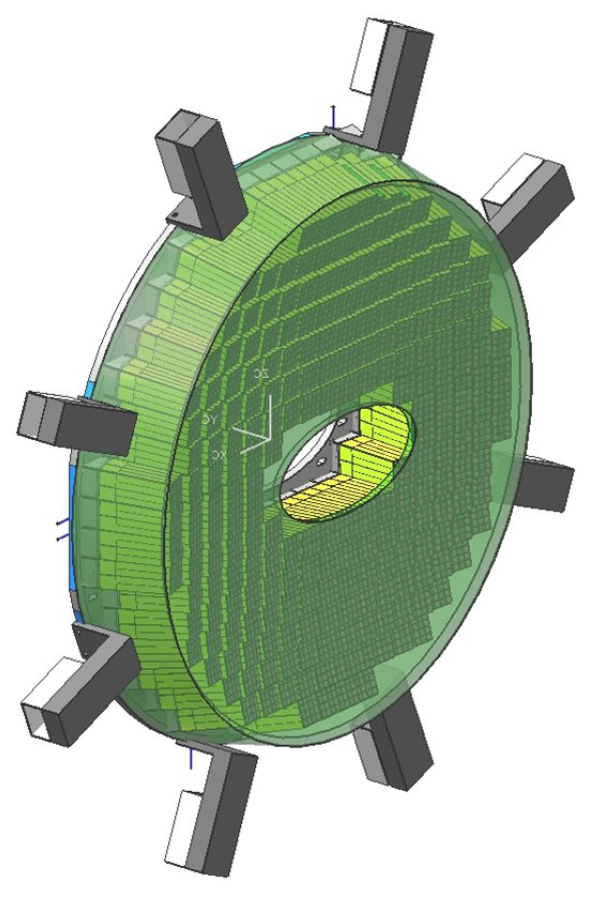}
\caption{PANDA Crystal Endcap Calorimeter~\cite{PANDA:2008uqa}. The
  PWO crystal modules are shown in green color, which is projective
  towards the target.}
\label{fig:PANDA_Endcap}
\end{figure}

The calorimeter on the electron\nobreakdash-going side consists of an
array of lead tungstate (PbWO$_4$) crystals (commonly known as PWO),
similar to the PANDA endcap crystal calorimeter shown in
Figure~\ref{fig:PANDA_Endcap} ~\cite{PANDA:2008uqa}. An enhanced light
output version of lead tungstate (PWO-II) was chosen to provide high
light yield ($\sim 20$~p.e./MeV at room temperature) at a moderate
cost ($\sim$ \EUR{5}/cm$^3$). It will provide an energy resolution
$\sim 1.5$\%/$\sqrt{E}$ and position resolution better than
$3~\mathrm{mm}/\sqrt{E}$ in order to measure the scattered electron energy
and angle in the \egoing direction down to low momentum with high precision.

The ePHENIX PWO calorimeter will consist of $\sim 5000$ crystals,
compared with 4400 crystals for the PANDA endcap, and will have a
similar size and shape to the PANDA crystals. They will be $\sim 2~\mathrm{cm}
\times 2~\mathrm{cm}$ (corresponding to one $R_M^2$) and will be read out with
four SiPMs. This is different than the PANDA readout, which uses large
area ($\sim 1$~cm$^2$) APDs. The SiPMs will provide higher gain, thus
simplifying the readout electronics, and will utilize the same readout
electronics as the other calorimeter systems in sPHENIX. It is also
expected that the cost of SiPMs will be less than that of APDs
covering the same area by the time they are needed for ePHENIX.

\subsection {Lead-scintillator electromagnetic calorimeter}

The electromagnetic calorimeter in the \hgoing direction consists of a
lead-scintillating fiber sampling configuration, similar to the
tungsten-scintillating fiber calorimeter in the central sPHENIX
detector. Lead is used instead of tungsten in order to reduce the
cost, but it is otherwise assumed to be of a similar geometry. It will
cover the rapidity range from $1<\eta<4$ and have 0.3~$X_0$ sampling
(2~mm lead plates) with 1~mm scintillating fibers, which will give an
energy resolution $\sim 12$\%/$\sqrt{E}$.  The segmentation and
readout will also be similar to the central tungsten calorimeter,
with $\sim 3~\mathrm{cm} \times 3~\mathrm{cm}$ towers (roughly 1~$R_M^2$) that are read
out with SiPMs. This segmentation leads to $\sim 26$K towers. By using
the same type of readout as the central calorimeter, the front end
electronics and readout system will also be similar, resulting in an
overall cost savings for the combined calorimeter systems.

\section{Hadron calorimeter}

The hadron calorimeter in the \hgoing direction consists of a
steel-scintillating tile design with wavelength shifting fiber
readout, similar to the central sPHENIX hadron calorimeter. It will be
$\sim 5$ $L_{abs}$ thick and cover a rapidity range from $1<\eta<5$.
The steel in the absorber will also serve as part of the flux return
for the solenoid magnet. The segmentation will be $\sim 10~\mathrm{cm}
\times 10~\mathrm{cm}$, resulting in $\sim 3000$ towers. The readout will also be
with SiPMs, similar to the central sPHENIX HCAL, which will again
provide an advantage in being able to use a common readout for all of
the calorimeter systems.

\section{Hadron PID detectors}
\label{sec:HadronPIDDetectors}

\begin{figure}
\centering
\subfloat[Aerogel and RICH gas radiators for hadron-going
direction]{\includegraphics[width=.4\textwidth]{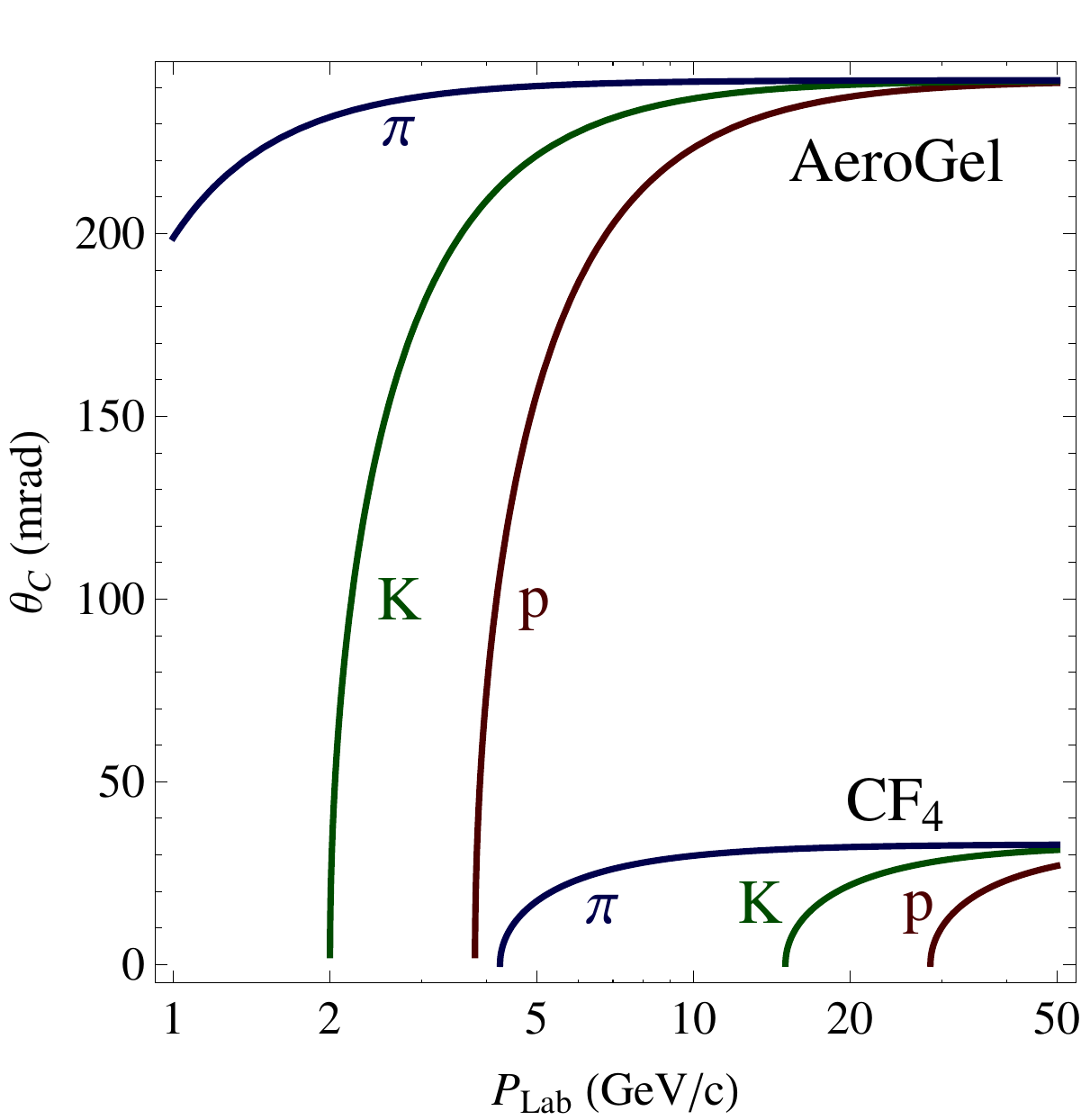}}
\subfloat[Fused silica radiator for barrel DIRC detector. Data are
measured by the BaBar DIRC~\cite{Adam:2004fq}]{\includegraphics[bb=0bp
  7bp 227bp 227bp,clip,width=0.45\textwidth]{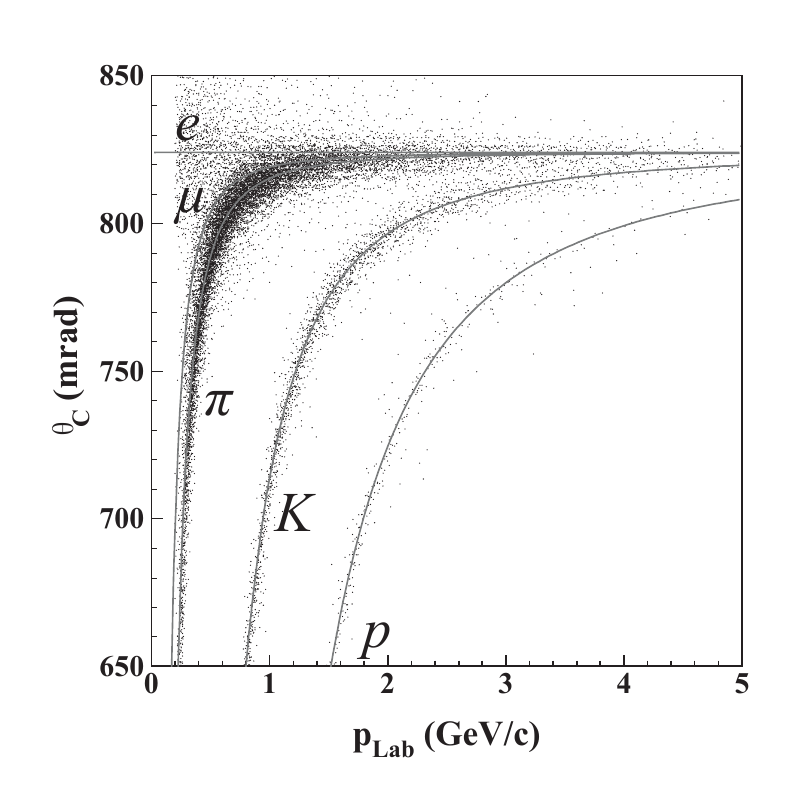}}
\caption{\v{C}erenkov angle versus momentum for various particle
  species.}
\label{fig:RICH_Angles}
\end{figure}

Hadron PID is planned for the \hgoing and barrel regions, covering
$-1.2 < \eta < 4$. In the \hgoing direction, two PID detectors cover
complementary momentum range: a gas-based RICH detector for the higher momentum
tracks and an aerogel-based RICH detector for the lower momentum region.  As in
the BaBar experiment~\cite{Adam:2004fq}, a DIRC detector identifies
hadron species in the central barrel.  In addition, the TPC detector assists
with PID by providing $dE/dx$ information for the low momentum region.

\subsection {\label{sec:RICH}Gas RICH detector}

\begin{figure}
\centering
\includegraphics[width=0.65\textwidth]{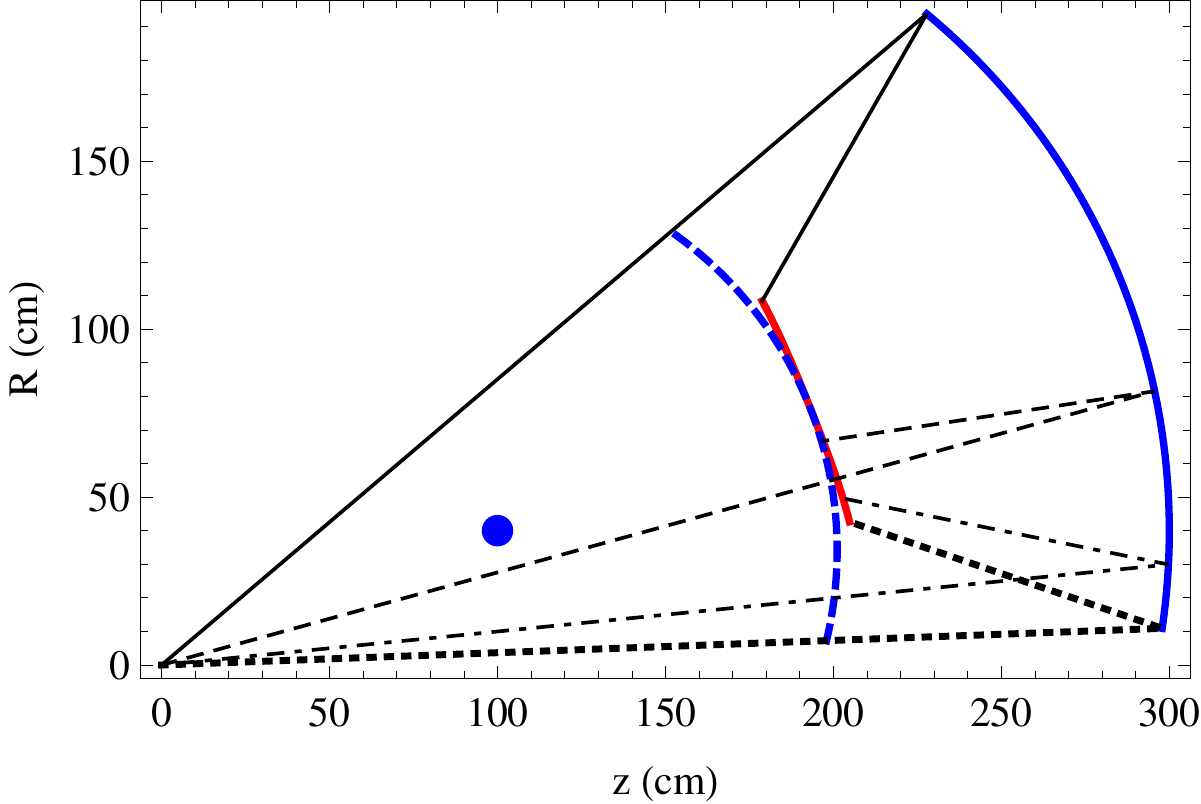}
\caption{The cross-section of the gas-based RICH detector in the $r$-$z$ plane
  that crosses the mirror center. The interaction point is
  centered at $(0,0)$. The geometric center of the mirror
  is shown as the blue dot at $(r, z) = (40~\mathrm{cm}, 100~\mathrm{cm})$. The mirror and
  RICH entrance window are shown by the solid and dashed blue curves,
  respectively.  Several example tracks and the central axis of their
  \v{C}erenkov light cone are illustrated by the black lines. The
  \v{C}erenkov photons are reflected by the mirror to the focal plane,
  shown in red. }
\label{fig:RICH_setup}
\end{figure}

High momentum hadron PID is provided by an optically focused RICH
detector using a gas radiator. The main features for this RICH setup
are
\begin{itemize}
\item One meter of CF$_4$ gas is used as the \v{C}erenkov radiator.
  The pion, kaon and proton thresholds are 4, 15 and 29 GeV,
  respectively.
\item \v{C}erenkov photons are focused to an approximately flat focal
  plane using spherical mirrors of 2~m radius, as shown in
  Figure~\ref{fig:RICH_setup}. The geometric center of the mirror is
  at $(r,z) = (40~\mathrm{cm}, 100~\mathrm{cm})$, as highlighted by the blue dot.
\item There are six azimuthal segmented RICH sextants.
\item The photon detector consists of CsI-coated GEM
  detectors~\cite{Anderson:2011jw}, which are installed on the focal
  plane. The CsI coating converts the \v{C}erenkov photons into electrons
  which are then amplified by the GEM layers and readout through
  mini-pads. The photon detector for each RICH sextant assumes a
  roughly triangle shape and covers an area of 0.3~m$^2$.
\end{itemize}

\begin{figure}
\centering
\includegraphics[width=0.65\textwidth]{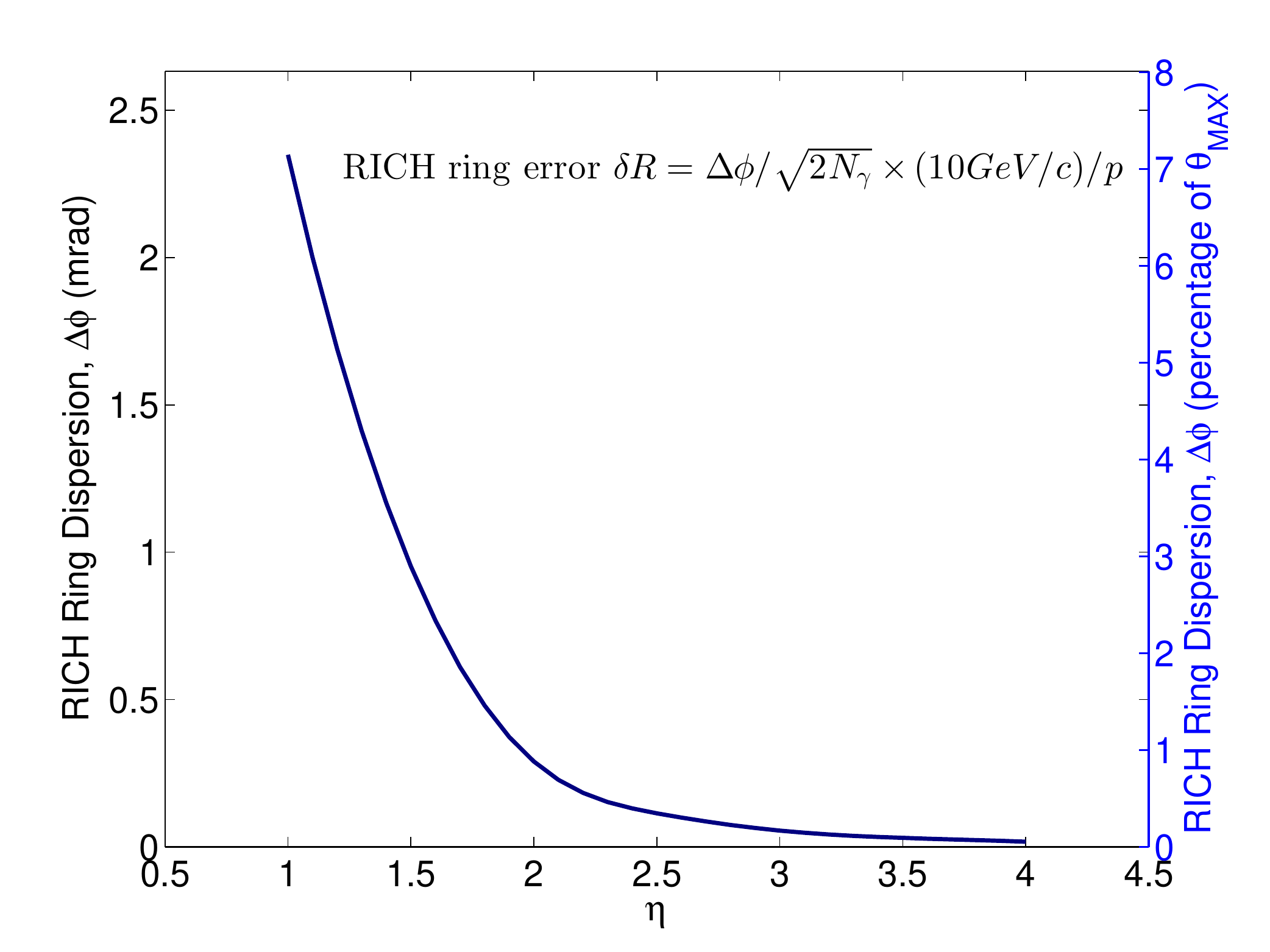}
\caption{Azimuthal angular dispersion of gas-based RICH ring due to fringe magnetic field for a $p=10$~GeV$/c$ track. It is compared to the maximum RICH ring angle as shown on the right vertical axis.}
\label{fig:RICH_field_effect}
\end{figure}
Two distortion effects were estimated to be sub-dominant in error contributions for most cases:
\begin{itemize}
\item Strong residual magnetic field ($\sim0.5$~Tesla) are present in the
  RICH volume. This field will bend the tracks as they radiate
  photons, and therefore smear the \v{C}erenkov ring in the azimuthal
  direction. However, the field design ensures that the field
  component is mostly parallel to the track inside RICH and therefore
  this smearing effect is minimized. The RMS size of the smearing,
  $\Delta\phi$, is evaluated as in Figure~\ref{fig:RICH_field_effect}.
  The uncertainty contribution to the RICH ring angular radius is
  $\delta R=\Delta\phi/\sqrt{2N_{\gamma}}(10\mbox{ GeV}/c)/p$, which
  is sub-dominant comparing to the photon measurement error for
  $\eta>1.5$. The field contribution was included in the RICH
  performance estimation.
\item For tracks that originate from an off-center vertex, their focal
  point may be offset from the nominal focal plane as shown in
  Figure~\ref{fig:RICH_setup}. The effect is $\eta$ dependent. 
  For the most extreme case, that a track of $\eta = 1$ originates from
  the vertex of $z = 40$~cm (1.5 sigma of expected vertex width), an
  additional relative error of $5\%/\sqrt{N_{\gamma}}$ 
  is contributed to the ring radius measurement, which averages over 
  all vertices to below $2\%/\sqrt{N_{\gamma}}$ contribution.
  For high $\eta$ tracks, the
  difference is negligible comparing to the nominal RICH error. 
\end{itemize}

We simulated the RICH performance with a radiator gas CF$_{4}$
(index of refraction 1.00062). We use \pythia to generate 
the momentum distributions for pions, kaons, and protons.  
For each particle species, we use the momentum resolution 
and RICH angular resolution, to calculate the particle 
mass $m(p,\theta_{Crk})$ distribution. For higher momentum tracks 
the combined information from tracking system and energy deposit in HCal 
helps to improve momentum resolution particularly at higher rapidities, 
where momentum resolution from tracking degrades. 
For example, at pseudorapidity $\eta$=4, the tracking momentum 
resolution for 50~GeV/c tracks is $\sim$50\% 
(see Figure~\ref{fig:momentum_resolution}), while HCal can provide 
energy measurements with precision $100\%/\sqrt{50\mathrm{[GeV]}} \sim 14\%$. 
Our simulation showed that the HCal is very effective in improving 
the resolution for high momentum track measurements even when 
this and other tracks (usually with lower momenta) are merged in a single 
cluster of deposited energy in HCal. 
In such a case, the contribution of lower energy tracks in HCal can be 
evaluated and subtracted based on momentum measurements in tracking system. 

\begin{figure}
  \centering
  \includegraphics[bb=0bp 0bp 550bp 262bp,clip,width=1.0\textwidth]{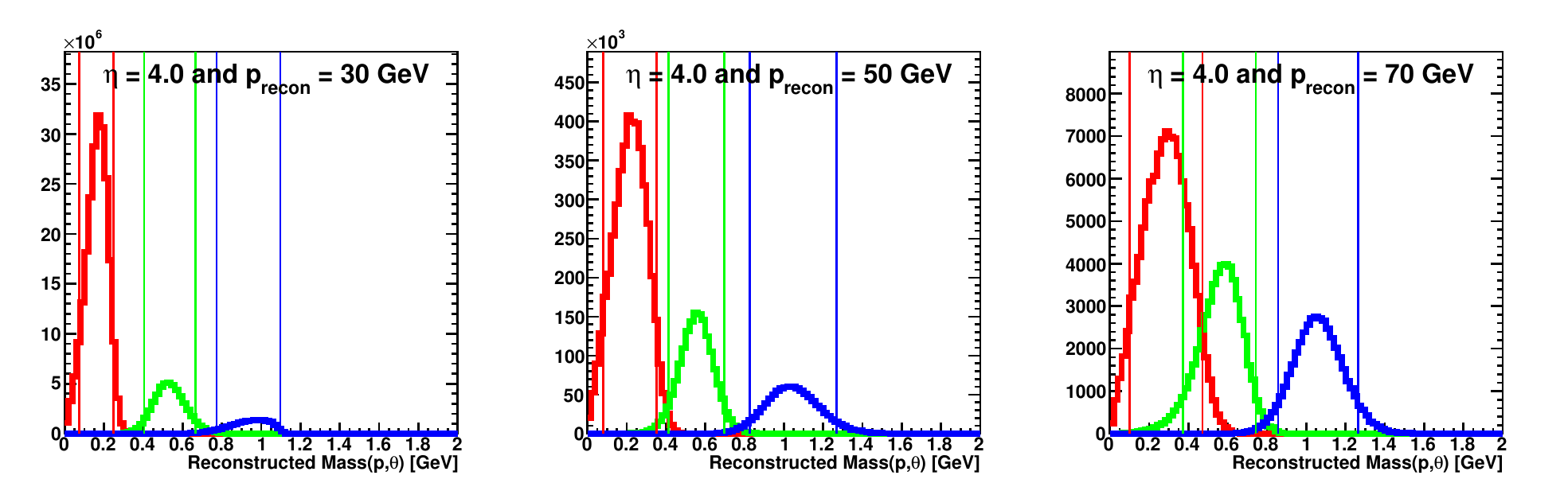}
  \caption{Reconstructed mass distribution via m(p,$\theta_{Crk}$) at $\eta=4$
    for reconstructed momenta 30 GeV/c (left), 50 GeV/c (middle) and 70 GeV/c 
    (right), for pions (red), kaons (green) and protons (blue), 
    with the parent momentum and particle
    abundances from the \pythia generator.  
    Vertical lines indicate the symmetric mass cuts corresponding 
    to 90\% efficiency. 
    Note that particle true momentum is on the average smaller than 
    reconstructed momentum, see Figure~\ref{fig:rich_purity}. 
  }
  \label{fig:rich_mreco}
\end{figure}

\begin{figure}
  \centering
  \includegraphics[width=1.0\textwidth]{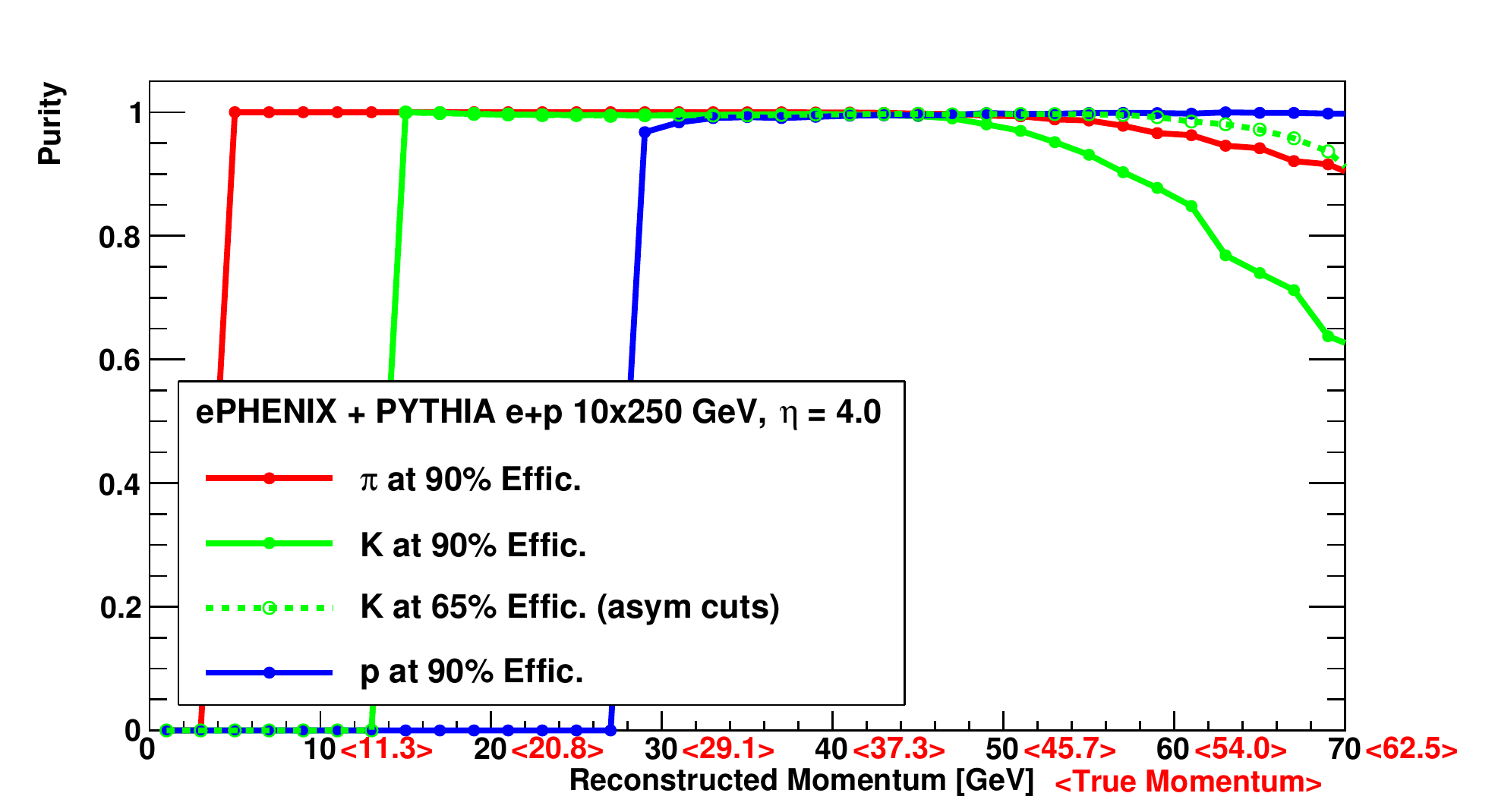}
  \caption{ $\pi, K, p$ purities at
    pseudorapidity 4.0 as a function of reconstructed momentum, 
    based on symmetric cut on reconstructed mass corresponding to 90\% 
    efficiency (solid lines), and asymmetric cut with stricter selection 
    on the kaons with efficiency 65\% (dashed line);
    Also indicated in angle brackets are the values of the average 
    true momentum at each reconstructed momentum, which are different due 
    to momentum smearing and sharply falling momentum spectra.
    }
  \label{fig:rich_purity}
\end{figure}

Figure~\ref{fig:rich_mreco} shows mass distributions for the 
most challenging high rapidity region $\eta$=4 for different reconstructed 
track momenta. We make a symmetric 90\% efficiency cut on the mass 
distributions, and calculate the purity for $\pi, K, p$, 
shown in Figure~\ref{fig:rich_purity}. One can see high purity for all 
particle species up to momenta $\sim$50 GeV/c. 
Introducing asymmetric cuts on the mass distributions (and sacrificing some 
efficiency) extends further our capabilities for high purity hadron 
identification. 

It is notable that the limitation on the mass resolution comes 
from the estimated 2.5\%
radius resolution per photon for the RICH from the EIC R\&D RICH
group.  Our calculation includes the effect of the magnetic field
distortion mentioned above, which is sub-dominant.  This is a somewhat
conservative estimate and LHCb and COMPASS have quoted values near 1\%
per photon.  The R\&D effort is working towards the best radius
resolution, though there are challenges in having the light focus and
readout within the gas volume in this configuration compared with LHCb
or COMPASS.

\subsection {Aerogel RICH detector}

The aerogel detector will provide additional particle ID for kaons in
the momentum range $\sim 3$--15~GeV/c when used in conjunction with
the gas RICH. Pions can be identified by the signal they produce in
the gas RICH starting at a threshold of $\sim$ 4 GeV/c,\ and kaons
will begin producing a signal in the aerogel at a threshold $\sim$ 3
GeV. Reconstructing a \v{C}erenkov ring in the aerogel enables one to
separate kaons from protons up to a momentum $\sim 10$~GeV/c with
reduced efficiency above that.

\begin{figure}
\centering
\includegraphics[width=0.6\textwidth]{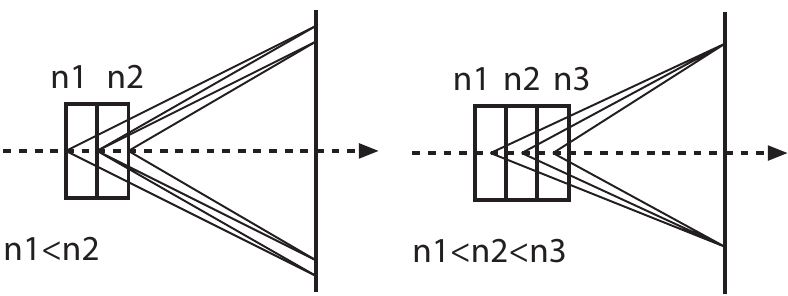}
\caption{Approximate focusing method using two (left) and three
  (right) layers of aerogel with slightly different indicies of
  refraction proposed by Belle~II~\cite{Iijima:2005qu}}
 \label{fig:Aerogel_setup}
\end{figure}

Measuring a ring in the aerogel detector is a challenging technical
problem for a number of reasons. Due to its relatively low light
output, it will require detecting single photons in the visible
wavelength range with high efficiency inside the rather strong fringe
field of the superconducting solenoid. Also, due to the limited space
available, it is difficult to have a strong focusing element in the
RICH to focus the light into a ring in a short distance. One
possibility for how this might be accomplished has been proposed by
the Belle II experiment~\cite{Iijima:2005qu} and is shown in
Figure~\ref{fig:Aerogel_setup}. It uses several layers of aerogel with
slightly different indices of refraction to achieve and approximate
focusing of the light onto an image plane located behind the radiator.
It should be possible to add additional layers of aerogel and optimize
their thickness for producing the best quality ring for kaons using
this technique, and therefore achieve good kaon-proton separation up
to the highest momentum.
One possibility for the photon detector would be large area
Microchannel Plate detectors (MCPs), such as those being developed by
the Large Area Picosecond Photodetector (LAPPD)
Collaboration~\cite{Adams:2013}. This effort is based on utilizing flat
panel screen technology to produce large area MCPs at very low cost,
while also preserving their excellent timing resolution
(typically $\sim$ 20-30 ps). These devices would use multi-alkali
photocathodes, which would be suitable for detecting the Cherenov light
from aerogel with high efficiency, and also provide high gain for detecting
single photoelectons. The excellent time resolution would also provide
additional time of flight capability when used in conjunction with the
BBC to further enhance hadron particle ID. While this is still an R\&D effort,
it has already produced very encouraging results and has substantial support
within the high energy physics community, and we feel that this would offer
an attractive low cost, high performance readout for the aerogel detector.

\subsection {DIRC}

The main form of particle ID in the central region will be provided by
a DIRC (Detection of Internally Reflected \v{C}erenkov Light). The
DIRC will be located at a radius of $\sim 80$~cm and extend $\sim
8$--10~cm in the radial direction. As we will be using the BaBar
magnet for ePHENIX, it would be a major benefit to also acquire the
BaBar DIRC, which was specifically designed to fit inside this magnet,
and would completely satisfy the physics requirements for ePHENIX.
However, since it is not certain at this time that the BaBar DIRC will
be available for ePHENIX, we consider it more as a model for the type
of DIRC that would be required in terms of its construction and
performance.

\begin{figure}
  \centering
  \subfloat[Nominal elevation view]{
    \includegraphics[width=0.64\textwidth]{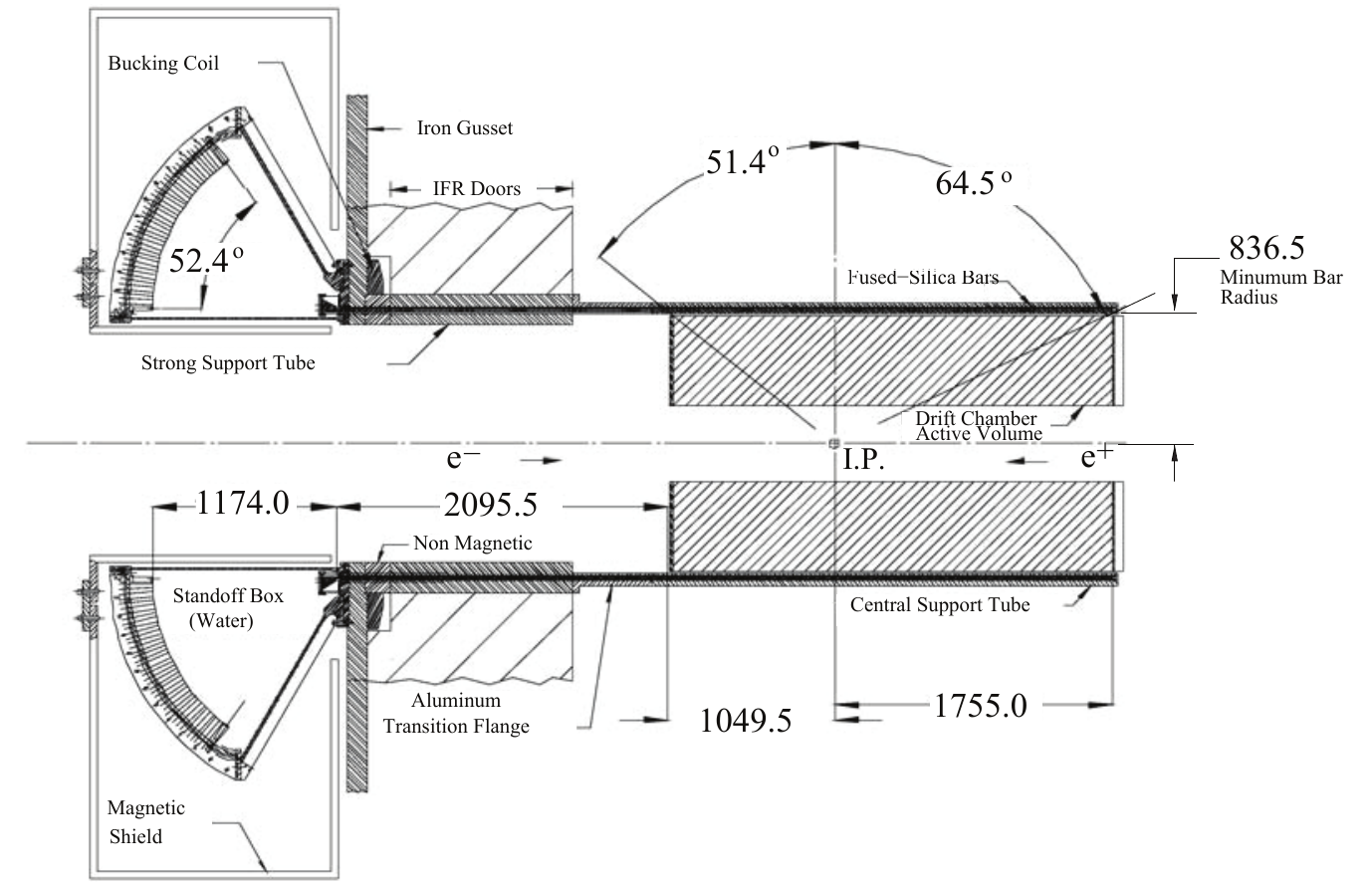}
  }\subfloat[Cross section through the central tube]{
    \includegraphics[width=0.35\textwidth]{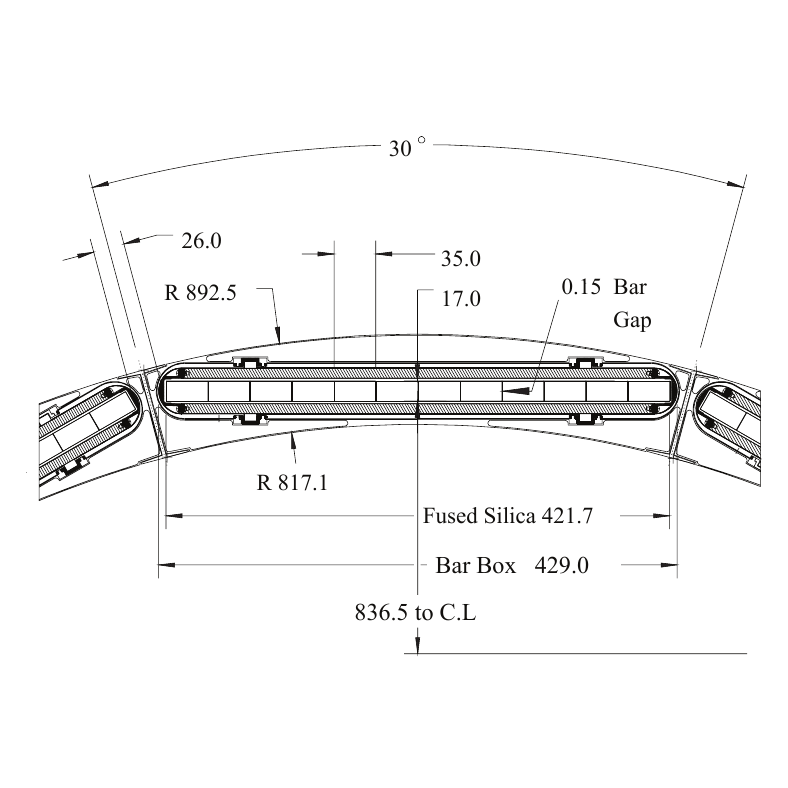}
  }
  \caption{BaBar DIRC geometry~\cite{Adam:2004fq}. All dimensions are given in mm.}
  \label{fig:BaBar_DIRC}
\end{figure}

The BaBar DIRC, shown in Figure~\ref{fig:BaBar_DIRC}, consists of 144
precision fabricated quartz radiator bars that collect \v{C}erenkov light
produced by charged particles traversing the bars. In the BaBar DIRC,
the quartz bars were read out on one end utilizing a large water
filled expansion volume to allow the light to spread out and be read
out using a large number (over 10,000) 28~mm diameter photomultiplier
tubes.

The BaBar design, while allowing for a conventional PMT readout
without the use of any focusing elements, requires a large expansion
volume and this places stringent demands on the mechanical
specifications for the detector. After the shutdown of BaBar at SLAC,
it was proposed to use the DIRC in the SuperB Experiment in Italy. In
doing so, it was also proposed to convert the original DIRC into a
Focusing DIRC (FDIRC)~\cite{Grauges:2010fi}, which would utilize
mirrors at the end of the radiator bars, allowing for a considerable
reduction in the size of the expansion region,
more highly pixellated PMTs, and an
overall expected improvement in performance. We would therefore
propose the same modification of the BaBar DIRC for ePHENIX, or would
construct a similar FDIRC if the BaBar DIRC were not available.

\begin{figure}
\centering
\includegraphics[width=.7\textwidth]{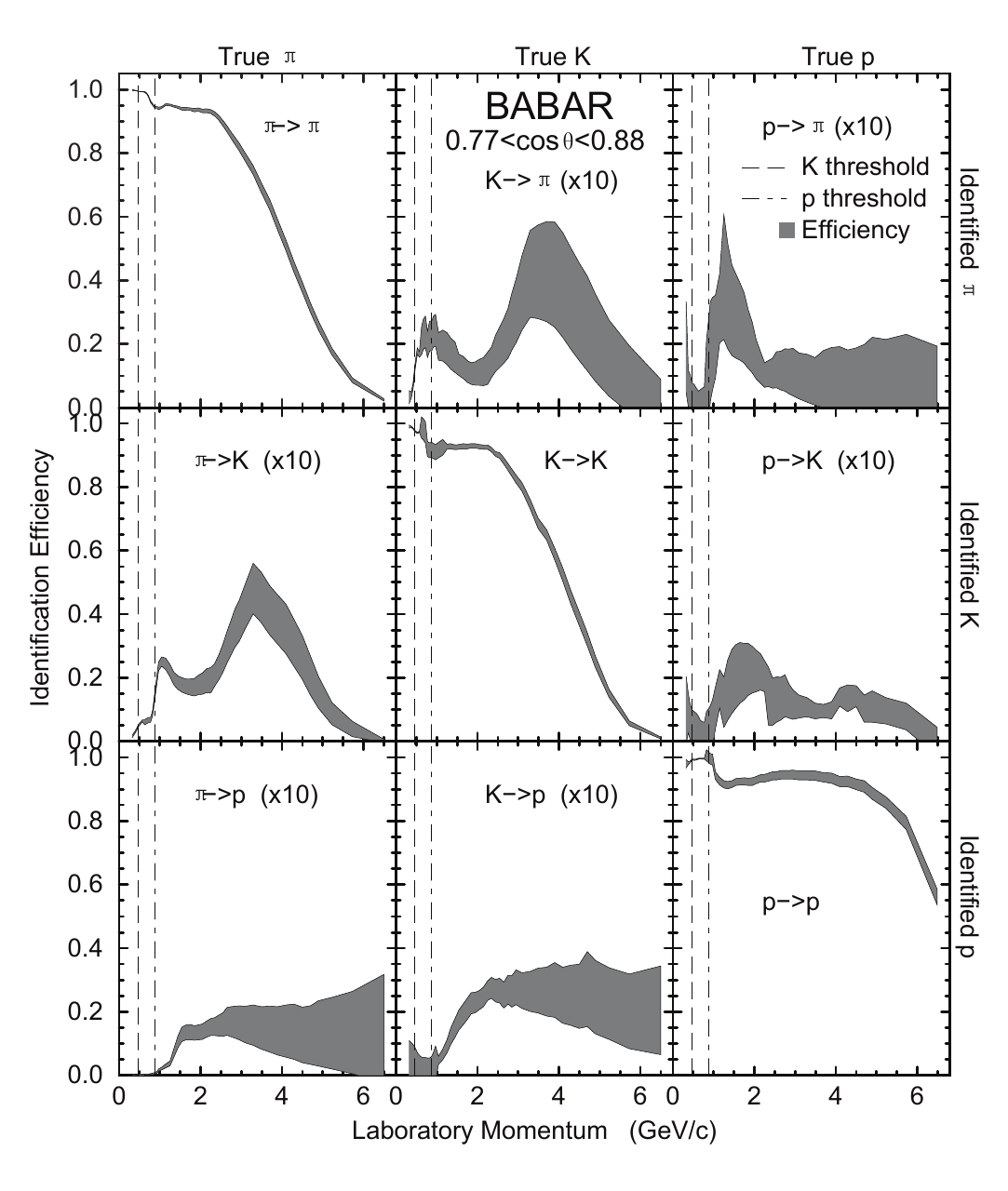}
\caption{Simulated PID Efficiency matrix and its uncertainty for
  $1.0<\eta<1.4$ region, utilizing combined information of the BaBar
  DIRC and $dE/dx$ measured in the tracking
  detector~\cite{Adam:2004fq}.  Note that the off-diagonal efficiency
  values are scaled by a factor of 10.}
\label{fig:FDIRC_PID}
\end{figure}

Similar to the BaBar technique~\cite{Adam:2004fq}, the hadron PID in
the barrel will be analyzed using an event likelihood analysis with
the DIRC and TPC $dE/dx$ information simultaneously.  A $dE/dx$
measurement in the TPC gives very good hadron separation for very low
momentum particles.  But the ability of that technique to separate
K-$\pi$ and p-K drops off quickly around 0.5~GeV$/c$ and 0.8~GeV$/c$,
respectively.  Meanwhile, the pions and kaons exceed their respective
DIRC \v{C}erenkov thresholds in this momentum region, as shown in
Figure~\ref{fig:RICH_Angles}. Therefore, the DIRC sensitivity for
K-$\pi$ and p-K turns on sharply. A combined analysis of both pieces
of information can give high PID purity up to a few GeV$/c$, as shown
by the BaBar experiment~\cite{Adam:2004fq}.  At higher momenta, the
DIRC ring resolution limits the separation capability. As shown in
Figure~\ref{fig:FDIRC_PID}, the K-$\pi$ and p-K separation gradually
drops below plateau above momentum of 2 and 5~GeV$/c$, respectively. A
$\sim20\%$ pion and kaon efficiency can still be maintained at
5~GeV$/c$.  The vast majority of hadron kinematics in SIDIS can be
covered in the $5\times100$~GeV$/c$ collisions. In the
$10\times250$~GeV$/c$ collisions, the low to intermediate-$z$ region
in SIDIS are still well covered by this design.

\section{\label{sec:beamline}Beamline detectors}

\begin{figure}
\centering
\includegraphics[bb=0bp 100bp 720bp 400bp,clip,width=1.0\textwidth]{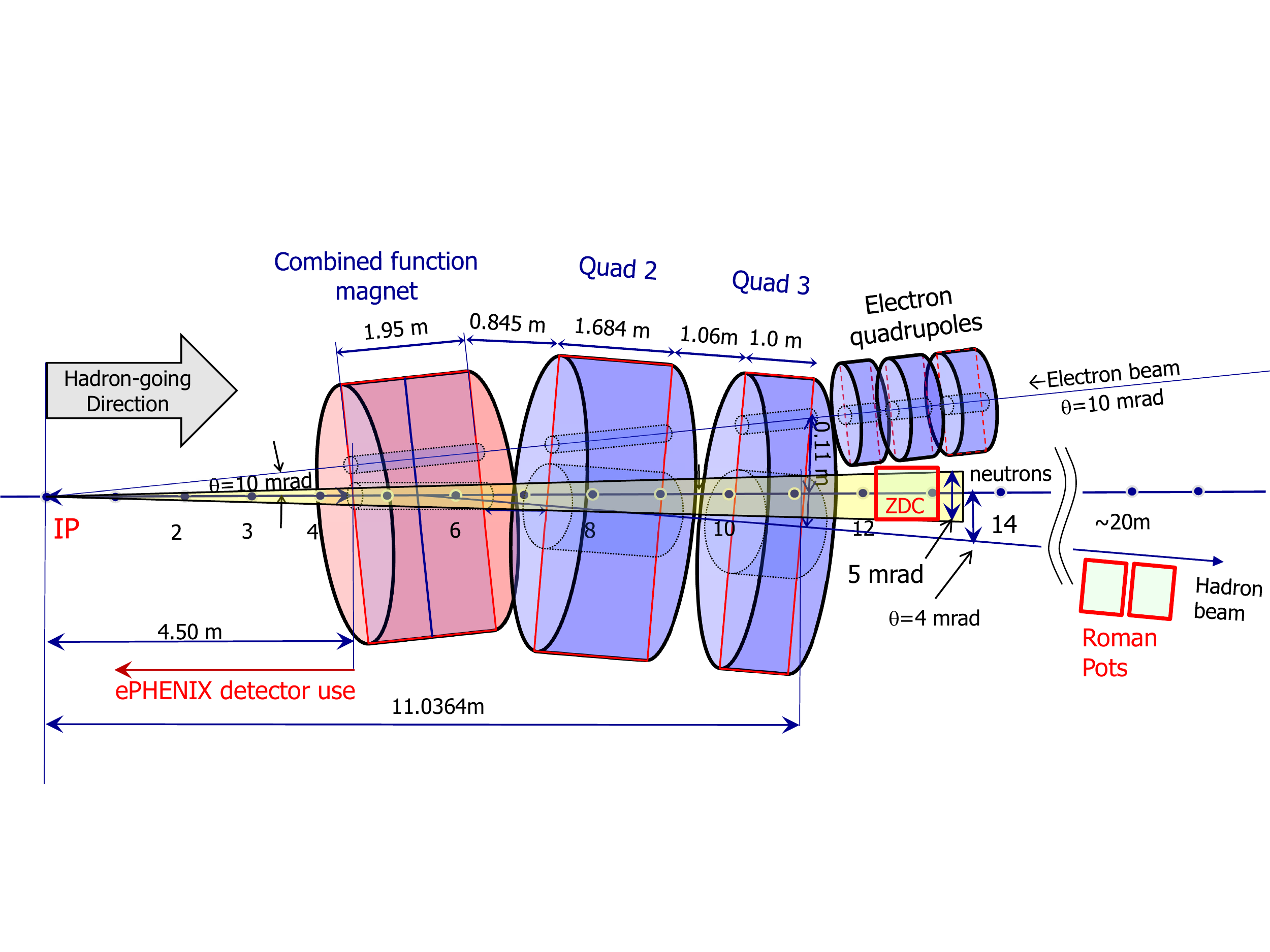}
\caption{Floor plan show the locations for ZDC and Roman Pots
 relative to the ePHENIX interaction
  point. One layout of the interaction point magnets are also
  shown~\cite{Trbojevic:2013}.}
\label{fig:ZDC}
\end{figure}

Two detectors will be installed near the outgoing hadron beam,
downstream of the ePHENIX detector. They will be included in the eRHIC
machine lattice design~\cite{Accardi:2012hwp}.

\begin{description}
\item[Zero Degree Calorimeter:] A Zero Degree Calorimeter (ZDC) is
  planned for the \hgoing direction for the ePHENIX IP. Consistent with the eRHIC IR design (Figure~\ref{fig:ZDC}),
  the ZDC will be installed about 12 meters downstream of the
  IP centered on the hadron direction at the IP. A  5~mrad cone
  opening of the IP is guaranteed by the ePHENIX detector and beam
  line magnets. The ZDC for the current PHENIX
  experiment~\cite{Adler:2000bd} and its design can be reused for this
  device.
\item[Roman Pots:] In exclusive deep inelastic $e$$+$$p$ scattering,
  the final state proton will have a small scattering angle and escape
  the main ePHENIX detector. Two silicon tracking stations (also
  called the Roman Pot spectrometer) will be installed close to the beam,
  inside the beam pipe,
  downstream in the \hgoing direction to capture such protons. Each of
  the ePHENIX Roman Pot stations utilizes four tracking modules to
  cover the full azimuthal angles. Each of the tracking modules can use
  the design of the existing STAR Roman Pots~\cite{Adamczyk:2012kn}.
  Depending on the eRHIC lattice and magnet design, their location
  will be around 20~meters from the IP. This Roman Pot spectrometer will
  provide high efficiency for the exclusive DIS events in the
  $e$$+$$p$ collisions.

\end{description}



\clearpage

\fancyhead{}
\chapter*{Acknowledgments}

We would like to acknowledge very helpful discussions and input from
many groups, including the EIC tracking consortium, and BNL's eRHIC
Task Force, Collider-Accelerator Department, and Superconducting
Magnet Division.

\backmatter

\clearpage
\phantomsection
\addcontentsline{toc}{chapter}{References}

\bibliographystyle{unsrturl}
\bibliography{references}

\begin{thebibliography}{10}

\bibitem{Aidala:2012nz}
C.~Aidala et~al.
\newblock {sPHENIX: An Upgrade Concept from the PHENIX Collaboration}.
\newblock 2012.
\newblock \href {http://arxiv.org/abs/1207.6378} {\path{arXiv:1207.6378}}.

\bibitem{decadalPlan}
PHENIX Collaboration.
\newblock {The PHENIX Decadal Plan}, 2010.
\newblock URL:
  \url{http://www.phenix.bnl.gov/phenix/WWW/docs/decadal/2010/phenix_decadal10%
_full_refs.pdf}.

\bibitem{Accardi:2012hwp}
A.~Accardi et~al.
\newblock {Electron Ion Collider: The Next QCD Frontier - Understanding the
  glue that binds us all}.
\newblock 2012.
\newblock \href {http://arxiv.org/abs/1212.1701} {\path{arXiv:1212.1701}}.

\bibitem{NSAC_LRP:2007}
{2007 Long Range Plan: The Frontiers of Nuclear Science}, 2007.
\newblock URL:
  \url{http://science.energy.gov/~/media/np/nsac/pdf/docs/NuclearScienceHighRe%
s.pdf}.

\bibitem{deFlorian:2009vb}
D.~de~Florian, R.~Sassot, M.~Stratmann, and W.~Vogelsang.
\newblock {Extraction of Spin-Dependent Parton Densities and Their
  Uncertainties}.
\newblock {\em Phys.~Rev.}, {\bf D}80:034030, 2009.
\newblock \href {http://arxiv.org/abs/0904.3821} {\path{arXiv:0904.3821}},
  \href {http://dx.doi.org/10.1103/PhysRevD.80.034030}
  {\path{doi:10.1103/PhysRevD.80.034030}}.

\bibitem{Anselmino:2007fs}
M.~Anselmino et~al.
\newblock {Transversity and Collins functions from SIDIS and $e^+e^-$ data}.
\newblock {\em Phys.~Rev.}, {\bf D}75:054032, 2007.
\newblock \href {http://arxiv.org/abs/hep-ph/0701006}
  {\path{arXiv:hep-ph/0701006}}, \href
  {http://dx.doi.org/10.1103/PhysRevD.75.054032}
  {\path{doi:10.1103/PhysRevD.75.054032}}.

\bibitem{Ji:1996ek}
X.-D. Ji.
\newblock {Gauge invariant decomposition of nucleon spin and its spin - off}.
\newblock {\em Phys.~Rev.~Lett.}, 78:610--613, 1997.
\newblock \href {http://arxiv.org/abs/hep-ph/9603249}
  {\path{arXiv:hep-ph/9603249}}, \href
  {http://dx.doi.org/10.1103/PhysRevLett.78.610}
  {\path{doi:10.1103/PhysRevLett.78.610}}.

\bibitem{Airapetian:2007vu}
A.~Airapetian et~al.
\newblock {Hadronization in semi-inclusive deep-inelastic scattering on
  nuclei}.
\newblock {\em Nucl.~Phys.}, B780:1--27, 2007.
\newblock \href {http://arxiv.org/abs/0704.3270} {\path{arXiv:0704.3270}},
  \href {http://dx.doi.org/10.1016/j.nuclphysb.2007.06.004}
  {\path{doi:10.1016/j.nuclphysb.2007.06.004}}.

\bibitem{Boer:2011fh}
D.~Boer et~al.
\newblock {Gluons and the quark sea at high energies: Distributions,
  polarization, tomography}.
\newblock 2011.
\newblock \href {http://arxiv.org/abs/1108.1713} {\path{arXiv:1108.1713}}.

\bibitem{Daniel:2011nq}
A.~Daniel et~al.
\newblock {Measurement of the nuclear multiplicity ratio for $K^0_s$
  hadronization at CLAS}.
\newblock {\em Phys.~Lett.}, B706:26--31, 2011.
\newblock \href {http://arxiv.org/abs/1111.2573} {\path{arXiv:1111.2573}},
  \href {http://dx.doi.org/10.1016/j.physletb.2011.10.071}
  {\path{doi:10.1016/j.physletb.2011.10.071}}.

\bibitem{Iancu:2002tr}
E.~Iancu, K.~Itakura, and L.~McLerran.
\newblock {Geometric scaling above the saturation scale}.
\newblock {\em Nucl.~Phys.}, A708:327--352, 2002.
\newblock \href {http://arxiv.org/abs/hep-ph/0203137}
  {\path{arXiv:hep-ph/0203137}}, \href
  {http://dx.doi.org/10.1016/S0375-9474(02)01010-2}
  {\path{doi:10.1016/S0375-9474(02)01010-2}}.

\bibitem{EIC-TF}
{eRHIC Task Force wiki}.
\newblock \url{https://wiki.bnl.gov/eic/index.php/Main_Page}.

\bibitem{Sauli:1997qp}
F.~Sauli.
\newblock {GEM: A new concept for electron amplification in gas detectors}.
\newblock {\em Nucl.~Instrum.~Meth.}, A386:531--534, 1997.
\newblock \href {http://dx.doi.org/10.1016/S0168-9002(96)01172-2}
  {\path{doi:10.1016/S0168-9002(96)01172-2}}.

\bibitem{Abbon:2007pq}
P.~Abbon et~al.
\newblock {The COMPASS experiment at CERN}.
\newblock {\em Nucl.~Instrum.~Meth.}, A577:455--518, 2007.
\newblock \href {http://arxiv.org/abs/hep-ex/0703049}
  {\path{arXiv:hep-ex/0703049}}, \href
  {http://dx.doi.org/10.1016/j.nima.2007.03.026}
  {\path{doi:10.1016/j.nima.2007.03.026}}.

\bibitem{Bell:1999vf}
R.~Bell et~al.
\newblock {The BaBar superconducting coil: Design, construction and test}.
\newblock {\em Nucl.~Phys.~Proc.~Suppl.}, 78:559--564, 1999.
\newblock \href {http://dx.doi.org/10.1016/S0920-5632(99)00603-9}
  {\path{doi:10.1016/S0920-5632(99)00603-9}}.

\bibitem{An:2008zzc}
S.~An et~al.
\newblock {A 20-ps timing device: A Multigap Resistive Plate Chamber with 24
  gas gaps}.
\newblock {\em Nucl.~Instrum.~Meth.}, A594:39--43, 2008.
\newblock \href {http://dx.doi.org/10.1016/j.nima.2008.06.013}
  {\path{doi:10.1016/j.nima.2008.06.013}}.

\bibitem{Adams:2013}
B.~Adams et~al.
\newblock {Measurements of the Gain, Time Resolution, and Spatial Resolution of
  a 20x20cm MCP-based Picosecond Photo-Detector}.
\newblock {\em Proceedings of the Vienna Conference on Instrumentation}, 2013.
\newblock URL:
  \url{http://psec.uchicago.edu/library/doclib/documents/222/sendit}.

\bibitem{Geronimo:2005}
B.~Yu et~al.
\newblock A gem based tpc for the legs experiment.
\newblock In {\em Nuclear Science Symposium Conference Record, 2005 IEEE},
  volume~2, pages 924--928, 2005.
\newblock \href {http://dx.doi.org/10.1109/NSSMIC.2005.1596405}
  {\path{doi:10.1109/NSSMIC.2005.1596405}}.

\bibitem{ALICE_TPC:2012}
L.~Musa et~al.
\newblock {Letter of intent for the upgrade of the alice experiment}.
\newblock Technical Report LHCC-I-022, CERN, 2012.

\bibitem{Ketzer:2013laa}
B.~Ketzer.
\newblock {A Time Projection Chamber for High-Rate Experiments: Towards an
  Upgrade of the ALICE TPC}.
\newblock 2013.
\newblock \href {http://arxiv.org/abs/1303.6694} {\path{arXiv:1303.6694}}.

\bibitem{Abe:2010aa}
T.~Abe et~al.
\newblock {The International Large Detector: Letter of Intent}.
\newblock 2010.
\newblock \href {http://arxiv.org/abs/1006.3396} {\path{arXiv:1006.3396}}.

\bibitem{Schade:2011zz}
P.~Schade and J.~Kaminski.
\newblock {A large TPC prototype for a linear collider detector}.
\newblock {\em Nucl.~Instrum.~Meth.}, A628:128--132, 2011.
\newblock \href {http://dx.doi.org/10.1016/j.nima.2010.06.300}
  {\path{doi:10.1016/j.nima.2010.06.300}}.

\bibitem{EIC_Tracking_RD:2013}
C.~Woody.
\newblock {Future Applications of GEM Detectors at BNL}.
\newblock Technical report, 2013.
\newblock Talk on RD51 Collaboration Meeting.

\bibitem{CMS_GEM:2012}
D.~Abbaneo et~al.
\newblock {Technical Proposal A GEM Detector System for an Upgrade of the CMS
  Muon Endcaps}.
\newblock Technical report, 2012.

\bibitem{PANDA:2008uqa}
{Technical Design Report for PANDA Electromagnetic Calorimeter (EMC)}.
\newblock 2008.
\newblock \href {http://arxiv.org/abs/0810.1216} {\path{arXiv:0810.1216}}.

\bibitem{Adam:2004fq}
I.~Adam et~al.
\newblock {The DIRC particle identification system for the BaBar experiment}.
\newblock {\em Nucl.~Instrum.~Meth.}, A538:281--357, 2005.
\newblock \href {http://dx.doi.org/10.1016/j.nima.2004.08.129}
  {\path{doi:10.1016/j.nima.2004.08.129}}.

\bibitem{Anderson:2011jw}
W.~Anderson et~al.
\newblock {Design, Construction, Operation and Performance of a Hadron Blind
  Detector for the PHENIX Experiment}.
\newblock {\em Nucl.~Instrum.~Meth.}, A646:35--58, 2011.
\newblock \href {http://arxiv.org/abs/1103.4277} {\path{arXiv:1103.4277}},
  \href {http://dx.doi.org/10.1016/j.nima.2011.04.015}
  {\path{doi:10.1016/j.nima.2011.04.015}}.

\bibitem{Iijima:2005qu}
T.~Iijima et~al.
\newblock {A Novel type of proximity focusing RICH counter with multiple
  refractive index aerogel radiator}.
\newblock {\em Nucl.~Instrum.~Meth.}, A548:383--390, 2005.
\newblock \href {http://arxiv.org/abs/physics/0504220}
  {\path{arXiv:physics/0504220}}, \href
  {http://dx.doi.org/10.1016/j.nima.2005.05.030}
  {\path{doi:10.1016/j.nima.2005.05.030}}.

\bibitem{Grauges:2010fi}
E.~Grauges et~al.
\newblock {SuperB Progress Reports -- Detector}.
\newblock 2010.
\newblock \href {http://arxiv.org/abs/1007.4241} {\path{arXiv:1007.4241}}.

\bibitem{Trbojevic:2013}
D.~Trbojevic, 2013.
\newblock Private communication.

\bibitem{Adler:2000bd}
C.~Adler et~al.
\newblock {The RHIC zero degree calorimeter}.
\newblock {\em Nucl.~Instrum.~Meth.}, A470:488--499, 2001.
\newblock \href {http://arxiv.org/abs/nucl-ex/0008005}
  {\path{arXiv:nucl-ex/0008005}}, \href
  {http://dx.doi.org/10.1016/S0168-9002(01)00627-1}
  {\path{doi:10.1016/S0168-9002(01)00627-1}}.

\bibitem{Adamczyk:2012kn}
L.~Adamczyk et~al.
\newblock {Single Spin Asymmetry $A_N$ in Polarized Proton-Proton Elastic
  Scattering at $\sqrt{s}=200$ GeV}.
\newblock {\em Phys.~Lett.}, B719:62--69, 2013.
\newblock \href {http://arxiv.org/abs/1206.1928} {\path{arXiv:1206.1928}},
  \href {http://dx.doi.org/10.1016/j.physletb.2013.01.014}
  {\path{doi:10.1016/j.physletb.2013.01.014}}.

\end{thebibliography}

\end{document}